\documentstyle{report}

\def\emline#1#2#3#4#5#6{%
       \put(#1,#2){\special{em:moveto}}%
       \put(#4,#5){\special{em:lineto}}}
\def\newpic#1{}

\newcommand{\bmat}{\left(\begin{array}}
\newcommand{\emat}{\end{array}\right)}
\newcommand{\be}{\begin{equation}}
\newcommand{\ee}{\end{equation}}
\newcommand{\ba}{\begin{array}}
\newcommand{\ea}{\end{array}}
\newtheorem{dfn}{Definition}[section]
\newtheorem{lemma}{Lemma}[section]
\newtheorem{remark}{Remark}[section]
\newtheorem{theorem}{Theorem}[section]
\makeatletter
\def\cl@chapter{}
\@addtoreset{equation}{section}

\def\thesection{\arabic{section}\protect\@blinkpoint}
\def\thesubsection{\arabic{section}.\arabic{subsection}\protect\@blinkpoint}
\def\thedfn{\arabic{section}.\arabic{dfn}\protect\@blinkpoint}
\def\thelemma{\arabic{section}.\arabic{lemma}\protect\@blinkpoint}
\def\theremark{\arabic{section}.\arabic{remark}\protect\@blinkpoint}
\def\thetheorem{\arabic{section}.\arabic{theorem}\protect\@blinkpoint}
\def\@blinkpoint{.}
\let\@blinkref=\ref
\def\ref#1{{\def\@blinkpoint{}\@blinkref{#1}}}
\long\def\@makecaption#1#2{%
   \vskip 10\p@
   \setbox\@tempboxa\hbox{#1. \ #2}%
   \ifdim \wd\@tempboxa >\hsize
       #1. \ #2\par
     \else
       \hbox to\hsize{\hfil\box\@tempboxa\hfil}%
   \fi}
\def\@begintheorem#1#2{\trivlist \item[\hskip \labelsep{\bf #1\ #2}]\sl}
\def\@opargbegintheorem#1#2#3{\trivlist
      \item[\hskip \labelsep{\bf #1\ #2\ (#3)}]\sl}

\def\cpages#1#2{\@ifundefined{r@#1}{\@warning
   {Reference `#1' on page \thepage \space
    undefined}}{\edef\@tempa{\@nameuse{r@#1}}\edef\@tempb{\expandafter
    \@cdr\@tempa\@nil}}\@ifundefined{r@#2}{\@warning
   {Reference `#2' on page \thepage \space
    undefined}}{\edef\@tempc{\@nameuse{r@#2}}\edef\@tempd{\expandafter
    \@cdr\@tempc\@nil}}\@ifundefined{r@#1}{{\reset@font\bf ?--?}}
    {\@ifundefined{r@#2}{{\reset@font\bf ?--?}}
    {\ifx\@tempb\@tempd\@tempb\else\@tempb--\@tempd\fi\null}}}

\def\vs{\vspace{3\unitlength}}
\def\hvs{\vspace{1.5\unitlength}}

\def\ombf{\mathchoice{\hbox{\boldmath$\displaystyle\omega$}}
{\hbox{\boldmath$\textstyle\omega$}}
{\hbox{\boldmath$\scriptstyle\omega$}}
{\hbox{\boldmath$\scriptscriptstyle\omega$}}}

\def\iint_#1{\intop\s@tiint@width\iint@indexshift=0pt
 \mathchoice{\global\iint@indexshift=1.6\iint@width}{}{}{}
 \iint@displaykern \kern-0.7\iint@width \intop_{#1\kern\iint@indexshift\null}}
\def\iint@displaykern{\mathchoice{\kern-0.7\iint@width}{}{}{}}
\newdimen\iint@width
\newdimen\iint@indexshift
\def\s@tiint@width{\mathchoice
 {\setbox0=\hbox{$\m@th\displaystyle{\intop}$} \global\iint@width=\wd0 }
 {\setbox0=\hbox{$\m@th\textstyle{\intop}$} \global\iint@width=\wd0 }
 {\setbox0=\hbox{$\m@th\scriptstyle{\intop}$} \global\iint@width=\wd0 }
 {\setbox0=\hbox{$\m@th\scriptscriptstyle{\intop}$} \global\iint@width=\wd0 }
 }

\makeatother

\newcommand{\bfig}{\begin{figure}}
\newcommand{\efig}{\end{figure}}

\begin{document}

\advance\hoffset-1.5cm \relax
\advance\voffset-3cm \relax
\setcounter{page}{0}
\thispagestyle{empty}
\begin{center}
\Large
Algebraic integrable dynamical systems,
2+1-dimensional models in wholly discrete space-time,
and inhomogeneous models in 2-dimensional statistical physics
\vspace{0.1\textheight}\\
\normalsize I.G. Korepanov,\\[\bigskipamount]
\small Chelyabinsk University of technology,\\
\small 76 Lenin prosp., Chelyabinsk 454080, Russia
\end{center}
\vspace{0.1\textheight}

\centerline{\bf Abstract}
\medskip

This paper is devoted to constructing and studying exactly solvable
dynamical systems in discrete time obtained from some algebraic
operations on matrices, to reductions of such systems leading to
classical field theory models in 2+1-dimen\-sional wholly discrete
space-time, and to connection between those field theories and
inhomogoneous models in 2-dimen\-sional statistical physics.

\clearpage
\tableofcontents
\clearpage
\section*{To the reader}
\addcontentsline{toc}{chapter}{To the reader}

This is an English version of my recent dissertation, or at least of its
substantial part. The Russian version contained also a long Introduction.
Here I decided to drop it: it seemed not very easy for me to write
an exciting enough Introduction in English. I hope that it will be clear
from the Contents and the text itself what this all is about.
I think also that it is enough to read a very short Section~\ref{1gensys}
in order to decide whether this paper is any interesting.
However, the Bibliography retains all references from the Russian text.

\bigskip\noindent

{\bf Acknoledgements.}
I owe to A.B.~Shabat the idea of ``local
reductions'', i.e.\ reductions to multidimensional field theories,
for dynamical systems of algebraic nature.
L.D.~Faddeev drew my attension to the work~\cite{BMV2} and urged me
to do more research in its direction.
V.V.~Sokolov gave me a
valuable consultation on dynamical systems associated with
orthogonal or symplectic matrices.
I express my deepest gratitude to them all.

\chapter{Dynamical system connected with the dimer model}

\section{Definition of the dynamical system. Gauge invariance}
\label{1gensys}

In this chapter we consider the following dynamical system in
discrete time.
Let
$$ L=\left(\ba{cc} A&B\\C&D\ea\right)$$
be a block matrix, $A, \ldots, D$ being $n\times n$ matrices consisting of
complex numbers. Consider the following two operations: construction of the
inverse matrix
$$L\rightarrow L^{-1}$$
and the block transposing
$$L=\left(\ba{cc} A&B\\C&D\ea\right)\rightarrow L^t=\left(\ba{cc}
A&C\\B&D\ea\right).$$
Now let a (birational) mapping $f$ be a composition of these two operations:
\be f(L)=(L^{-1})^t.\label{11.1}\ee
Let us introduce the discrete integer-valued time $\tau$, and let the matrix
$L$ depend on $\tau$ so that
\be L(\tau+1)=f(L(\tau)).\label{11.2}\ee

This dynamical system has been already mentioned in literature \cite{BMV2}.
In the present chapter, the integrability of this system is demonstrated,
assuming that the
``motion'' is considered up to a ``gauge transformation'' (see below).
We also demonstrate what conditions (reduction) must be imposed on the
matrix $L$ in order to obtain a meaningful evolution equation in
2+1-dimen\-sional space-time. This equation is a 2+1-dimen\-sional
version of Toda lattice in discrete time. The evolution is of hyperbolic
character: perturbations propagate not faster than fixed ``light speed''.
The solution to a Cauchy problem can be constructed in theta functions
according to a rather simple, in principle, scheme. The remarkable
property of the 2+1-dimen\-sional model is that its ``integrals of
motion'' are nothing else than a statistical sum of the well-known
flat dimer model (in our case, the statistical sum of dimer model
depends on two ``spectral parameters'').

Returning then back to the general dynamical system~(\ref{11.2}), we
consider its connections with a discrete analogue of the Lax $L,A$ pair
and possible generalizations coming therefrom.

\begin{dfn}\label{1gauge-dfn}
We will call the gauge
transformation of the matrix $L$  the following transformation of its
blocks:
\be
A\rightarrow GAH,\,\ldots, \,D\rightarrow GDH,
\label{11.3}
\ee
with $G$ and $H$  non-degenerate $n\times n$ matrices.
\end{dfn}

It is clear that the transformation~(\ref{11.3}) commutes with
the shift by two time units. Two matrices $L$ and $L'$ connected by the
transformation~(\ref{11.3}) will be
called gauge equivalent.
Thus, dynamics~(\ref{11.2}) induces a dynamics on the
set of gauge invariance classes of matrices $L$.

\section{Vacuum curves and vacuum vectors}
\label{1vacuum}

It turns out that the dynamics~(\ref{11.2}) preserves the so-called vacuum
curve $\Gamma$ of the operator $L$ (the bases being fixed, we make no
difference between a linear operator and its matrix). To be exact, $\Gamma$
remains unchanged under the transformation $f\circ f$, and undergoes a simple
transformation under $f$. The curve $\Gamma$ together with the class of linear
equivalence of the pole divisor of the vacuum vectors (see below) determines
the matrix $L$ up to a gauge transformation. The set of those classes of
linear equivalence is isomorphic to a complex torus---the Jacobian of the
curve $\Gamma$. The dynamics~(\ref{11.2}) linearizes on the Jacobian, i.e.\
the transformation $f$ corresponds to a constant shift on the torus. Now,
let us discuss these facts in detail.

The vacuum curve of the operator $L$ is an algebraic curve in the space
${\rm C}^2$ of two variables $u,v$. Here are two equivalent definitions of it
\cite{Krichever}.

\begin{dfn}
\label{1dfn1}
Consider the relation
\be L(U\otimes X)=V\otimes Y,\label{12.1}\ee
wherein
$$U=\left(\ba{c}u\\1\ea\right),\quad V=\left(\ba{c}v\\1\ea\right)$$
are two-dimensional vectors, $X$ and $\;Y$ are $n$-dimensional vectors. For a
generic matrix $L$, the non-zero solutions $(U,V,X,Y)$ of the
relation~(\ref{12.1}) are parametrized, up to a scalar factor in $X$ and $Y$,
by points of an algebraic curve $\Gamma$ of genus $g=(n-1)^2$ given by an
equation of the form
\be P(u,v)=0,\label{12.2}\ee
$P(u,v)$ being a polynomial of degree $n$ in each variable, i.e.
\be P(u,v)=\sum_{j,k=1}^n a_{jk}u^jv^k.\label{12.3}\ee
$\Gamma$ is called the vacuum curve of the operator $L$.
\end{dfn}

\begin{dfn}
\label{1dfn2}
The vacuum curve of the operator $L$ is the curve $\Gamma$ in ${\bf C}^2$
given by the equation
\be P(u,v)=\det(V^\perp LU)=\det(uA+B-uvC-vD)=0,\label{12.4}\ee
where
$$V^\perp=(1,-v).$$
\end{dfn}

Let us denote the points of the vacuum curve by the letter $z=(u,v)\in\Gamma$.
Then $U=U(z)$ and $V=V(z)$ are meromorphic vectors on $\Gamma$ with the pole
divisors $D_U$ and $D_V$ of degree $n$, while $X=X(z)$ and $Y=Y(z)$, if
normalized by, e.g.,\ the condition that their $n$th coordinates equal unity,
become meromorphic vectors with pole divisors $D_X$ and $D_Y$ of degree
$n^2-n$ \cite{Krichever}. Under this normalization, a meromorphic scalar
factor $h(z)$ must be added into~(\ref{12.1}):
\be L(U(z)\otimes X(z))=h(z)V(z)\otimes Y(z).\label{12.5}\ee
The linear equivalence of divisors
$$D_U+D_X\sim D_V+D_Y$$
holds and is provided by the function $h(z)$ in the sense that $h(z)$ has its
poles in the points of $D_U+D_X$ and zeros in the points of $D_V+D_Y$.

As is shown in the paper \cite{Krichever}, the vacuum curve equation $P(u,v)=
0$ and the class of linear equivalence of divisor $D_X$ or $D_Y$ determine a
generic matrix $L$ to within a gauge transformation, and vice versa, the
gauge transformations do not change the vacuum curve and the classes of linear
equivalence of divisors. In other words, the correspondence
$$(\mbox{class of gauge equivalence of }L)\leftrightarrow(\Gamma,\mbox{ the
class of }D_X)$$
is a birational isomorphism.

We will call $X(z)$ the vacuum vector and $Y(z)$ the covacuum vector in the
point $z$ of the curve $\Gamma$. $X(z)=X(u,v)$ generates the (one-dimensional)
kernel of the matrix
\be uA+B-uvC-vD.\label{12.6}\ee

The Definition~\ref{1dfn1} allows one to trace what happens with the vacuum
curve and vacuum vectors under the transformation $L\rightarrow L^{-1}$, while
the Definition~\ref{1dfn2} allows one to trace what happens under the
transformation $L\rightarrow L^t$. Namely, it is seen from the relation
$$L^{-1}\bigl(V(z)\otimes Y(z)\bigr)=h(z)^{-1}U(z)\otimes X(z)$$
that the vacuum curve equation for the matrix $L^{-1}$ is
$$P(v,u)=0,$$
while its vacuum vector in the point $(v,u)$ coincides with the covacuum
vector of the initial matrix $L$. As for the block transposing, the vacuum
curve equation for the matrix $L^t$
$$\det(uA+C-uvB-vD)=0$$
may be rewritten as
$$u^nv^n\det(v^{-1}A-B+u^{-1}v^{-1}C-u^{-1}D)=0,$$
i.e.
$$u^nv^nP(-v^{-1},-u^{-1})=0.$$
The vacuum vector of the matrix $L^t$ in the point $(-v^{-1},-u^{-1})$ of its
vacuum curve coincides with the vacuum vector $X(u,v)$ of the matrix $L$.

Combining these considerations, one finds out that the vacuum curve $\tilde
\Gamma$ of the matrix $(L^{-1})^t$ is given by equation
$$u^nv^nP(-u^{-1},-v^{-1})=0,$$
while the vacuum vector $\tilde X(-u^{-1},-v^{-1})$ coincides with the vector
$Y(u,v)$ of the matrix $L$.

Identifying the curves $\Gamma$ and $\tilde\Gamma$ by means of the isomorphism
$$(u,v)\leftrightarrow(-u^{-1},-v^{-1}),$$
one sees that
$$D_{\tilde X}\sim D_Y\sim D_X+D_U-D_V,$$
which means that, in essence, the transformation~(\ref{11.1}) results in adding
a fixed element of the Picard group, namely the equivalence class of the
divisor $D_U-D_V$, to the pole divisor $D_X$ of the vacuum vectors. It is
clear also that after two transformations one returns to the initial curve:
$$\tilde{\tilde\Gamma}=\Gamma.$$

\section{Reduction to evolution equation in the 2+1-dimensional space-time}
\label{1reduction}

The dynamical system of the previous section admits an interesting reduction,
i.e.\ some special choice of the matrices $A,\ldots, D$ that is in agreement
with the evolution. In this section, it will be convenient to treat the
matrices $A,\ldots, D$ as linear operators acting from the linear space
${\cal H}_1$ into the linear space ${\cal H}_2$ (of the same finite
dimension). This being the situation at the moment $\tau$, the operators act,
of course, from ${\cal H}_2$ into ${\cal H}_1$ at the moment $\tau+1$, and so
on.

Let each of the spaces ${\cal H}_1$, ${\cal H}_2$ be a direct sum of $lm/2$
identical subspaces of dimension $d$, where $l,m$ are even numbers. Let us
imagine these subspaces as situated
at the vertices of the square lattice on the torus
of the sizes $l\times m$ (which will mean the periodic boundary conditions in
both discrete space variables). Let the subspaces be arranged in checkerboard
fashion, as in Fig.~\ref{1Fig1}, where the empty circles correspond to
subspaces of the space ${\cal H}_1$, while the filled circles correspond to
those of the space ${\cal H}_2$.
\bfig
\begin{center}
\unitlength=0.209206\normalbaselineskip
\special{em:linewidth 1.00pt}
\linethickness{1.00pt}
\begin{picture}(80.00,80.00)
\put(20.00,20.00){\circle{4.00}}
\put(40.00,20.00){\circle*{4.00}}
\put(60.00,20.00){\circle{4.00}}
\put(60.00,40.00){\circle*{4.00}}
\put(40.00,40.00){\circle{4.00}}
\put(20.00,40.00){\circle*{4.00}}
\put(20.00,60.00){\circle{4.00}}
\put(40.00,60.00){\circle*{4.00}}
\put(60.00,60.00){\circle{4.00}}
\put(22.00,20.00){\vector(1,0){16.00}}
\put(20.00,22.00){\vector(0,1){16.00}}
\put(40.00,38.00){\vector(0,-1){16.00}}
\put(38.00,40.00){\vector(-1,0){16.00}}
\put(40.00,42.00){\vector(0,1){16.00}}
\put(42.00,40.00){\vector(1,0){16.00}}
\put(58.00,20.00){\vector(-1,0){16.00}}
\put(60.00,22.00){\vector(0,1){16.00}}
\put(60.00,58.00){\vector(0,-1){16.00}}
\put(58.00,60.00){\vector(-1,0){16.00}}
\put(22.00,60.00){\vector(1,0){16.00}}
\put(20.00,58.00){\vector(0,-1){16.00}}
\put(10.00,40.00){\vector(1,0){8.00}}
\put(40.00,70.00){\vector(0,-1){8.00}}
\put(70.00,40.00){\vector(-1,0){8.00}}
\put(40.00,10.00){\vector(0,1){8.00}}
\emline{20.00}{18.00}{1}{20.00}{10.00}{2}
\emline{18.00}{20.00}{3}{10.00}{20.00}{4}
\emline{10.00}{60.00}{5}{18.00}{60.00}{6}
\emline{20.00}{62.00}{7}{20.00}{70.00}{8}
\emline{60.00}{70.00}{9}{60.00}{62.00}{10}
\emline{62.00}{60.00}{11}{70.00}{60.00}{12}
\emline{70.00}{20.00}{13}{62.00}{20.00}{14}
\emline{60.00}{18.00}{15}{60.00}{10.00}{16}
\linethickness{0.40pt}
\put(0.00,0.00){\vector(1,0){80.00}}
\put(0.00,0.00){\vector(0,1){80.00}}
\put(30.00,61.00){\makebox(0,0)[cb]{\large$A$}}
\put(50.00,59.00){\makebox(0,0)[ct]{\large$D$}}
\put(19.00,50.00){\makebox(0,0)[rc]{\large$C$}}
\put(59.00,50.00){\makebox(0,0)[rc]{\large$C$}}
\put(30.00,39.00){\makebox(0,0)[ct]{\large$D$}}
\put(50.00,41.00){\makebox(0,0)[cb]{\large$A$}}
\put(61.00,30.00){\makebox(0,0)[lc]{\large$B$}}
\put(30.00,21.00){\makebox(0,0)[cb]{\large$A$}}
\put(50.00,19.00){\makebox(0,0)[ct]{\large$D$}}
\put(2.00,80.00){\makebox(0,0)[lc]{$\eta$}}
\put(80.00,2.00){\makebox(0,0)[cb]{$\xi$}}
\put(39.00,30.00){\makebox(0,0)[rc]{\large$C$}}
\put(21.00,30.00){\makebox(0,0)[lc]{\large$B$}}
\put(41.00,50.00){\makebox(0,0)[lc]{\large$B$}}
\special{em:linewidth 0.10pt}
\emline{23.00}{59.00}{17}{21.00}{57.00}{18}
\emline{21.00}{55.00}{19}{25.00}{59.00}{20}
\emline{27.00}{59.00}{21}{21.00}{53.00}{22}
\emline{29.00}{59.00}{23}{21.00}{51.00}{24}
\emline{21.00}{49.00}{25}{31.00}{59.00}{26}
\emline{33.00}{59.00}{27}{21.00}{47.00}{28}
\emline{21.00}{45.00}{29}{35.00}{59.00}{30}
\emline{37.00}{59.00}{31}{21.00}{43.00}{32}
\emline{22.00}{42.00}{33}{38.00}{58.00}{34}
\emline{39.00}{57.00}{35}{23.00}{41.00}{36}
\emline{25.00}{41.00}{37}{39.00}{55.00}{38}
\emline{39.00}{53.00}{39}{27.00}{41.00}{40}
\emline{29.00}{41.00}{41}{39.00}{51.00}{42}
\emline{39.00}{49.00}{43}{31.00}{41.00}{44}
\emline{33.00}{41.00}{45}{39.00}{47.00}{46}
\emline{39.00}{45.00}{47}{35.00}{41.00}{48}
\emline{37.00}{41.00}{49}{39.00}{43.00}{50}
\emline{43.00}{39.00}{51}{41.00}{37.00}{52}
\emline{41.00}{35.00}{53}{45.00}{39.00}{54}
\emline{47.00}{39.00}{55}{41.00}{33.00}{56}
\emline{49.00}{39.00}{57}{41.00}{31.00}{58}
\emline{41.00}{29.00}{59}{51.00}{39.00}{60}
\emline{53.00}{39.00}{61}{41.00}{27.00}{62}
\emline{41.00}{25.00}{63}{55.00}{39.00}{64}
\emline{57.00}{39.00}{65}{41.00}{23.00}{66}
\emline{42.00}{22.00}{67}{58.00}{38.00}{68}
\emline{59.00}{37.00}{69}{43.00}{21.00}{70}
\emline{45.00}{21.00}{71}{59.00}{35.00}{72}
\emline{59.00}{33.00}{73}{47.00}{21.00}{74}
\emline{49.00}{21.00}{75}{59.00}{31.00}{76}
\emline{59.00}{29.00}{77}{51.00}{21.00}{78}
\emline{53.00}{21.00}{79}{59.00}{27.00}{80}
\emline{59.00}{25.00}{81}{55.00}{21.00}{82}
\emline{57.00}{21.00}{83}{59.00}{23.00}{84}
\emline{17.00}{61.00}{85}{19.00}{63.00}{86}
\emline{19.00}{65.00}{87}{15.00}{61.00}{88}
\emline{13.00}{61.00}{89}{19.00}{67.00}{90}
\emline{19.00}{69.00}{91}{11.00}{61.00}{92}
\emline{10.00}{62.00}{93}{18.00}{70.00}{94}
\emline{16.00}{70.00}{95}{10.00}{64.00}{96}
\emline{10.00}{66.00}{97}{14.00}{70.00}{98}
\emline{12.00}{70.00}{99}{10.00}{68.00}{100}
\emline{68.00}{10.00}{101}{70.00}{12.00}{102}
\emline{70.00}{14.00}{103}{66.00}{10.00}{104}
\emline{64.00}{10.00}{105}{70.00}{16.00}{106}
\emline{70.00}{18.00}{107}{62.00}{10.00}{108}
\emline{61.00}{11.00}{109}{69.00}{19.00}{110}
\emline{67.00}{19.00}{111}{61.00}{13.00}{112}
\emline{61.00}{15.00}{113}{65.00}{19.00}{114}
\emline{63.00}{19.00}{115}{61.00}{17.00}{116}
\emline{42.00}{62.00}{117}{50.00}{70.00}{118}
\emline{48.00}{70.00}{119}{41.00}{63.00}{120}
\emline{41.00}{65.00}{121}{46.00}{70.00}{122}
\emline{44.00}{70.00}{123}{41.00}{67.00}{124}
\emline{41.00}{69.00}{125}{42.00}{70.00}{126}
\emline{52.00}{70.00}{127}{43.00}{61.00}{128}
\emline{45.00}{61.00}{129}{54.00}{70.00}{130}
\emline{56.00}{70.00}{131}{47.00}{61.00}{132}
\emline{49.00}{61.00}{133}{58.00}{70.00}{134}
\emline{59.00}{69.00}{135}{51.00}{61.00}{136}
\emline{53.00}{61.00}{137}{59.00}{67.00}{138}
\emline{59.00}{65.00}{139}{55.00}{61.00}{140}
\emline{57.00}{61.00}{141}{59.00}{63.00}{142}
\emline{39.00}{11.00}{143}{38.00}{10.00}{144}
\emline{39.00}{13.00}{145}{36.00}{10.00}{146}
\emline{34.00}{10.00}{147}{39.00}{15.00}{148}
\emline{39.00}{17.00}{149}{32.00}{10.00}{150}
\emline{30.00}{10.00}{151}{38.00}{18.00}{152}
\emline{37.00}{19.00}{153}{28.00}{10.00}{154}
\emline{26.00}{10.00}{155}{35.00}{19.00}{156}
\emline{33.00}{19.00}{157}{24.00}{10.00}{158}
\emline{22.00}{10.00}{159}{31.00}{19.00}{160}
\emline{29.00}{19.00}{161}{21.00}{11.00}{162}
\emline{21.00}{13.00}{163}{27.00}{19.00}{164}
\emline{25.00}{19.00}{165}{21.00}{15.00}{166}
\emline{21.00}{17.00}{167}{23.00}{19.00}{168}
\emline{62.00}{42.00}{169}{70.00}{50.00}{170}
\emline{70.00}{48.00}{171}{63.00}{41.00}{172}
\emline{65.00}{41.00}{173}{70.00}{46.00}{174}
\emline{70.00}{44.00}{175}{67.00}{41.00}{176}
\emline{69.00}{41.00}{177}{70.00}{42.00}{178}
\emline{61.00}{43.00}{179}{70.00}{52.00}{180}
\emline{70.00}{54.00}{181}{61.00}{45.00}{182}
\emline{61.00}{47.00}{183}{70.00}{56.00}{184}
\emline{70.00}{58.00}{185}{61.00}{49.00}{186}
\emline{61.00}{51.00}{187}{69.00}{59.00}{188}
\emline{67.00}{59.00}{189}{61.00}{53.00}{190}
\emline{61.00}{55.00}{191}{65.00}{59.00}{192}
\emline{63.00}{59.00}{193}{61.00}{57.00}{194}
\emline{18.00}{38.00}{195}{10.00}{30.00}{196}
\emline{10.00}{32.00}{197}{17.00}{39.00}{198}
\emline{15.00}{39.00}{199}{10.00}{34.00}{200}
\emline{10.00}{36.00}{201}{13.00}{39.00}{202}
\emline{11.00}{39.00}{203}{10.00}{38.00}{204}
\emline{10.00}{28.00}{205}{19.00}{37.00}{206}
\emline{19.00}{35.00}{207}{10.00}{26.00}{208}
\emline{10.00}{24.00}{209}{19.00}{33.00}{210}
\emline{19.00}{31.00}{211}{10.00}{22.00}{212}
\emline{11.00}{21.00}{213}{19.00}{29.00}{214}
\emline{19.00}{27.00}{215}{13.00}{21.00}{216}
\emline{15.00}{21.00}{217}{19.00}{25.00}{218}
\emline{19.00}{23.00}{219}{17.00}{21.00}{220}
\end{picture}
\end{center}
\caption{Integrable dynamical system in the 2+1-dimen\-sional space-time}
\label{1Fig1}
\efig

Let then the operators $A,\ldots, D$ be such that the image of each of the
mentioned $d$-dimensional subspaces with respect to, say, operator $A$ lies in
the $d$-dimensional subspace of ${\cal H}_2$ at which points the arrow marked
``$A$'' that links these two subspaces (Fig.~\ref{1Fig1}). Analogously, the
restrictions on $B$, $C$, $D$ are depicted in Fig.~\ref{1Fig1} (see also
formula~(\ref{13.8}) for non-degenerate $A,\ldots, D$). Thus, to each link of
the lattice a $d\times d$ matrix is attached that is a block of one of the
``large'' matrices $A,\ldots, D$. Let us shade half of the squares of the
lattice in a checkerboard way, as in Fig.~\ref{1Fig1}. One can verify that the
evolution of the system may be described as follows.

At the first step, each of the four $d\times d$ matrices that correspond to
the arrows surrounding each shaded square is transformed into a matrix
expressed through just these four matrices. This goes according to the
following formulae, in which the $d\times d$ blocks are somewhat freely
denoted by the same letters $A,\ldots, D$ as the ``large'' matrices:
\be A\longrightarrow(A-BD^{-1}C)^{-1},\label{13.1a}\ee
\be B\longrightarrow(B-AC^{-1}D)^{-1},\label{13.1b}\ee
\be C\longrightarrow(C-DB^{-1}A)^{-1},\label{13.1c}\ee
\be D\longrightarrow(D-CA^{-1}B)^{-1}.\label{13.1d}\ee
However, the formulae~(\ref{13.1a}--\ref{13.1d}) apply equally to the ``large''
matrices.

After the transformation~(\ref{13.1a}--\ref{13.1d}),
all the arrows reverse, and
at the second step the non-shaded squares are engaged in the same way
according to
the same formulae~(\ref{13.1a}--\ref{13.1d}).
Then everything is repeated. Thus,
the evolution is of hyperbolic nature: each local perturbation spreads not
faster than one unit of length per unit of time.

Let us clarify the symmetries of vacuum curves and divisors $D_X$ in this
``reduced'' model. Let us introduce two integer-valued coordinates $\xi,\eta$
for the vertices of the lattice, so that $\xi$ increases by 1 in passing from
a vertex one step to the right, and $\eta$ increases by 1 in passing one step
upwards. $\xi$ and $\eta$ are defined modulo $l$ and $m$ respectively. A
$d$-dimentional subspace of ${\cal H}_1$ or ${\cal H}_2$ will be denoted
${\cal H}_{\xi\eta}$ if it corresponds to a vertex with coordinates
$\xi,\eta$. Consider a linear transformation in spaces ${\cal H}_1$ and
${\cal H}_2$ consisting in multiplying the vectors of each subspace
${\cal H}_{\xi\eta}$ by $\omega^\xi_1,\quad$ $\omega_1$ being a fixed
primitive root of the $l$-th degree of unity:
$$
\omega_1^l=1.
$$
This corresponds to the following transformation of the operators
$A,\ldots, D$ (from now on we speak of each of these operators ``as a whole'',
not of their blocks):
\be
A\rightarrow\omega_1A,\quad B\rightarrow B,\quad C\rightarrow C,
\quad D\rightarrow
\omega_1^{-1}D.
\label{13.2}
\ee
Consider also another linear transformation in ${\cal H}_1$ and ${\cal H}_2$,
consisting in multiplying the vectors of each subspace ${\cal H}_{\xi\eta}$
by $\omega_2^\eta,\;$ $\omega_2$ being a fixed primitive root of the $m$-th
degree of unity:
$$
\omega_2^m=1.
$$
This corresponds to the following transformation:
\be
A\rightarrow A,\quad B\rightarrow\omega_2B,\quad C\rightarrow\omega_2^{-1}C,
\quad D\rightarrow D.
\label{13.3}
\ee

The vacuum curve of the operator $L$, which is given by equation~(\ref{12.4})
$$
P(u,v)=\det(uA+B-uvC-vD)=0,
$$
must be invariant under the transformations~(\ref{13.2}), (\ref{13.3}). This
leads to the invariance of the polynomial $P(u,v)$ with respect to the
following transformations $g_1$ and $g_2$:
\be
 g_1(u,v)=(\omega_1u,\omega_1^{-1}v),
\label{13.4}
\ee
\be
g_2(u,v)=(\omega_2^{-1}u,\omega_2^{-1}v).
\label{13.5}
\ee
This invariance, then, leads to the following statement: only those
coefficients $a_{jk}$ are non-zero in the vacuum curve equation
(see~(\ref{12.2}), (\ref{12.3})) for the ``reduced'' model, for which
\be
\left. \ba{l}j-k\equiv0({\rm mod}\;l),\smallskip\\j+k\equiv0({\rm mod}\;m).
\ea\right\}
\label{13.6}
\ee

As for the divisor $D_X$, let us recall that it consists of such points in the
curve $\Gamma$ in which vanishes the last coordinate of the vector $X$ (see
\cite{Krichever}), the latter being an eigenvector of the matrix~(\ref{12.6})
with zero eigenvalue:
\be
(uA+B-uvC-vD)X(u,v)=0.
\label{13.7}
\ee
This immediately leads to the conclusion: the divisor $D_X$ is invariant with
respect to the transformations~(\ref{13.4},~\ref{13.5}).

Under some additional condition, the inverse statement also holds: if the
curve $\Gamma$ and divisor $D_X$ are invariant under the
transformations~(\ref{13.4}), (\ref{13.5}), then the corresponding $L$-operator
comes from a ``reduced'' model described in this section. For instance, this
is true if $l/2$ and $m/2$ are relatively prime numbers. If these numbers are
not relatively prime, some conditions are to be imposed on the divisor $D_X$.
To avoid going into details of this latter case, let us not consider it here.

Thus, let an operator $L=\left(\!\ba{cc}A&B\\C&D\ea\!\right)$ be given, $A,
\ldots,
D$ being $n\times n$ matrices, $n=(lm/2)d,\quad l$ and $m$ even, and $l/2$ and
$m/2$ being relatively prime. Let the vacuum curve $\Gamma$ of the operator
$L$ and the divisor $D_X$ be invariant under the action of the group $\cal G$
generated by its elements $g_1,g_2$~(\ref{13.4},~\ref{13.5}), $\;\omega_1$ and
$\omega_2$ being primitive roots of degrees $l$ and $m$ of unity. Then the
linear space in which operators $A,\ldots, D$ act decomposes into a direct sum
of $lm/2\;\;$ $d$-dimensional subspaces ${\cal H}_{\xi\eta},\;\;$ $\xi$ and
$\eta$ being integers modulo $l$ and $m$ respectively and such that $\xi+\eta$
is an even number, and the following equalities between the images of these
subspaces hold (in a ``generic'' case of non-degenerate $A,\ldots, D$):
\be
A{\cal H}_{\xi-1,\eta+1}=B{\cal H}_{\xi\eta}=C{\cal H}_{\xi,\eta+2}=
D{\cal H}_{\xi+1,\eta+1}.
\label{13.8}
\ee
The equalities~(\ref{13.8}) mean exactly that one is in the situation of
Fig.~\ref{1Fig1}.

Let us prove the above statements. First, the natural projection from the
curve $\Gamma$ to its factor $\Gamma/\cal G$ has no branch points (here the
fact that $l/2$ and $m/2$ are relatively prime is used to demonstrate that
ramification does not occur when $u$ or $v$ equals zero or infinity). Thus,
the $n$-dimensional linear space of meromorphic functions $x(z)=x(u,v)$ whose
pole divisor is $D_X$ decomposes into a direct sum of subspaces of equal
dimensions corresponding to the characters of (commutative) group $\cal G$.
Each of these subspaces consists of functions $x(z)$ satisfying relations
$$
x(gz)=\chi_{\xi\eta}(g)x(z),
$$
the character $\chi_{\xi\eta}$ being a scalar factor
$$
\chi_{\xi\eta}(g)=\omega_1^{\xi a}\omega_2^{\eta b},
$$
where
$$
g=g_1^ag_2^b.
$$
The equality $g_1^{l/2}g_2^{m/2}=1$ means that $\xi+\eta$ must be an even
number.

The components of the vector $X(z)$ are exactly the functions $x(z)$. In an
appropriate basis, $d$ components correspond to each character
$\chi_{\xi\eta}$. Let us denote ${\cal H}_{\xi\eta}$ the set of vectors with
other components equal to zero. Now, the equalities~(\ref{13.8}) are to be
proved to end this section.

Consider the decomposition of vector $X(u,v)$ into a sum
$$
X(u,v)=\sum_{\xi,\eta}X_{\xi,\eta}(u,v),
$$
where $X_{\xi,\eta}\in{\cal H}_{\xi,\eta}$. Then
$$
X_{\xi,\eta}\bigl( g(u,v)\bigr)=\chi_{\xi,\eta}(g)X_{\xi,\eta}(u,v).
$$
Consider the sum
\be
 \sum_{g\in{\cal G}}\chi_{\xi,\eta}(g^{-1})g\{(uA+B-uvC-vD)X(u,v)\}=0
\label{13.9}
\ee
(which is equal to zero because of~(\ref{13.7})). The action of $g$ upon the
braces in~(\ref{13.9}) means that each $u$ and $v$ in the braces is
transformed according to~(\ref{13.4}), (\ref{13.5}), i.e. $u$ changes into
$\chi_{1,-1}(g)u$, and $v$ changes into $\chi_{-1,-1}(g)v$. The
equality~(\ref{13.9}) gives thus, after cancelling a factor equal to
the order of group $\cal G$,
\be
uAX_{\xi-1,\eta+1}(u,v)+BX_{\xi,\eta}(u,v)-uvCX_{\xi,\eta+2}(u,v)-
vX_{\xi+1,\eta+1}(u,v)D=0.
\label{13.10}
\ee
Let us set $u=0$ in~(\ref{13.10}). Then $v$ can take $n$ different values $v_j$
satisfying relation $P(0,v_j)=0$. To these values $v_j$ correspond $d$
linearly independent vectors $X_{\xi\eta}(0,v_j)$, and also $d$ vectors
$X_{\xi+1,\eta+1}(0,v_j)$. Thus, the equalities
$$
BX_{\xi\eta}(0,v_j)=v_jDX_{\xi+1,\eta+1}(0,v_j)
$$
that result from~(\ref{13.10}) give
$$
B{\cal H}_{\xi\eta}=D{\cal H}_{\xi+1,\eta+1}.
$$
Analogously, one can as well obtain the rest of equalities~(\ref{13.8}).

\section{Connection to dimer model}\label{1dimers}

As has been demonstrated, the integrals of motion of the dynamical system
of Section~\ref{1gensys} and its reductions (if the {\em even} degrees
of the transformation~(\ref{11.1}) are considered)
are the coefficients $a_{jk}$ of the vacuum curve~(\ref{12.3}). These
coefficients are determined up to a common factor, so they may be divided by
$a_{00}$. As one can see, the resulting coefficients are those of the
polynomial
\be\det\bigl(1+uAB^{-1}-vDB^{-1}-uvCB^{-1}\bigr).\label{1determ}\ee
In other words, the determinant~(\ref{1determ}) is an integral of motion for
any $u,v$.

Let us turn now to the model from section~\ref{1reduction}, that is to the
model in 2+1-dimensional discrete space-time with periodic boundary
conditions, and let the dimension $d$ of the linear space corresponding to
each vertex be equal to 1. Each of the ``small'' matrices $A,B,C,D$
corresponding to the links will then be a single (depending on the link)
number $a,b,c$ or $d$. It is well known that the determinant of any
$N\times N$ matrix is a sum of its matrix elements products corresponding in
a certain way to the permutations of $N$ objects, while each permutation
decomposes into a product of the cyclic ones. In our situation, the cyclic
permutations correspond to the non-selfintersecting closed paths (contours)
going along the arrows of the following diagram~(Fig.~\ref{1arrows}) (thus,
general permutations correspond to the sets of non-intersecting paths).
To each closed path corresponds the product of the weights $ua,-uvc,-vd,
b^{-1}$ on its links, and, to get right signs for the terms of which the
determinant~(\ref{1determ}) is made up, one should add a minus sign to each
such product containing an {\em even} number of the factors $b^{-1}$.
\bfig
\begin{center}
\unitlength=0.209206\normalbaselineskip
\special{em:linewidth 0.4pt}
\linethickness{0.4pt}
\begin{picture}(57.00,42.00)
\put(21.00,6.00){\circle{2.00}}
\put(6.00,6.00){\circle*{2.00}}
\put(6.00,21.00){\circle{2.00}}
\put(36.00,21.00){\circle{2.00}}
\put(21.00,21.00){\circle*{2.00}}
\put(36.00,6.00){\circle*{2.00}}
\put(21.00,36.00){\circle{2.00}}
\put(6.00,36.00){\circle*{2.00}}
\put(36.00,36.00){\circle*{2.00}}
\put(51.00,21.00){\circle*{2.00}}
\put(51.00,6.00){\circle{2.00}}
\put(51.00,36.00){\circle{2.00}}
\put(20.00,6.00){\vector(-1,0){13.00}}
\put(22.00,6.00){\vector(1,0){13.00}}
\put(50.00,6.00){\vector(-1,0){13.00}}
\put(21.00,20.00){\vector(0,-1){13.00}}
\put(6.00,20.00){\vector(0,-1){13.00}}
\put(36.00,20.00){\vector(0,-1){13.00}}
\put(51.00,20.00){\vector(0,-1){13.00}}
\put(6.00,35.00){\vector(0,-1){13.00}}
\put(21.00,35.00){\vector(0,-1){13.00}}
\put(51.00,35.00){\vector(0,-1){13.00}}
\put(7.00,21.00){\vector(1,0){13.00}}
\put(35.00,21.00){\vector(-1,0){13.00}}
\put(37.00,21.00){\vector(1,0){13.00}}
\put(50.00,36.00){\vector(-1,0){13.00}}
\put(22.00,36.00){\vector(1,0){13.00}}
\put(20.00,36.00){\vector(-1,0){13.00}}
\put(14.00,37.00){\makebox(0,0)[cb]{$-vd$}}
\put(28.00,37.00){\makebox(0,0)[cb]{$ua$}}
\put(44.00,37.00){\makebox(0,0)[cb]{$-vd$}}
\put(14.00,22.00){\makebox(0,0)[cb]{$ua$}}
\put(28.00,22.00){\makebox(0,0)[cb]{$-vd$}}
\put(44.00,22.00){\makebox(0,0)[cb]{$ua$}}
\put(44.00,7.00){\makebox(0,0)[cb]{$-vd$}}
\put(28.00,7.00){\makebox(0,0)[cb]{$ua$}}
\put(14.00,7.00){\makebox(0,0)[cb]{$-vd$}}
\put(7.00,29.00){\makebox(0,0)[lc]{$b^{-1}$}}
\put(22.00,29.00){\makebox(0,0)[lc]{$-uvc$}}
\put(37.00,29.00){\makebox(0,0)[lc]{$b^{-1}$}}
\put(52.00,29.00){\makebox(0,0)[lc]{$-uvc$}}
\put(52.00,14.00){\makebox(0,0)[lc]{$b^{-1}$}}
\put(37.00,14.00){\makebox(0,0)[lc]{$-uvc$}}
\put(22.00,14.00){\makebox(0,0)[lc]{$b^{-1}$}}
\put(7.00,14.00){\makebox(0,0)[lc]{$-uvc$}}
\put(6.00,42.00){\vector(0,-1){5.00}}
\put(21.00,42.00){\vector(0,-1){5.00}}
\put(36.00,42.00){\vector(0,-1){5.00}}
\put(51.00,42.00){\vector(0,-1){5.00}}
\put(0.00,36.00){\vector(1,0){5.00}}
\put(0.00,6.00){\vector(1,0){5.00}}
\put(57.00,21.00){\vector(-1,0){5.00}}
\emline{0.00}{21.00}{1}{5.00}{21.00}{2}
\emline{6.00}{5.00}{3}{6.00}{0.00}{4}
\emline{21.00}{0.00}{5}{21.00}{5.00}{6}
\emline{36.00}{5.00}{7}{36.00}{0.00}{8}
\emline{51.00}{0.00}{9}{51.00}{5.00}{10}
\emline{52.00}{6.00}{11}{57.00}{6.00}{12}
\emline{57.00}{36.00}{13}{52.00}{36.00}{14}
\put(36.00,35.00){\vector(0,-1){13.00}}
\end{picture}
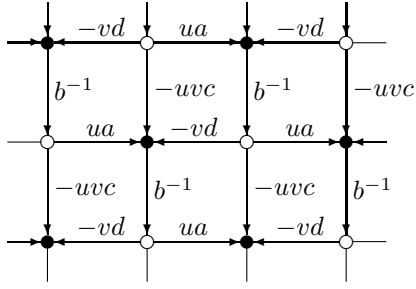
\caption{The ways along these arrows are connected with both the vacuum curve
and the dimer model}
\label{1arrows}
\end{center}
\efig

\begin{remark} \label{1r1}
Another way to obtain right signs is: to multiply each $b$ by $-1$ and then
multiply each product corresponding to a closed path (and containing
{\em any} number of $b$'s) by $-1$.
\end{remark}

It turns out that the determinant~(\ref{1determ}) is connected with the
statistical sum of the well known dimer model~\cite{dimersbook}.
Let us define the correspondence between the sets of paths and the dimer
configurations as follows. Let the empty set of paths correspond to the
``standard'' dimer configuration, the dimers being placed on the
``$B$-links'' (Fig.~\ref{1standard}). For a non-empty set of paths, let us
change the standard configuration along all the paths, replacing each dimer by
a free link and vice versa. One can verify that this is a bijective
correspondence.

The statistical sum being considered, let the weights $-b$ (not $b^{-1}$)
correspond to the ``$B$-links'', while to the other links correspond
the unchanged weights $ua,-vd,-uvc$. Then one can see that the statistical
sum, if multiplied by $\prod\nolimits_{\mbox{\footnotesize over all links}}
(-b^{-1})$
(let us call the result the normalized statistical sum), consists of the same
terms as the determinant~(\ref{1determ}), up to different signs of some of
them. Let us emphasize that the dimer model is, of course, {\em
inhomogeneous}: the weights $a,b,c,d$ are different for different links.

\bfig
\begin{center}
\unitlength=0.209206\normalbaselineskip
\special{em:linewidth 0.2pt}
\linethickness{0.2pt}
\begin{picture}(57.00,42.00)
\put(21.00,6.00){\circle{2.00}}
\put(6.00,21.00){\circle{2.00}}
\put(36.00,21.00){\circle{2.00}}
\put(21.00,21.00){\circle*{2.00}}
\put(36.00,6.00){\circle*{2.00}}
\put(21.00,36.00){\circle{2.00}}
\put(6.00,36.00){\circle*{2.00}}
\put(36.00,36.00){\circle*{2.00}}
\put(51.00,21.00){\circle*{2.00}}
\put(51.00,6.00){\circle{2.00}}
\put(51.00,36.00){\circle{2.00}}
\put(6.00,6.00){\circle*{2.00}}
\emline{7.00}{36.00}{1}{20.00}{36.00}{2}
\emline{22.00}{36.00}{3}{35.00}{36.00}{4}
\emline{37.00}{36.00}{5}{50.00}{36.00}{6}
\emline{52.00}{36.00}{7}{57.00}{36.00}{8}
\emline{0.00}{36.00}{9}{5.00}{36.00}{10}
\emline{5.00}{21.00}{11}{0.00}{21.00}{12}
\emline{0.00}{6.00}{13}{5.00}{6.00}{14}
\emline{7.00}{6.00}{15}{20.00}{6.00}{16}
\emline{22.00}{6.00}{17}{35.00}{6.00}{18}
\emline{37.00}{6.00}{19}{50.00}{6.00}{20}
\emline{52.00}{6.00}{21}{57.00}{6.00}{22}
\emline{57.00}{21.00}{23}{52.00}{21.00}{24}
\emline{50.00}{21.00}{25}{37.00}{21.00}{26}
\emline{35.00}{21.00}{27}{22.00}{21.00}{28}
\emline{20.00}{21.00}{29}{7.00}{21.00}{30}
\emline{6.00}{37.00}{31}{6.00}{42.00}{32}
\emline{36.00}{42.00}{33}{36.00}{37.00}{34}
\emline{21.00}{35.00}{35}{21.00}{22.00}{36}
\emline{51.00}{22.00}{37}{51.00}{35.00}{38}
\emline{36.00}{20.00}{39}{36.00}{7.00}{40}
\emline{6.00}{7.00}{41}{6.00}{20.00}{42}
\emline{21.00}{5.00}{43}{21.00}{0.00}{44}
\emline{51.00}{0.00}{45}{51.00}{5.00}{46}
\special{em:linewidth 2pt}
\linethickness{2pt}
\emline{6.00}{5.00}{47}{6.00}{0.00}{48}
\emline{21.00}{7.00}{49}{21.00}{20.00}{50}
\emline{6.00}{22.00}{51}{6.00}{35.00}{52}
\emline{36.00}{35.00}{53}{36.00}{22.00}{54}
\emline{21.00}{37.00}{55}{21.00}{42.00}{56}
\emline{51.00}{42.00}{57}{51.00}{37.00}{58}
\emline{51.00}{20.00}{59}{51.00}{7.00}{60}
\emline{36.00}{5.00}{61}{36.00}{0.00}{62}
\end{picture}
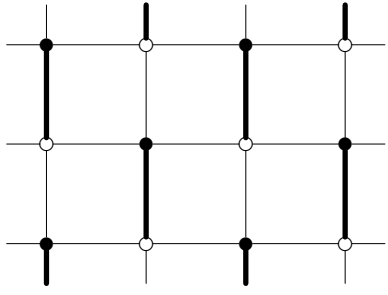
\caption{The standard dimer configuration}
\label{1standard}
\end{center}
\efig

Let us study these signs in detail. Note that the conditions of
non-inter\-secting and non-selfintersecting impose strong restrictions on the
possible path configurations. Every closed path on the torus is
homologically equivalent to a linear combination with integer coefficients of
two {\em basis cycles} $\bf a$ and $\bf b$ whose intersection number is~1
(I use the boldface font for cycles, because the letters $a,b\ldots$ are
already in use). If the torus is cut along a closed non-selfintersecting path
$\bf c$ not equivalent to zero, the result will be homeomorphic to the lateral
surface of a cylinder (this follows, e.g., from~\cite{rDNF}, chapter~1,
section~3). Then the contour~$\bf d$ going along a generatrix of the cylinder
in a properly chosen direction has the intersection number~1 with the
contour~$\bf c$. The intersection number being bilinear and
integer-valued, we find that if the contour~$\bf c$ is
homologically equivalent to a sum $l{\bf a}+m{\bf b}$, then $l$ and $m$
cannot have common divisors (not equal to~$\pm1$). Thus, the following lemma
is valid.

\begin{lemma} \label{1l2}
Every closed non-selfintersecting path on the torus is homologically
equivalent to a linear combination of the basis paths $\bf a$ and $\bf b$
with relatively prime integer coefficients.
\end{lemma}

Now let us pass to the case of several contours on the torus. If they do not
intersect, their intersection numbers equal~0 (of course) and thus their
homological classes must be proportional to one another. This together with
Lemma~\ref{1l2} leads to the following lemma.

\begin{lemma} \label{1l1}
Several closed non-intersecting and non-selfintersecting paths going along
the arrows on the torus, as in Fig.~\ref{1arrows}, are necessarily all
homologically equivalent to one another.
\end{lemma}

If two paths are homologically equivalent, then the terms of the same degrees
in $u$ and $v$ correspond to them (one can see in Fig.~\ref{1arrows} that the
different ways round an ``elementary square'' yield the same degrees of
$u$ and $v$). Let the basis paths $\bf a$ and $\bf b$ yield the terms
proportional to $x=u^{\alpha_1}v^{\beta_1},y=u^{\alpha_2}v^{\beta_2}$
correspondingly (with the factors of proportionality not depending on
$u,v$). According to Lemma~\ref{1l2}, the determinant~(\ref{1determ}) and the
statistical sum of the dimer model are polynomials in $x,y$. The following
lemma sums up this section.

\begin{lemma} \label{1l3}
Let $f(x,y)$ and $s(x,y)$ be the determinant~(\ref{1determ}) and the
normalized statistical sum of the dimer model considered as functions of
$x$ and $y$. Then
\be s(x,y)=\frac{1}{2}\bigl(-f(x,y)+f(-x,y)+f(x,-y)+f(-x,-y)\bigr),\label{1fs}
\ee
\be f(x,y)=\frac{1}{2}\bigl(-s(x,y)+s(-x,y)+s(x,-y)+s(-x,-y)\bigr).\label{1sf}
\ee
\end{lemma}

{\it Proof.} If the normalized statistical sum consists of the terms
$$
c_{jk}x^jy^k=c_{jk}(u^{\alpha_1}v^{\beta_1})^j(u^{\alpha_2}v^{\beta_2})^k,
$$
then the determinant consists of the same terms multiplied by
$$
(-1)^{\mbox{\footnotesize number of contours}}=(-1)^{\mbox{\footnotesize
g.c.d.}(j,k)}=(-1)^{jk+j+k}
$$
(here Remark~\ref{1r1} and Lemmas~\ref{1l2} and~\ref{1l1} are used).
This means
that the signs of all the terms must be changed except where both numbers
$j$ and $k$ are even. This is exactly what the formulae~(\ref{1fs},~\ref{1sf})
do. The lemma is proved.

\section{Equation of motion for ``physical'' variables}
\label{1seccirc}

For the dynamical system connected with the dimer model, to each link
of the square lattice
corresponds a complex number denoted
$a,b,c$ or $d$, as was explained in the beginning of Section~\ref{1dimers}.
Recall that the dependence of those numbers on a link
of the lattice is implied, but usually not indicated explicitely,
not to overload the formulae
and figures such as Fig.~\ref{1arrows}. We study the evolution of
the dynamical system up to the gauge transformations (\ref{11.3}).
It is easily seen that, for the dimer model, those transformations can
be described as follows.
Let us call the {\em elementary gauge transformation} corresponding to
a given vertex of the lattice and a non-zero
complex number $\lambda$ the following: take the numbers
$a,b,c$ and $d$ corresponding to the four
incoming and outgoing links
for the given vertex, and multiply them all by $\lambda$.
The {\em general gauge transformation\/} for the dimer model
will be a composition of any number of elementary transformations
(which, of course, commute with each other).

\begin{remark}
There exists also a discrete group of transformations
of the form~(\ref{11.3}) compatible
with the reduction of the system to the dimer model
which consists of translations
of the lattice: $G=H^{-1}$ in
(\ref{11.3}) is in this case the operator of translation
by an integer vector $(\xi,\eta)$, with
$\xi+\eta$ even. Here we, however, do not consider such transformations.
\end{remark}

In this section, we will introduce, instead of
$a,b,c,d$, new variables $\Omega$ which can be called
``physical'' in the sense that $\Omega$
does not change under (general) gauge transformations. We will write out
the ``equation of motion'' for $\Omega$, which turns out to be
a 2+1-dimensional variant of the Toda lattice in discrete time.
The remarkable and not {\em a priori\/} evident
property of this equation is the completely equal status
of the three coordinates: spatial
$\xi,\eta$ and the time $\tau$.
We will consider also the questions of how to come back from $\Omega$'s
to the variables $a,b,c,d$, and whether it is possible to
express
the statistical sum in terms of $\Omega$.

So, let a number $\Omega$
correspond to each two-dimen\-sional cell
(square) of the lattice, which is a
``multiplicative circulation''
composed from the numbers $a,b,c,d$ on the links bordering the cell
as follows:
$$
\Omega=a^{\varepsilon_a}b^{\varepsilon_b}c^{\varepsilon_c}d^{\varepsilon_d},
$$
where
$$
\varepsilon_a=-\varepsilon_b=-\varepsilon_c=\varepsilon_d=\pm 1,
$$
the sign ``$+$'' corresponding to non-shaded squares
on Fig.~\ref{1Fig1}, and the sign ``$-$'' corresponding to shaded squares.
In other words, let us move around the square anticlockwise,
and take the number on each link in the degree one or minus one
if we pass that link in or against
its direction correspondingly.

\begin{remark}\label{1circ-r-h}
This construction is described naturally
in terms of cohomology theory.
Namely, the torus ${\rm T}^2$ with the square lattice on it
and the given orientation (direction) for all links of the
lattice is a\/
{\em CW-complex\/}
(\cite{rDNF}, chapter~1, \S 4).
The ``field'' $a,b,c,d$ is a\/
{\em 1-cochain\/} with coefficients in the
multiplicative group ${\bf C}^{\ast}$ of non-zero
numbers (we assume that in the ``generic'' situation,
which we prefer to consider,
no one of the numbers $a,b,c,d$ equals zero), while its ``rotor''
$\Omega$ is the\/ {\em coboundary\/} of that cochain.
This cohomological interpretation will be of use for us later, when
we will consider the reconstruction of the ``field''
$a,b,c,d$ from~$\Omega$.
\end{remark}

Let us examine the change of circulations $\Omega$ in one step of
evolution. Here one must consider separately
the squares of the form
$\ \matrix{
\unitlength=0.09\normalbaselineskip
\special{em:linewidth 1.0pt}
\linethickness{1.0pt}
\begin{picture}(32.00,25.00)
\put(6.00,22.00){\vector(1,0){20.00}}
\put(5.00,21.00){\vector(0,-1){20.00}}
\put(26.00,0.00){\vector(-1,0){20.00}}
\put(27.00,1.00){\vector(0,1){20.00}}
\put(3.00,11.00){\makebox(0,0)[rc]{$c$}}
\put(16.00,20.00){\makebox(0,0)[ct]{$a$}}
\put(29.00,11.00){\makebox(0,0)[lc]{$b$}}
\put(16.00,2.00){\makebox(0,0)[cb]{$d$}}
\end{picture}
 }$ (i.e.\ shaded ones in
Fig.~\ref{1Fig1}) and the squares
$\ \matrix{
\unitlength=0.09\normalbaselineskip
\special{em:linewidth 1.0pt}
\linethickness{1.0pt}
\begin{picture}(32.00,25.00)
\put(6.00,0.00){\vector(1,0){20.00}}
\put(27.00,21.00){\vector(0,-1){20.00}}
\put(26.00,22.00){\vector(-1,0){20.00}}
\put(5.00,1.00){\vector(0,1){20.00}}
\put(3.00,11.00){\makebox(0,0)[rc]{$b$}}
\put(16.00,20.00){\makebox(0,0)[ct]{$d$}}
\put(29.00,11.00){\makebox(0,0)[lc]{$c$}}
\put(16.00,2.00){\makebox(0,0)[cb]{$a$}}
\end{picture}
 }$ (non-shaded in Fig.~\ref{1Fig1}).
The former we will call the {\em active\/} squares, the latter ---
the {\em non-active\/} ones.

The circulation around an active square is
\be
\Omega=a^{-1}c d^{-1}b.
\label{1circ3}
\ee
The transformation (\ref{13.1a}--\ref{13.1d}), which in our case
can be written as the transformation
\be
\matrix{
a\to d\,(ad-bc)^{-1}, & & b\to -c\,(ad-bc)^{-1}, \cr
c \to -b\,(ad-bc)^{-1}, & & d \to a\,(ad-bc)^{-1}
}
\label{1circ4}
\ee
with the simultaneous reversing of all the arrows,
changes the circulation in an obvious way as
\be
\Omega \to \Omega^{-1}.
\label{1circ5}
\ee

Convert
now the formulae~(\ref{1circ4}) for active squares to the following
form which we need for examining what happens in the neighboring
non-active squares:
\be
\matrix{
a \to a^{-1}(1-\Omega)^{-1},\hfill & &
b \to b^{-1}(1-\Omega^{-1})^{-1},\hfill \cr
c \to c^{-1}(1-\Omega^{-1})^{-1},\hfill & &
d \to d^{-1}(1-\Omega)^{-1}.\hfill
}
\label{1circ6}
\ee
This done, consider the formulae~(\ref{1circ6}) not for one active
square, but for four active squares adjacent to the chosen non-active
one. To be exact, let us consider the formula $a\to a^{-1}(1-\Omega)^{-1}$
for the active square having the common link $a$ with the chosen
non-active square, and similarly treat the transformation formulae
for $b$, $c$ and $d$,
so that $a,b,c,d$ are now the numbers corresponding to
the four links surrounding the {\em non-active\/} square,
while the four $\Omega$'s in (\ref{1circ6}) become different, namely
the circulations around the four active squares.
Now we see that the circulation around the non-active square itself
$$
\Omega = \Omega(\xi, \eta) = ac^{-1} db^{-1},
$$
$\xi, \eta$ being the integer-valued coordinates of, say,
south-western vertex of this square, are transformed by one step of
evolution as
\begin{eqnarray}
\Omega(\xi,\eta) \to  \Omega(\xi,\eta)\cdot
\left(1-\Omega(\xi,\eta-1)\right)\cdot
\left(1-\Omega(\xi,\eta+1)\right) \times \nonumber\\
\times\left(1-\Omega^{-1}(\xi-1,\eta)\right)^{-1}\cdot
\left(1-\Omega^{-1}(\xi+1,\eta) \right)^{-1}.
\label{1circ8}
\end{eqnarray}

After the step of evolution, the active squares become non-active
and vice versa. Thus, with the proper choice of
the origin for the integer-valued time coordinate
$\tau$, the active squares have the {\em odd\/} sum
$\xi+\eta+\tau$ of their three coordinates. As for the non-active squares,
they can be totally eliminated from consideration
when deriving the ``equation of motion'' for $\Omega$
because, according to (\ref{1circ5}), the circulation around a
non-active square is simply the inverse of the circulation around
the same square in the previous moment
$\tau$, when it was active. Writing out explicitely the dependence of
$\Omega$ on the time, we get from (\ref{1circ5}) and (\ref{1circ8})
the following ``equation of motion'' containing only the circulations
around active squares:
\be
\matrix{
\Omega(\xi,\eta,\tau-1) \Omega(\xi,\eta,\tau+1)=\hspace*{8em}\cr
 \noalign{\medskip}
\hspace*{4em}={\textstyle\left(1-\Omega(\xi,\eta-1,\tau)\right)
\left(1-\Omega(\xi,\eta+1,\tau)\right) \over  \textstyle
\left(1-\Omega^{-1}(\xi-1,\eta,\tau)\right)
\left(1-\Omega^{-1}(\xi+1,\eta,\tau)\right)}, \cr
 \noalign{\medskip}
\xi+\eta+\tau \hbox{\bf\ even}.
}
\label{1circ9}
\ee

Note the following compact form of equation~(\ref{1circ9}).
Introduce the ``discrete pseudo-Laplacians'' $\Delta_{\tau,\xi}$ and
$\Delta_{\eta,\xi}$ acting on functions  $F$ on the cubic
lattice, by the formulae
\be
\vbox{
\hbox{$\left(\Delta_{\tau,\xi} F\right)(\xi,\eta,\tau)=F(\xi,\eta,\tau-1)
+F(\xi,\eta,\tau+1)-$}
\hbox{$\phantom{\left(\Delta_{\tau,\xi} F\right)(\xi,\eta,\tau)=}
-F(\xi-1,\eta,\tau)-F(\xi+1,\eta,\tau);$}
}
\label{1circ10}
\ee
\be
\vbox{
\hbox{$\left(\Delta_{\eta,\xi}F\right)(\xi,\eta,\tau)=
F(\xi,\eta-1,\tau)+F(\xi,\eta+1,\tau)- $}
\hbox{$\phantom{\left(\Delta_{\eta,\xi}F\right)(\xi,\eta,\tau)=}
-F(\xi-1,\eta,\tau)-F(\xi+1,\eta,\tau)$.}
}
\label{1circ11}
\ee
By means of these operators, (\ref{1circ9}) is rewritten simply as
\be
\Delta_{\tau,\xi} \ln \Omega = \Delta_{\eta,\xi}\ln (\Omega-1).
\label{1circ12}
\ee

It is seen from (\ref{1circ12}) that the spatial coordinate
$\eta$ and the time $\tau$ in our model are present at an equal status.
Namely, the equation~(\ref{1circ12}) obviously does not change
under simultaneous changes
\be
\tau\leftrightarrow \eta, \quad
\Omega \leftrightarrow 1-\Omega.
\label{1circ13}
\ee
The equal rights
between the two spatial axes must exist, of course, too.
It is described, as one can easily verify, by the substitutions
\be
\xi\leftrightarrow \eta, \quad \Omega\leftrightarrow \Omega^{-1}.
\label{1circ14}
\ee
It is implied, of course, that the interchanges of coordinate axes
in (\ref{1circ13},
\ref{1circ14}) act only on the subscripts of pseudo-Laplacians
of the type (\ref{1circ10},
\ref{1circ11}). Those interchanges generate the symmetric group
$S_3$ acting on the set
$(\xi,\eta,\tau)$, and the corresponding transformations of
$\Omega$ generate a group of linear-fractional
transformations isomorphic to $S_3$.

\begin{remark}\label{1remToda}
The equation (\ref{1circ9}) or (\ref{1circ12}) can be obtained by
a simple transformation from the
``discrete Toda field'' equation from~\cite{Ward} (see also
\cite{Levi-P-S,Hirota-discrToda,Saito(h)}),
at least if we do not take into account the boundary conditions. That
discrete field in \cite{Ward} depends on an integer-valued coordinate
$k$ and two more coordinates $u$ and $v$, each
taking values with steps $h$,
and is written as $f_k(u,v)$. The field equation has the form
(formula~(8) from~\cite{Ward})
\begin{eqnarray}
\exp\bigl(f_k(u+h,v+h)-f_k(u+h,v)-f_k(u,v+h)+f_k(u,v)\bigr)=\nonumber\\
=\Bigl(1+h^2\exp\bigl(f_{k+1}(u+h,v)-f_k(u,v+h)\bigr)\Bigr)\times
\hspace*{5em}\nonumber\\
\times\Bigl(1+h^2\exp\bigl(f_k(u+h,v)-f_{k-1}(u,v+h)\bigr)
\Bigr).\hspace*{3em}
\label{1circ100}
\end{eqnarray}

Divide (\ref{1circ100}) by the equation obtained from (\ref{1circ100})
by the change
$$
k\to k-1,\qquad u\to u-h,\qquad v\to v+h,
$$
and introduce a new variable
$$
g_k(u,v)=-h^2\exp\bigl(f_k(u+h,v)-f_{k-1}(u,v+h)\bigr).
$$
We then get the equation
$$
{g_k(u,v+h)\,g_k(u-h,v)\over g_k(u,v)\,g_k(u-h,v+h)}=
{\bigl(1-g_{k+1}(u,v)\bigr)\bigl(1-g_{k-1}(u-h,v+h)\bigr)\over
\bigl(1-g_k(u,v)\bigr)\bigl(1-g_k(u-h,v+h)\bigr)}.
$$
Finally, setting
$$
\eta=k,\qquad \xi=-h^{-1}u+h^{-1}v,\qquad \tau=h^{-1}u+h^{-1}v,
$$
\nopagebreak
$$
\Omega(\xi,\eta,\tau)=g_k(u,v),
$$
we get for $\Omega$ equation (\ref{1circ9}).

Thus, the dynamical system connected with the inhomogeneous dimer model
is a variant of the Toda lattice.
\end{remark}

Let us now pay attention to
the question to what degree can the ``field'' $a,b,c,d$ be
reconstructed from its ``rotor''
$\Omega$. First of all, this can be done only if $\Omega$ is indeed
a coboundary (see Remark~\ref{1circ-r-h}), i.e.\  if the corresponding
to $\Omega$ cohomology class in the group
$H^2(T^2,{\bf C}^{\ast})$ vanishes.
The mentioned cohomology group, according to the book~\cite{rDNF}, is
isomorphic to
${\bf C}^{\ast}$. Constructively speaking, all this looks as follows:
two ``fields''
$\Omega$ belong to the same cohomology class
if and only if
the products of their values over
all two-dimen\-sional cells are equal (recall that the group
${\bf C}^{\ast}$ is multiplicative), and the field $\Omega$ is a coboundary
if this product equals 1.

Assuming that this last condition is satisfied, we can find a cochain
$a,b,c,d$ whose coboundary is $\Omega$. It is natural to want to
determine that cochain to within the
(general) gauge transformations introduced in the beginning
of this section.
It is easy to see that, in cohomological terms,
the gauge transformations are changes of the 1-cochain $a,b,c,d$ by
{\em coboundaries of 0-cochains}. From its coboundary, however, $\Omega$
can be restored only to within a {\em cocycle}.
To find $a,b,c,d$ up to a coboundary, one must fix two more
parameters  because, according to~\cite{rDNF}, the group
$H^1(T^2,{\bf C}^{\ast})$
is isomorphic to ${\bf C}^{\ast}\oplus{\bf C}^{\ast}$. Those two parameters
are nothing else than the
{\em values
of cochain $a,b,c,d$  on two (arbitrarily chosen)
basis cycles\/} on the torus, i.e.\ the products of numbers corresponding
to the links, taken in degrees $\pm 1$, over
all links entering a basis cycle. Here, as well as earlier,
the sign + is taken if the direction of our way around the cycle
coincides with the direction of a link, and the sign $-$ is taken if it
does not.

Comparing this situation to the representation of the motion integrals
(\ref{1determ}) as a sum over the sets of closed paths on the torus
given in Section~\ref{1dimers}, we see that the determinant (\ref{1determ})
and, consequently, the statistical sum are
{\em not\/} determined uniquely by a field $\Omega$.
They are determined only up to a two-parameter transformation which,
as one can see from the considerations of Section~\ref{1dimers},
can be interpreted as a renormalization of the variables $u,v$:
$$
u\to {\rm const}_1 \cdot u, \quad
v \to {\rm const}_2 \cdot v.
$$

\section{The discrete analog of Lax pair and a generalization
of the dynamical
system}
\label{1seclax}

Now let us return from the reduction of Section~\ref{1reduction} to general
matrices $L=\left(\ba{cc}A&B\\C&D\ea\right)$. Let us consider the evolution
described in Section~\ref{1gensys} from another viewpoint. Denote
$$
(L^{-1})^t=\left(\ba{cc}\tilde A&\tilde B\\\tilde C&\tilde D\ea\right).
$$
This means that
\be
\left(\ba{cc}\tilde A&\tilde C\\\tilde B&\tilde D\ea\right)
\left(\ba{cc}A&B\\C&D\ea\right)=\mbox{\large\bf1}.
\label{14.1}
\ee
It follows from the equality~(\ref{14.1}) that
\begin{eqnarray}
\tilde AA+\tilde CC&=&\tilde BB+\tilde DD,\label{14.2}\\
\tilde AB+\tilde CD&=&0,\nonumber\\
\tilde BA+\tilde DC&=&0.\nonumber
\end{eqnarray}
These three equations are equivalent to the fact that the following equality
holds for any complex $u$:
\be
-(\tilde A-u\tilde B)^{-1}(\tilde C-u\tilde D)=(uA+B)(uC+D)^{-1}.
\label{14.3}
\ee
Vice versa, from~(\ref{14.3}) follows
$$
\tilde L^tL=\left(\ba{cc}F&0\\0&F\ea\right),
$$
$F$ being equal to both sides of~(\ref{14.2}), i.e.
$$
\tilde L=\left(\ba{cc}F&0\\0&F\ea\right)(L^{-1})^t.
$$
It is clear that with any choice of $F$ the matrix $\tilde L$ belongs to the
same equivalence class. The formula~(\ref{14.3}) defines the same evolution
in the space of these classes as it was in Section~\ref{1gensys}, with the
agreement that the operators without a tilde correspond to the moment of time
$\tau$, while those with a tilde correspond to the moment $\tau+1$.

The formula~(\ref{14.3}) suggests the following generalization. Let, from now
on, $A(u)$ and $B(u)$ be matrices depending polynomially on $u$:
\be
 A(u)=A_0+A_1 u+\cdots+A_{m_A}u^{m_A},
\label{14.4}
\ee
\be
 B(u)=B_0+B_1 u+\cdots+B_{m_B}u^{m_B}.
\label{14.5}
\ee
We will look for matrices $\tilde A(u),\tilde B(u)$---the matrix polynomials
of the same degrees $m_A$ and $m_B$ in $u$---that satisfy, for any $u$, the
equation
\be
\tilde B(u)^{-1}\tilde A(u)=A(u)B(u)^{-1}.
\label{14.6}
\ee

The relation~(\ref{14.6}) provides what is called a discrete analog of the Lax
$L,A$-pair.
Let us forget for a moment that we are already using the letters
$L$ and $A$ for other purposes, and remind that the Lax $L,A$~pair
is a pair of operators depending on the time and other parameters
and having usually some special form, the evolution of operator $L$
in the case of discrete time being described by the formula
\be
L(\tau+1)=A(\tau)L(\tau)A(\tau)^{-1}.
\label{14.6a}
\ee
On the other hand, let us rewrite (returning to {\em our\/} notations)
(\ref{14.6}) in the following form:
\be
\tilde B(u)^{-1}\tilde A(u)=A(u)B(u)^{-1}A(u) A(u)^{-1}.
\label{14.6b}
\ee
Comparing (\ref{14.6a}) and (\ref{14.6b}), we see that, in our case,
$B(u)^{-1}A(u)$ plays the r\^ole of {\em Lax\/} $L$-operator, while
$A(u)$ --- the r\^ole of Lax $A$-operator.

Let $v$ be an eigenvalue of both sides of~(\ref{14.6}). Let $Y(u,v)$ be the
corresponding eigenvector normalized, as in Section~\ref{1vacuum}, so that its
last coordinate equals unity, and let $X(u,v)$ be the vector proportional to
$B(u)^{-1}Y(u,v)$ and normalized in the same way. One can verify that this may
be described by the following formula ($h(u,v)$ being a scalar factor):
\be
\left(\ba{c}A(u)\\B(u)\ea\right)X(u,v)=h(u,v)\left(\ba{c}v\\1\ea\right)
\otimes Y(u,v),
\label{14.7}
\ee
which is in obvious analogy to~(\ref{12.5}). The divisor equivalence is
\be
 mD_u+D_X\sim D_v+D_Y,
\label{14.8}
\ee
$D_u$ and $D_v$ being pole divisors of the functions $u$ and $v,\quad m={\rm
max}(m_A,m_B).$

For a given $u$, the eigenvalues $v$ come from the equation
$$P(u,v)=\det\bigl(A(u)-vB(u)\bigr)=0.$$
It defines an algebraic curve $\Gamma$---``generalized vacuum curve''. Let us
calculate the genus $g$ of the curve $\Gamma$. First, we need to know the
number of branch points of the projection
\be
(u,v)\longrightarrow u
\label{14.9}
\ee
of the curve $\Gamma$ onto the complex plane.

Consider $P(u,v)$ as a polynomial in $v$:
\be
P(u,v)=a_0(u)+a_1(u)+\cdots+a_n(u)v^n.
\label{14.10}
\ee
One can verify that $a_j(u)$ has a degree
\be
{\rm deg}\;a_j(u)=(n-j)m_A+j_B.
\label{14.11}
\ee
 From this one can deduce that the discriminant of $P(u,v)$ considered as a
polynomial in $v$ is a polynomial of degree
$$
b=(m_A+m_B)\,n\,(n-1)
$$
in $u$. The mapping~(\ref{14.9}) being $n$-sheeted and the number of branch
points equalling $b$, one obtains from the Riemann---Hurwitz formula that
\be
 g=(n-1)\left(\frac{m_A+m_B}{2}n-1\right).
\label{14.12}
\ee

So, the following construction has been described. Given two polynomial matrix
functions $A(u)$ and $B(u)$, one considers the meromorphic matrix function
$A(u)B(u)^{-1}$ (or else $B(u)^{-1}A(u)$), and from this function the
algebro-geometrical objects arise: the generalized vacuum curve $\Gamma$ and
the linear equivalence class of the pole divisor $D_Y$ (or, respectively,
$D_X$) of the eigenvectors of the mentioned meromorphic matrix function.
Instead of the pair $(A(u),B(u))$, it is sufficient to indicate its
equivalence class with respect to gauge transformations
\be
A(u)\rightarrow GA(u)H,\;B(u)\rightarrow GB(u)H;
\label{14.13}
\ee
instead of the function $A(u)B(u)^{-1}$, its equivalence class with respect to
transformations
$$
A(u)B(u)^{-1}\rightarrow GA(u)B(u)^{-1}G^{-1}
$$
will suffice. Then it turns out that the correspondence between such
equivalence classes (either of the pairs $(A(u),B(u))$ or the functions
$A(u)B(u)^{-1}$) and the abovementioned algebro-geometrical objects is a
birational isomorphism, the divisors $D_X$ and $D_Y$ being of degree $g+n-1$,
as in Section~\ref{1vacuum}.

The easiest way to show this is to start from a given curve $\Gamma$ defined
by the equation
$$
P(u,v)=\sum_{j=0}^n\;\sum_{k=0}^{(n-j)m_A+jm_B}\;a_{jk}v^ju^k=0
$$
(compare with~(\ref{14.10}, \ref{14.11})) and a divisor $D_X$ in it of degree
$g+n-1$. The number of coefficients $a_{jk}$ minus one common factor equals
\be(n+1)\left(\frac{m_A+m_B}{2}n+1\right)-1.\label{14.14}\ee
The linear equivalence class of divisor $D_X$ is defined, as is known, by $g$
parameters. Adding up the expressions~(\ref{14.14}) and~(\ref{14.12}), one gets
the total of
\be
(m_A+m_B)n^2+1
\label{14.15}
\ee
parameters.

Then, the gauge equivalence class of the pair $(A(u),B(u))$
is constructed out
of relation~(\ref{14.7}). To give more details, one must at first choose a
divisor $D_Y$ satisfying the equivalence~(\ref{14.8}).
Then the poles and zeros
of the function $h(u,v)$ are determined. For $X(u,v)$ and $Y(u,v)$ one must
take columns consisting each of $n$ linearly independent meromorphic functions
with corresponding pole divisors. The arbitrariness in these constructions
leads exactly to the fact that $A(u)$ and $B(u)$ are determined up to a
transformation~(\ref{14.13}).

The pair $(A(u),B(u))$, up to a scalar common factor, is determined by
$(m_A+m_B+2)n^2-1$ parameters (see~(\ref{14.4}, \ref{14.5})). In taking the
gauge equivalence class, the number of parameters is reduced by $2(n^2-1)$.
The result is again~(\ref{14.15}). This means that, indeed, to a generic pair
$(A(u),B(u))$ corresponds a divisor $D_X$ of degree $g+n-1$ and the
correspondence
$$
\bigl(\mbox{gauge equivalence class of the pair }(A(u),B(u))
\bigr)\longleftrightarrow\bigl(\Gamma,\mbox{ class of }D_X\bigr)
$$
is a birational isomorphism.

Now let us recall that $Y(u,v)$ was defined as an eigenvector of the operator
$A(u)B(u)^{-1}$, while $X(u,v)$, as is easily seen, is an eigenvector of
$B(u)^{-1}A(u)$. The relation~(\ref{14.6}) means that for the pair
$(\tilde A(u),\tilde B(u))$ its vector $\tilde X(u,v)$ is nothing else than
$Y(u,v)$, i.e.\ the equivalence holds
\be
D_{\tilde X}\sim D_X+(mD_u-D_v).
\label{14.16}
\ee

Now, assuming that if a quantity without a tilde corresponds to the moment of
time $\tau$ then that with a tilde corresponds to $\tau+1$, one comes to a
conclusion that to the adding of unity to the time corresponds a constant
shift~(\ref{14.16}) in the Jacobian of the curve~$\Gamma$. Thus, the dynamics
of the system in this section, as well as in Section~\ref{1vacuum},
linearizes.

\chapter{Dynamical system connected with the transposition of
three matrices}

\section{Definition of dynamical system}
\label{2secdef}

Let
\be
{\cal A}=\bmat{ccc}A&B&C\\D&F&G\\H&J&K \emat
\label{201}
\ee
be a block matrix acting in the linear space of $m+n+r$--dimensional
complex column vectors, so that, for example,
$A,F$ and $K$ are square matrices
of sizes $m \times m,\; n \times n,\; r \times r$ respectively.
Consider the problem of factorization of matrix $\cal A$ into a
product of the form
\be
{\cal A} = {\cal A}_1 {\cal A}_2 {\cal A}_3,
\label{202}
\ee
where
\be
\ba {c} {\cal A}_1 = \bmat {ccc}A_1&B_1&{\bf 0}\\C_1&D_1&{\bf 0} \\
{\bf 0}&{\bf 0}&{\bf 1} \emat, \quad
{\cal A}_2= \bmat {ccc} A_2&{\bf 0}&B_2\\{\bf 0}&{\bf 1}&{\bf 0}\\
C_2&{\bf 0}&D_2 \emat, \medskip   \\
{\cal A}_3=\bmat {ccc}{\bf 1}&{\bf 0}&{\bf 0}\\
{\bf 0}&A_3&B_3\\{\bf 0}&C_3&D_3 \emat. \ea
\label{203}
\ee
The factorization~(\ref{202}) may be seen as a generalization of the
factorization of an  orthogonal rotation in the 3-dimensional space
into rotations through the ``Euler angles''. However, at the moment we
consider no orthogonality conditions.

One can obtain from the factorization~(\ref{202}) other factorizations
of the same kind by using the following transformation of the triple
${\cal A}_1,{\cal A}_2,{\cal A}_3$:
\be
\ba {c}{\cal A}_1 \rightarrow {\cal A}_1 \bmat {ccc}
M_1^{-1}&{\bf 0}&{\bf 0}\\
{\bf 0}&M_2^{-1}&{\bf 0}\\
{\bf 0}&{\bf 0}&{\bf 1}
\emat ,
\quad {\cal A}_3 \rightarrow
\bmat {ccc}
{\bf 1}&{\bf 0}&{\bf 0}\\
{\bf 0}&{M_2}&{\bf 0}\\
{\bf 0}&{\bf 0}&{M_3}
\emat {\cal A}_3,  \medskip \\
{\cal A}_2 \rightarrow
\bmat {ccc}
{M_1}&{\bf 0}&{\bf 0}\\
{\bf 0}&{\bf 1}&{\bf 0}\\
{\bf 0}&{\bf 0}&{\bf 1}
\emat
{\cal A}_2
\bmat {ccc}
{\bf 1}&{\bf 0}&{\bf 0}\\
{\bf 0}&{\bf 1}&{\bf 0}\\
{\bf 0}&{\bf 0}&M_3^{-1}
\emat,  \ea
\label{204}
\ee
$ M_1, M_2$ and $M_3$ being arbitrary non-degenerate matrices of proper
sizes.

\begin{lemma} \label{2lem1}  For a generic matrix $\cal A$
factorization~(\ref{202}) is (if exists) unique to within the
transformations~(\ref{204}).
\end{lemma}

{\it Proof}. First, let us specify that matrix $\cal A$ will be called
generic with respect to this lemma if it a)~is non-degenerate and
b)~{\em allows}\/ a factorization~(\ref{202}) with matrices ${\cal A}_2$
and ${\cal D}_2$ non-degenerate. Let
\be {\cal A}= {\cal A}'_1{\cal A}'_2{\cal A}'_3\label{205} \ee
be another factorization of the same kind.
 From~(\ref{202}) and (\ref{205}) one finds
\be
{\cal A}'_2= {\cal A}''_1 {\cal A}_2 {\cal A}''_3,
\label{206}
\ee
where
\be
{\cal A}''_1= {({\cal A}'_1)}^{-1} {\cal A}_1, \qquad
{\cal A}''_3=  {\cal A}_3 {({\cal A}'_3)}^{-1}.
\label{207}
\ee
Let us denote the blocks in the dashed matrices by the same letters
$ A_1, \ldots,$ $D_3 $ as in equalities~(\ref{203}) with proper number of
dashes added to them. The relation~(\ref{206}) is rewritten as
\be
\bmat{ccc} A'_2&{\bf 0}&B'_2\\{\bf 0}&{\bf 1}&{\bf 0}\\C'_2&{\bf 0}&D'_2
\emat=
\bmat{ccc} A''_1 A_2 & A''_1B_2C''_3+B''_1A''_3 & A''_1B_2D''_3+B''_1B''_3\\
C''_1A_2 & C''_1B_2C''_3+D''_1A''_3 & C''_1B_2D''_3+D''_1B''_3\\
C_2 & D_2C''_3 & D_2D''_3 \emat  .
\label{208} \ee
 From here, one obtains at once the equalities
$$C'_2=C_2,\quad C'_1={\bf 0},\quad C''_3={\bf 0}.$$
Taking this into account, one finds from the block in 2nd row and 2nd
column that
$$D''_1A''_3={\bf 1}.$$
Thus, $D''_1$ and $A''_3$ are non-degenerate. Now the blocks just above
the main diagonal yield
$$ B''_1={\bf 0},\quad B''_3={\bf 0}. $$
So, the matrices ${\cal A}''_1$ and ${\cal A}''_3$~(\ref{207}) are
block--diagonal. In is easy to see that this means exactly that
${\cal A}'_1, {\cal A}'_2, {\cal A}'_3$ are obtained from
${\cal A}_1, {\cal A}_2, {\cal A}_3$ by the transformation~(\ref{204}).
The lemma is proved.
\medskip

Now let us construct, starting from the block matrix~${\cal A}$,
new matrix ${\cal B}$  by following means: factorize  ${\cal A}$
into the product~(\ref{202}) and set
\be
{\cal B}={\cal A}_3{\cal A}_2{\cal A}_1.
\label{209}
\ee
 From the above considerations it is seen that the matrix ${\cal B}$
is determined to within the transformations
\be
{\cal B}\rightarrow
\bmat{ccc}
{M_1}&{\bf 0}&{\bf 0}\\
{\bf 0}&{M_2}&{\bf 0}\\
{\bf 0}&{\bf 0}&{M_3}
\emat
{\cal B}
\bmat {ccc}
M_1^{-1}&{\bf 0}&{\bf 0}\\
{\bf 0}&M_2^{-1}&{\bf 0}\\
{\bf 0}&{\bf 0}&M_3^{-1}
\emat.
\label{210}
\ee
Let us call such transformations, as applied to the block matrices
here, the {\em gauge transformations}. The following simple but important
observation is valid: {\em if matrix ${\cal A}$ itself undergoes a gauge
transformation, this in no way affects the set of matrices ${\cal B}$
obtained from formula~(\ref{209})}.

It will be shown in Section~\ref{2secevd} that factorization~(\ref{202})
does exist for a generic matrix {\cal A}. This factorization will be
constructed by means of algebraic geometry. Taking this into account,
we are ready now to define the dynamic system
that we are going to examine. Let ${\cal M}$ be the set of block
matrices~(\ref{201}) taken to within gauge transformations~(\ref{210}),
or, using stricter language, the set of equivalence classes of such
matrices with respect to transformations~(\ref{210}). The set ${\cal M}$
will be our ``phase space''. Then, the birational mapping $f$ is defined
on the set ${\cal M}$ that brings into correspondence with a matrix
${\cal A}$, factorized into the product~(\ref{203}),
the matrix ${\cal B}$ factorized into the product~(\ref{209}).
Let us now bring into consideration the ``discrete time'' $\tau$
taking integer values and say that to the trasition from time $\tau$ to
time $\tau+k$ corresponds the mapping
$$\underbrace{f\circ \cdots\circ f}_{k\ \rm times}.$$

\section{Connection with Clifford algebras and ``twisted''
Yang--Baxter equation}
\label{5sec}

The dynamical system described in Section~\ref{2secdef}
deals with linear operators acting in the
{\em  direct sum\/} of three vector spaces. Now we will show that
one can pass on from the direct sum to a tensor product of (other)
spaces, and connect our dynamical system with the
``twisted'', or generalized, Yang--Baxter equation
(about which see~\cite{MN,KashStrog,Tarasov}). The arising solutions of
that equation belong to the ``free fermion'' type. This will be also seen
in Section~\ref{2seckag}, where we, basically, will also arrive at a
``twisted'' Yang--Baxter equation. In the present section, we
consider the most general situation, and use algebraic means,
while in Section~\ref{2seckag} we will consider some reduction of our
model, and will use some topological considerations as well.

Let us begin with the constructing, out of a given square matrix $R$ of size
$(n_1+n_2)\times (n_1+n_2)$, of the operator $\cal R$ acting in the
tensor product of spaces ${\cal E}_1$ and ${\cal E}_2$ having
dimensions $2^{n_1}$ and $2^{n_2}$ correspondingly.

Consider the {\em Clifford algebra\/} generated by fermionic creation
and annihilation operators $a_j^{\pm}$, where $1\leq j \leq N$, i.e.\
the algebra with relations
\be
\matrix {
a_j^+a_k^+ + a_k^+a_j^+ = a_j^-a_k^- + a_k^-a_j^-=0,\cr
\noalign{\smallskip}
a_j^-a_k^+ + a_k^+a_j^- = \delta_{jk},
}
\label{516}
\ee
$\delta_{jk}$ being the Kronecker
delta symbol. Such algebra can be realized by operators in
a $2^N$-dimensional space, and than there exists in that space
a non-zero vector (``vacuum'') $\Omega $ such that
$$
a_j^-\Omega =0 \quad \hbox{ for all } j.
$$

Let us have constructed two such algebras, with $N=n_1$ and $n_2$, acting
in spaces ${\cal E}_1$ and ${\cal E}_2$. We are going to explain
how to obtain out of them, in the spirit of~\cite{JW},
one ``large'' algebra acting in
${\cal E}_1\otimes {\cal E}_2$.

Consider the
{\em  particle number operator\/}
\be
{\cal N}_1=\sum^{n_1}_{j=1} a_j^+a_j^-
\label{518}
\ee
acting in ${\cal E}_1$. The eigenvalues of this operator are integers,
so one can introduce the operator
\be
{\cal Z}_1=(-1)^{{\cal N}_1}.
\label{519}
\ee
Then let us define the operators $b_j^{\pm}$ in
${\cal E}_1\otimes {\cal E}_2$ by formulae
\be
b_j^{\pm}=a_j^{\pm} \otimes {\bf 1}\quad \hbox{if \ } 1\leq j\leq n_1,
\label{520}
\ee
\be
b_j^{\pm}={\cal Z}_1 \otimes a_{j-{n_1}}^{\pm}\quad \hbox{if \ }
n_1+1\leq j \leq n_1+n_2.
\label{521}
\ee
(it is understood that
the left factor in the tensor products acts in ${\cal E}_1$, while the
right one---in ${\cal E}_2$). A simple check shows that the operators
(\ref{520}--\ref{521}) satisfy the same commutation relations
(\ref{516}), and the vacuum is now the tensor product of vacua
in  ${\cal E}_1$ and ${\cal E}_2$.

Let us now associate to a matrix $R$ the automorphism
$\varphi$ of the Clifford algebra acting in
${\cal E}_1\otimes {\cal E}_2$ which acts on the creation and
annihilation operators in the following way:
$$
\varphi (b_j^-)=\sum_{k=1}^{{n_1}+{n_2}} R_{jk}b_k^-,
$$
$$
\varphi (b_j^+)=\sum_{k=1}^{{n_1}+{n_2}}\left( \left(R^{-1}\right)^{\rm T}
\right)_{jk}b_k^+.
$$

According to the theory of Clifford algebras (as it can be found,
e.g., in~\cite{Jimbo-et-al}),  $\varphi$ is an
{\em inner\/} automorphism, i.e.\ it can be represented in the form
\be
\varphi ({\bf A})=\tilde{\cal R}{\bf A}{\tilde{\cal R}}^{-1},
\label{524}
\ee
$\bf A$ being an arbitrary element of the algebra,
$\tilde{\cal R}\in {\rm Aut}\,{\cal E}_1\otimes {\cal E}_2$.
To conclude the construction of operator
${\cal R}$, it remains to define, similarly to~(\ref{518}), the
particle number operator
$$
{\cal N}_2=\sum^{n_2}_{j=1} a^+_j a^-_j
$$
in the space ${\cal E}_2$, and to set
\be
{\cal R}=\tilde{\cal R}\cdot \left(-1\right)^{{\cal N}_1\otimes {\cal N}_2}.
\label{525}
\ee

\begin{theorem}\label{5th}
Let the matrices ${\cal A}_1$, ${\cal A}_2$, ${\cal A}_3$ be as
in Section~\ref{2secdef} (formulae~(\ref{203})), and let there be given
also the ``primed''
matrices ${\cal A}'_1$, ${\cal A}'_2$, ${\cal A}'_3$ of similar form,
e.g.
$$
{\cal A}'_1=
\pmatrix{A'_1& B'_1&\bf 0\cr
C'_1&D'_1&\bf 0 \cr
\bf 0&\bf 0&\bf 1}
$$
etc. Let the equality
\be
{\cal A}_1{\cal A}_2{\cal A}_3={\cal A}'_3{\cal A}'_2{\cal A}'_1.
\label{527}
\ee
hold. We will associate to the matrix ${\cal A}_1$ or, to
be exact, to its ``nontrivial'' part
$R=\pmatrix{A_1&B_1\cr C_1&D_1}$, the operator ${\cal R}={\cal R}_{12}$
acting in  ${\cal E}_1\otimes {\cal E}_2$ according to the construction
described by formulae (\ref{516}--\ref{525}), setting $n_1=m$, $n_2=n$.
In the same way, let us associate to the matrix
${\cal A}'_1$  the operator
${\cal R}'={\cal R}'_{12}$.

Consider also a space ${\cal E}_3$ of dimension $2^r$ and, replacing
${\cal E}_2$ by ${\cal E}_3$ and $n_2$ by $r$ in the
construction (\ref{516}--\ref{525}), associate to the matrices
${\cal A}_2$  and ${\cal A}'_2$ the operators
${\cal L}={\cal L}_{13}$ and ${\cal L}'={\cal L}'_{13}$, correspondingly
(the letter ${\cal L}$ plays the r\^ole of ${\cal R}$ from (\ref{525}),
the subscripts of ${\cal L}$  and ${\cal L}'$ denote the numbers of
spaces ${\cal E}_j$ in whose tensor product an operator acts).
Finally, to the matrices ${\cal A}_3$ and ${\cal A}'_3$ we will associate
similarly the operators ${\cal M}={\cal M}_{13}$
and ${\cal M}'={\cal M}'_{13}$.

Under these conditions, the ``twisted'' Yahg--Baxter equation holds:
\be
{\cal R}{\cal L}{\cal M}={\cal M}'{\cal L}'{\cal R}',
\label{528}
\ee
where we imply, as usual, that each operator is multiplied tensorly
by the unity operator in the lacking space, e.g.\
${\cal L}$ is multiplied by
${\bf 1}\in {\rm Aut}\, {\cal E}_2$.
\end{theorem}

The equality (\ref{527}) must be interpreted as the decomposition~(\ref{202})
of the operator $\cal A$ obtained at the {\em previous\/}  step of
evolution from Section~\ref{2secdef} as the product
$$
{\cal A}={\cal A}'_3{\cal A}'_2{\cal A}'_1
$$
of type (\ref{209}). Thus, (\ref{527}) can be regarded as a description
of a step of that evolution, while (\ref{528}) is a reformulation
of (\ref{527}) in which the direct sums of vector spaces are replaced
by tensor products.
\medskip

{\it Proof of Theorem~\ref{5th}}.
To begin with, introduce operators
$\hat{\cal R}$, $\hat{\cal L}$, $\hat{\cal M}$,
$\hat{\cal M}'$, $\hat{\cal L}'$, $\hat{\cal R}'$ as follows.
Generalize the construction (\ref{520}--\ref{521}) for the tensor product
${\cal E}_1\otimes{\cal E}_2\otimes{\cal E}_3$ of {\em three\/} spaces,
joining the creation and annihilation operators acting in them
in a single algebra by multiplying them, if necessary, tensorly
by operators of type (\ref{519}), so that the upper relations~(\ref{516})
hold. We get
\begin{eqnarray}
c_j^{\pm}=a_j^{\pm}\otimes{\bf 1}\otimes{\bf 1}
 & \hbox{ if }&1\leq j\leq m,
\label{529}\\
c_j^{\pm}={\cal Z}_1\otimes a_j^{\pm}\otimes{\bf 1}
&\hbox{ if }&
m+1\leq j\leq m+n,
\label{530}\\
c_j^{\pm}={\cal Z}_1\otimes{\cal Z}_2\otimes a_j^{\pm}&
\hbox{ if }& m+n+1\leq j\leq m+n+r.
\label{531}
\end{eqnarray}
The operators  $\hat{\cal R}, \ldots , \hat{\cal R}'$ in space
${\cal E}_1\otimes{\cal E}_2\otimes{\cal E}_3$ are determined by the
following conditions: 1)~each of them doesn't change the vacuum
$\Omega_1\otimes
\Omega_2\otimes\Omega_3$ and 2)~consider a column of height $m+n+r$ made
of the {\em annihilation\/} operators $c_j^-$ (\ref{529}--\ref{531}).
The action of each of the operators
$\hat{\cal R},\ldots,\hat{\cal R}'$ on that column
by conjugation of every element, e.g.
$$
c_j^-\to \hat{\cal R}c_j^-\hat{\cal R}^{-1},
$$
is equivalent to the action of the corresponding matrix
${\cal A}_1$, ${\cal A}_2$,
${\cal A}_3$, ${\cal A}'_3$, ${\cal A}'_2$ or ${\cal A}'_1$ on the whole
column, e.g.
$$
\pmatrix{c_1\cr\vdots\cr c_{m+n+r}} \to
{\cal A}_1
\pmatrix{c_1\cr\vdots\cr c_{m+n+r}}.
$$

Such operators $\hat{\cal R},\ldots, \hat{\cal R}'$, according to the
theory of Clifford algebras~\cite{Jimbo-et-al}, exist and are unique.
Besides, the equality
\be
\hat{\cal R}\hat{\cal L}\hat{\cal M}=\hat{\cal M}'\hat{\cal L}'\hat{\cal R}',
\label{534}
\ee
holds, because both sides of (\ref{534}) preserve the vacuum and act
identically on the annihilation operator column, namely:
$$
\pmatrix{c_1\cr \vdots\cr c_{m+n+r}} \to
{\cal A}_1{\cal A}_2{\cal A}_3
\pmatrix{c_1\cr \vdots\cr c_{m+n+r}}
={\cal A}'_3{\cal A}'_2{\cal A}'_1
\pmatrix{c_1\cr \vdots\cr c_{m+n+r}}.
$$

The following lemma shows how the operators in (\ref{534}) (they
don't have, generally, the form required in Yang--Baxter equation)
are connected with the operators in relation~(\ref{528}) that we are
proving. More exactly, we will connect the operators
``with hats'' with the operators ``with tildes'' which,
we remind, were defined each in its own tensor product of
{\em two\/} spaces.

\begin{lemma} \label{5lem}
\be
\hat{\cal R}=\tilde{\cal R}, \quad \hat{\cal M}=\tilde{\cal M},\quad
\hat{\cal M}'=\tilde{\cal M}',\quad \hat{\cal R}'=\tilde{\cal R},
\label{536}
\ee
while $\hat{\cal L}$  and $\hat{\cal L}'$ are connected with
$\tilde {\cal L}$ and $\tilde{\cal L}'$ as follows:
\be
\hat{\cal L}^{(\prime)}=(-1)^{{\cal N}_1{\cal N}_2}\,
\tilde{\cal L}^{(\prime)}\,
(-1)^{{\cal N}_1{\cal N}_2},
\label{537}
\ee
where\/ $(\prime)$ denotes the prime or its absence, and\/
${\cal N}_1{\cal N}_2$
is, strictly speaking, ${\cal N}_1\otimes{\cal N}_2\otimes{\bf 1}$.
\end{lemma}

{\it Proof of Lemma \ref{5lem}\/}  is based on the fact that LHS's and RHS's
of (\ref{536}) and (\ref{537}) define (by conjugations) the same
automorphisms of the Clifford algebra acting in
${\cal E}_1\otimes{\cal E}_2\otimes {\cal E}_3$. This is checked
by direct calculations for each of the equalities
(\ref{536}, \ref{537}) and for each set of operators
(\ref{529}, \ref{530}, \ref{531})  separately. We will remark only
that the r\^ole of factors
$(-1)^{{\cal N}_1{\cal N}_2}$ in (\ref{537}) is seen from equalities
of type
\be
(-1)^{{\cal N}_1{\cal N}_2} \left({\cal Z}_1\otimes a_k^{\pm}
\otimes{\bf 1}\right)=
\left({\bf 1}\otimes a_k^{\pm}\otimes {\bf 1}\right)
(-1)^{{\cal N}_1{\cal N}_2},
\label{538}
\ee
which demonstrate how an operator, in this case, ${\cal Z}_1$, disappears
or appears after a conjugation with $(-1)^{{\cal N}_1{\cal N}_2}$.
\medskip

To conclude the proof of Theorem~\ref{5th}, let us write out
the equality~(\ref{534}) in terms of operators ${\cal R},\ldots,{\cal R}'$
defined by equalities of type (\ref{525}), using (\ref{536}) and (\ref{537}):
\be
\vbox{\hbox{$\displaystyle
{\cal R}{\cal L}(-1)^{{\cal N}_1({\cal N}_2+{\cal N}_3)}
{\cal M}(-1)^{{\cal N}_2{\cal N}_3} = $}
\medskip
\hbox{$\displaystyle\qquad\qquad
={\cal M}'(-1)^{{\cal N}_2({\cal N}_1+{\cal N}_3)}
{\cal L}'(-1)^{{\cal N}_1({\cal N}_2+{\cal N}_3)}
{\cal R}'(-1)^{{\cal N}_1{\cal N}_2}.$}}
\label{539}
\ee
All the degrees of minus one in LHS and RHS of (\ref{539}) can be
conveyed through other factors to the right using the fact that ${\cal M}$
commutes with
${\cal N}_2+{\cal N}_3$ (i.e.\ ``preserves the total  number of
particles of second and third kinds'') and with ${\cal N}_1$, similarly
${\cal L}'$ commutes with ${\cal N}_1+
{\cal N}_3$ and with ${\cal N}_2$,
and ${\cal R}'$  commutes with ${\cal N}_1+{\cal N}_2$ and with ${\cal N}_3$.
This done, the factor
$(-1)^{{\cal N}_1{\cal N}_2 +{\cal N}_1{\cal N}_3+{\cal N}_2{\cal N}_3},$
arising in both LHS and RHS, is canceled, and we come to (\ref{528}).
The theorem is proved.

\section{Invariant algebraic curve of matrix $\protect\cal A$
and some divisors in it}
\label{2seccrv}

The dynamical system of Section~\ref{2secdef} turns out to be
completely integrable.
To be exact, {\em an invariant curve}  $\Gamma $ can be constructed out
of matrix $\cal A$, together with a divisor  $\cal D$ on it.
In terms of these algebro-geometrical objects, the evolution is as follows:
$\Gamma$ does not change, while  $\cal D$---more precisely, its linear
equivalence class---depends linearly on the discrete time $\tau$.

Let us start from the definition of the curve $\Gamma$. The word
``invariant'' in this definition will be justified in Section~\ref{2secevd}.

\begin{dfn} \label{2dfn1} The invariant curve $\Gamma$ of the operator
$\cal A$ of the form~(\ref{201}) is an algebraic curve in
${\bf C}P^1\times{\bf C}P^1\times{\bf C}P^1$ (i.e.\ in the space of
three complex variables $u,v,w$, each allowed also to take value $\infty$)
given by equations
\be \det({\cal A}-\bmat{ccc} u{\bf 1}_m &  {\bf 0} & {\bf 0}\\
{\bf 0} & v{\bf 1}_n & {\bf 0}\\ {\bf 0} & {\bf 0} & w{\bf 1}_r \emat )=0,
\label{211} \ee
\be v=uw. \label{212} \ee
\end{dfn}

Here a subscript of each {\bf 1} means the size of corresponding unity
martix, while {\bf 0} denotes rectangular zero matrices of
different sizes. Strictly speaking, equations (\ref{211}, \ref{212})
define the ``finite part'' of the curve $\Gamma$, the whole curve $\Gamma$
being its closure in Zariski topology.

The equality~(\ref{211}) means that a column vector
${\cal X}= \bmat{c} X\\Y\\Z \emat$ exists, with $X,Y,Z$ column vectors
of dimensions $m,n$ and $r$, correspondingly, such that
\be {\cal A} \bmat{c}X\\Y\\Z \emat = \bmat{c}uX \\ vY \\ wZ \emat.
\label{213} \ee
Such vectors  $\cal X$ form a one-dimensional holomorfic bundle over
$\Gamma$.

The next lemma shows the structure of the zero and pole divisors of
functions $u,v,w$. For these divisors, the notations $(u)_0,(u)_{\infty}$
etc.\ are used.

\begin{lemma}\label{2lem2} There exist such effective divisors (i.e.\
finite sets of points)  ${\cal D}_1, \ldots , {\cal D}_6$ in the curve
$\Gamma$ that
\be \ba{lll} (u)_{\infty}= {\cal D}_1+{\cal D}_2,&
(v)_{\infty}= {\cal D}_1+{\cal D}_3,& (w)_{\infty}= {\cal D}_3+{\cal D}_4,\\
(u)_0= {\cal D}_4+{\cal D}_6,& (v)_0= {\cal D}_5+{\cal D}_6, &
(w)_0= {\cal D}_5+{\cal D}_2. \ea  \label{214} \ee
${\cal D}_3$ and ${\cal D}_5$ are of degree   $m, \quad {\cal D}_2$ and
${\cal D}_4$ are of degree $n, \quad {\cal D}_1$ and ${\cal D}_6$
are of degree $r$. Generally, {\em all} points included in divisors
${\cal D}_1, \ldots , {\cal D}_6$ are different from each other.
\end{lemma}

{\it Proof}. Consider, e.g., the case $u=0, w \not= 0,
w \not= \infty$. Then, according to~(\ref{212}), $v=0$.
The equality~(\ref{213}) turns into the following system:
$$ \left\{ \ba{l} AX+BY+CZ=0, \\ DX+FY+GZ=0, \\ HX+JY+KZ=wZ. \ea \right.$$

One can express $X$ and  $Y$ through $Z$ (e.g.,
$Y=-(F-DA^{-1}B)^{-1}(G-DA^{-1}C)Z)$
and then substitute these expressions into the third one. One will come to
an equation of the form
\be \tilde K Z=wZ \label{215} \ee
which has $r$ characretistic roots  $w_1, \ldots ,w_r$, different from
each other in general case. This is how  $r$ points $(0,0,w_1),
\ldots ,(0,0,w_r)$ of divisor ${\cal D}_6$ are obtained. The other
divisors in~(\ref{214}) arise in a similar way. The lemma
is proved.
\medskip

The vector $\bmat{c}X\\Y\\Z \emat $ in (\ref{213}) is determined
up to a scalar factor which may depend on the point in the curve
$\Gamma$.  So, this vector  can be normalized by setting its first
coordinate identically equal to unity~(cf. \cite{Krichever}).
$\bmat{c} X\\Y\\Z \emat$ becomes then a meromorphic vector on $\Gamma$
with a certain pole divisor $\cal D$. However,  $X,Y$ and $Z$ taken
separetely satisfy stronger restrictions, as the following lemma shows.
In the lemma, $(f)$ denotes the divisor of a function $f$ (zeros enter
with the $+$ sign, poles with the $-$ sign, as usual).

\begin{lemma} \label{2lem3} The column vector $X$ consists of
functions $f$ such that
\be (f)+{\cal D}-(u)_{\infty}\geq0; \label{216} \ee
the column vector $Y$ consists of functions $f$ such that
\be (f)+{\cal D}-(v)_{\infty}\geq0; \label{217} \ee
the column vector $Z$ consisits of functions $f$ such that
\be (f)+{\cal D}-(w)_{\infty}\geq0. \label{218} \ee
\end{lemma}
{\it Proof}. One can see immediately from the formula~(\ref{213}) that
the vector  $uX$  entering into R.H.S. cannot grow faster than the vector
$\bmat{c} X\\Y\\Z \emat$ in L.H.S. in such points where $u=\infty$.
This is exactly what the inequality~(\ref{216}) states.
The inequalities~(\ref{217}) and (\ref{218}) are proved similarly.
\medskip

Now the time has come to make it sure that the curve $\Gamma$,
for a generic matrix $\cal A$, is a smooth irreducible curve.
One may wish also to calculate its genus in some simple way.
To do that, we are now going to examine a relatively simple
particular case of the matrix $\cal A$, although at the time ``generic''
enough to make sure that such its features as genus and the degree of
divisors are the same for matrices in some Zariski neighborhood.

Thus, let all the matrix elements of $\cal A$ equal zero except the ones
lying, first, in the main diagonal and, second, in the ``broken''
diagonal parallel to the main one (for these latter matrix elements,
the difference between the numbers of a column and a row must be
some constant modulo $m+n+r$). The elements in the main diagonal will
be denoted as $a_1, \ldots, a_m, f_1, \ldots, f_n,$ $k_1, \ldots, k_r$;
and let the elements in the broken diagonal be all equal to the
same complex number $s$:
\be
{\cal A} = \left(
{\linethickness{0.05pt}
\ba{ccc|ccc|ccc}
a_1 & & & & s & & & & \\
&\ddots &&&& \ddots &&&\\
&&a_m &&&& \ddots && \\ \hline
&&&f_1&&&& \ddots& \\
&&&& \ddots &&&& s \\
s&&&&&f_n&&&\\ \hline
& \ddots &&&&& k_1 && \\
&& \ddots &&&&& \ddots & \\
&&& s &&&&&k_r \ea
}
\right).
\label{219}
\ee
It does not matter through which blocks exactly the ``broken'' diagonal
passes.

For the finite  $u,w$, the curve $\Gamma$ now examined is given by
equation (resulting from the substitution of (\ref{219}) and
(\ref{212}) into (\ref{211}))
\be
F(uw) \equiv \prod^m_{\alpha =1}(a_{\alpha}-u) \cdot
\prod^n_{\beta =1}(f_{\beta}-uw)\cdot
\prod^r_{\gamma =1}(k_{\gamma}-w)\pm s^{m+n+r}=0. \label{220} \ee
As is known, in singular points
\be \left\{ \ba{l} \partial F / \partial u=0, \quad \\
\partial F / \partial w=0. \ea \right.
\label{221}
\ee
The system~(\ref{221}) has a finite number of solutions, and, changing $s$
in~(\ref{220}), one can make these solutions not to lie in the curve
$\Gamma$, which thus will be free of singularities for finite $u,w$.
It is an easy exercise to show that there are no singullarities when $u$
or $w$ is infinite as well.

Returning now to general matrices $\cal A$  and curves $\Gamma$, let us
note that it cannot be that the system~(\ref{221}) or its substitute
in the neighborhood of infinite $u$ or $w$ possesses solutions
{\em in the curve}\/ $\Gamma$ in general case and does not possess them
in a particular case. Thus, the smoothness of $\Gamma$ for a generic
$\cal A$ is clear. As for irreducibility, to prove it let us examine the
natural projection of $\Gamma$ onto the Rimann sphere ${\bf C}P^1$
of the variable $u$. This projection is an $(n+r)$-sheet cover,
and if $\Gamma$ consisted of two or more components, the sheets of the cover
would split into groups belonging to each component.
To prove that it is not so in the general case, it is enough to
present an example where it is not so. To do this, take  $\cal A$ of the
form~(\ref{219}) and, moreover, put $f_1=\cdots =f_n=k_1=\cdots =k_r=0$.
Equation~(\ref{220}) then becomes
$$ w^{n+r}u^n \prod^m_{\alpha =1} (a_{\alpha}-u) \pm s^{m+n+r}=0.$$
Let  $a_1 \not= 0$ and not coinside with other $a_{\alpha}$.
Then in a neighborhood of the point $(u,w)=(a_1,\infty)$ the variable~$w$
behaves, up to a nonzero factor, like
$$ w^{-1}\sim (a_1-u)^{1/(n+r)}.$$
 From here one sees that all the mentioned $n+r$ sheets belong to a
{\em single} component, i.e.\ the irreducibility of $\Gamma$ is proved.
\medskip

Now let us denote the number of branch points of the cover
$\Gamma \rightarrow {\bf C} \ni u$ \/ as  $b$. Then  the {\em genus\/}
of the curve \/ $\Gamma$, according to the Riemann---Hurwitz formula, is
\be g=1-n-r+\frac{b}{2}. \label{222} \ee
Our next aim is to express  $b$  and  $g$ through $m,n$ and $r$.

\begin{lemma}\label{2lem4} {The degree of the vector
$\bmat{c}X\\Y\\Z \emat $ pole divisor $\cal D$ is $m+b/2$.
}
\end{lemma}

{\it Proof}. Write out the equation~(\ref{213}) ``explicitly'':
\be
\left. \ba{c}  AX+BY+CZ=uX,\\ DX+FY+GZ=vY,\\
HX+JY+KZ=wZ. \ea \right\}
\label{223}
\ee
Expressing  $X$ through $Y$ and $Z$ by means of the first of these
equations and substituting into the rest, one finds:
\be
\bmat{cc}\! u^{-1} \left( D(u-A)^{-1}B+F\right) & u^{-1}
\left( D(u-A)^{-1}C+G\right)\! \\
\!H(u-A)^{-1}B+J & H(u-A)^{-1}C+K\! \emat\! \bmat{c}\!Y\!\\\!Z\! \emat\! =
\!w\! \bmat{c}\! Y\!\\\!Z\! \emat\!.
\label{224}
\ee
To a generic $u$ correspond  $n+r$ different $w=w_1, \ldots, w_{n+r}$,
and the corresponding $n+r$ vectors $\bmat{c} Y\\Z \emat $
are linearly independent as eigenvectors of the matrix in L.H.S.
of~(\ref{224}). An easy check shows that in the ``suspicious'', from
the standpoint of equation~(\ref{224}), points  $u=0, \infty$, and also
in points where  $\det (u-A)=0$, there exist as well
$n+r$ linearly independent
vectors $\bmat{c} Y\\Z \emat$ ---solutions of limit cases of the
system~(\ref{224}).

Consider a determinant
\be
d= \left|\ba{ccc} Y(w_1) & \ldots & Y(w_{n+r})\\
Z(w_1) & \ldots & Z(w_{n+r}) \ea \right|.
\label{224a}
\ee
Given $u$, it changes its sign under odd permutations of $w$'s.

However, $d\/^2$ is a function of $u$ only. From the above one sees
that $d\/^2(u)$ vanish in branch points where to a given $u$
correspond less than $n+r$ values of $v$ or $w$. This yields $b$
zeros of the function $d\/^2(u)$. There is, however, one more cause for
this function to vanish. According to Lemmas \ref{2lem3} and
\ref{2lem2}, $Y$ and $Z$ must vanish as a whole in the points of divisor
${\cal D}_3$ ---the common part of divisors $(v)_{\infty}$ and
$(w)_{\infty}$.
The degree of divisor ${\cal D}_3$ is $m$, so we get $2m$ more
zeros of $d\/^2(u)$.

The number of function $d\/^2(u)$ poles equals the number of its zeros,
i.e.\ $b+2m$. Thus, the meromorphic vector $\bmat{c} Y\\Z \emat$ has
$b/2+m$ poles, as desired. It remains just to note that vector~$X$,
in accord with formula~(\ref{216}), has no poles that $\bmat{c} Y\\ Z \emat$
doesn't have, and on the other hand doesn't vanish in points of
divisor ${\cal D}_3$. The lemma is proved.
\medskip

Let us turn again to matrices  $\cal A$ of the form~(\ref{219}). For such
matrix, it is easy to find the vector $\bmat{c}X \\ Y \\Z \emat=\cal X$  in
a given point $(u,v,w) \in \Gamma$. Let us assume that the ``broken''
diagonal is the one adjacent to the main diagonal, so that there is only
one letter $s$ in the lower left corner. Then the following holds
for the vector $\cal X$ coordinates:
$$ \ba{l}
(a_1-u)X_1+sX_2=0,\\
(a_2-u)X_2+sX_3=0,\\ \dotfill \\
(a_m-u)X_m+sY_1=0,\\
(f_1-v)Y_1+sY_2=0,\\ \dotfill \\
(f_n-v)Y_n+sZ_1=0, \\
(k_1-w)Z_1+sZ_2=0, \\ \dotfill \\
(k_n-w)Z_n+sX_1=0. \ea $$

 From here the ratios between vector $\cal X$ coordinates are readily seen.
Assuming the normalization condition $X_1\equiv 1$, one finds out that
$\cal X$ has the poles a) of the order $m$ in $n$
points $(u,v,w)=(\infty, f_{\beta}, 0)$ and b) of the order $m+n$
in $r$ points $(u,v,w)=(\infty, \infty, k_{\gamma})$. In all, $\cal X$
possesses thus $mn+mr+nr$ poles, taking their multiplicities into account.
Recalling Lemma~\ref{2lem4} and formula~(\ref{222}), one can now find
the genus $g$
of the curve as well. As a matrix $\cal A$ of the form~(\ref{219}) is
``generic enough'', the results on the degree of divisor $\cal D$ of the
vector $\cal X$ and genus $g$ of the curve apply also to curves corresponding
to generic matrices $\cal A$. Let us formulate them as the following lemma.

\begin{lemma} \label{2lem5}
For a generic matrix $\cal A$, the genus of the curve  $\Gamma$ is
\be g=mn+mr+nr-m-n-r+1, \label{225} \ee
while the degree of divisor $\cal D$ of the meromorphic vector
$\cal X=\bmat{c}X \\ Y \\Z \emat$  is
\be mn+mr+nr=g+m+n+r-1.\label{226} \ee
\end{lemma}

Thus, in this section we have constructed, for a given matrix $\cal A$,
an algebraic curve $\Gamma$ and a bundle of vectors $\cal X$ over it,
and calculated the genus $g$ of the curve and the degree of the bundle
(i.e.\ the divisor $\cal D$ degree). As a helpful tool, a matrix $\cal A$
of special simple form~(\ref{219}) was used which, from many viewpoints,
was ``generic enough''. In Section~\ref{2secevd} we will study how
these objects behave under evolution introduced in Section~\ref{2secdef}.

\section{Evolution in terms of divisors}
\label{2secevd}

In this section it is shown, at first, that there exists a one-to-one
correspondence
(more precisely, a birational isomorphism) between the set of block
matrices $\cal A$ (\ref{201}) taken up to gauge transformations (\ref{204}),
and the set of pairs (an algebraic curve, a linear equivalence class of
divisors on it) of a certain kind. This correspondence has, in essence,
been constructed in Section~\ref{2seccrv}, and here are some missing details.
Then, it is explained which divisors and why correspond to the factors
${\cal A}_1, {\cal A}_2$ and
${\cal A}_3 $ in (\ref{202}) taken separately. Finally, it is demonstrated
that to the matrix $\cal B$ \/ (\ref{209}) obtained from $\cal A$
by reversing the order of its factors, the same curve $\Gamma$ corresponds,
but the divisor undergoes some constant shift. Thus, the motion linearizes
in the Jacobian of curve $\Gamma$. Let us proceed to a detailed
consideration.

Equations~(\ref{211}, \ref{212}) define, for a block matrix
$\cal A$, an algebraic curve $\Gamma$. Those equations can obviously be
written as
\begin{eqnarray} \sum^{m}_{i=0} \sum^{n}_{j=0} \sum^{r}_{k=0} a_{ijk}
u^i v^j w^k =0, \vspace{-1.8ex} \nonumber \\
\vspace{-2.8ex} \label{227} \\
v=uw. \nonumber \phantom{u^i v^j w^k =0,} \end{eqnarray}
Besides, a linear bundle over $\Gamma$ has been constructed in
Section~\ref{2seccrv}---the bundle of vectors
$\bmat{c} X\\Y\\Z \emat$
(\ref{213}).
That means that the divisor
$\cal D$ of the bundle is determined, up to linear equivalence, whose
degree is $g+m+n+r-1$, \ $g$ being the curve's genus (\ref{226}).
Gauge transformations (\ref{204}) do not change a pair
($\Gamma$, class of divisor  $\cal D$~).

Now let us show how to construct the matrix $\cal A$ starting from
coefficients
$a_{ijk}$ of the curve (\ref{227}) (arbitrary complex numbers in
``general position'') and a divisor $\cal D$  of degree $g+m+n+r-1$. Note
that
genus $g$ of the curve $\Gamma$ defined by formulae (\ref{227}) without any
(a priori)
connection with block matrices is given by the same formula~(\ref{225}).
This can be seen, e.g., by starting again from the ``simple'' curve
(\ref{220}) of Section~\ref{2seccrv} whose genus is known.
{\em Define} now the meromorphic column vectors
$X,Y$ and $Z$,
guided by Lemma~\ref{2lem3}: for components of vector $X$,
take $m$ linearly independent meromorphic functions on $\Gamma$
satisfying relation~(\ref{216}), and for $Y$ and $Z$
take, similarly, $n$ functions satisfying (\ref{217}) and $r$  functions
satisfying (\ref{218}).

Note also that Lemma~\ref{2lem2} about divisors $(u)_{\infty}, (v)_{\infty},
(w)_{\infty}$ entering in formulae (\ref{216}--\ref{218}) remains valid
for curves defined by an
``abstract'' system (\ref{227}), which is immediately seen on
substituting zero or infinity for $u, v$, or $w$ in
(\ref{227}).

It is clear now that relation (\ref{213})   {\em determines\/}
unambiguously the matrix $\cal A$ (cf.\ a similar construction in
paper~\cite{Krichever}).
Another choice of linearly independent functions for components of
$X,Y$ and $Z$ leads, of course, to a gauge transformation~(\ref{210}).
The vectors $X,Y,Z$ change under it to $M_1X, M_2Y, M_3Z$. On the other
hand,
if divisor~$\cal D$ is changed to another divisor belonging to the same
linear equivalence class,
the vectors $X, Y$ and $Z$ are just multiplied by
a scalar meromorphic function~$h$ having zeros in the points of the first
divisor and poles---in the points of the second one.

Hence, the following theorem has been proved.

\begin{theorem}\label{2thvstav}
If block matrices $\cal A$ and $\hat{\cal A}$  of the form (\ref{201})
are connected by a gauge transformation
\be \hat{\cal A}=\pmatrix{M_1 &\bf 0&\bf 0 \cr
\bf 0&M_2&\bf 0 \cr \bf 0&\bf 0&M_3} {\cal A}
\pmatrix{{M_1}^{-1}&\bf 0&\bf 0 \cr
\bf 0&{M_2}^{-1}&\bf 0 \cr
\bf 0&\bf 0&{M_3}^{-1}
} \label{227a} \ee
then they have a common invariant curve\/ $\Gamma$ given by
equations of the form (\ref{227}), and the holomorphic bundles
of the vectors~$\cal X$ corresponding to them by formula~(\ref{213})
are isomorphic.

Conversely, if two matrices $\cal A$ and $\hat{\cal A}$ of the form
(\ref{201})
have the same invariant curve\/~$\Gamma$ and the corresponding vector
$\cal X$ and $\hat{\cal X}$ bundles are isomorphic, then
(\ref{227a}) holds. Being properly normalized (say, by a condition
that the first coordinate identically equals unity),
$\cal X$ and $\hat{\cal X}$ become meromorphic vectors with
linear equivalent pole divisors $\cal D$ and $\hat{\cal D}$
of degree $g+m+n+r-1$, and
\be
\hat{\cal X}= \pmatrix{M_1 X \cr M_2Y \cr M_3 Z}
\cdot h(u,v,w),
\label{227b}
\ee
where $h$ is a scalar meromorphic function whose divisor
$(h)=(h)_0-(h)_{\infty}$ satisfies equality
$$
(h)={\cal D} -\hat{\cal D}.
$$
\end{theorem}

Examine now each multiplier in factorization~(\ref{202}) separately.
Lemma~\ref{2lem1}  shows that factorization~(\ref{202}), if exists, is
unique to within the transformations~(\ref{204}).
Let us demonstrate how to construct this factorization by
algebro-geometrical means.

Consider the following figure (Fig.~\ref{2figdecom}).

\bfig
\begin{center}
\unitlength=0.209206\normalbaselineskip
\special{em:linewidth 0.4pt}
\linethickness{0.4pt}
\begin{picture}(76.00,77.42)
\put(19.00,5.17){\vector(0,1){6.88}}
\put(19.00,60.22){\line(0,0){0.00}}
\put(19.00,60.22){\line(0,0){0.00}}
\put(19.00,60.22){\line(0,0){0.00}}
\put(19.00,60.22){\line(0,0){0.00}}
\put(19.00,60.22){\line(0,0){0.00}}
\put(19.00,12.05){\vector(0,1){55.05}}
\put(19.00,67.10){\line(0,1){8.17}}
\put(4.00,60.22){\vector(1,0){7.00}}
\put(11.00,60.22){\vector(1,0){55.00}}
\put(66.00,60.22){\line(1,0){8.00}}
\put(4.00,5.17){\vector(1,1){7.00}}
\put(11.00,12.05){\vector(1,1){54.00}}
\put(65.00,66.24){\line(1,1){9.00}}
\put(3.00,60.22){\makebox(0,0)[rc]{$X$}}
\put(11.00,62.37){\makebox(0,0)[cb]{12}}
\put(16.00,71.40){\makebox(0,0)[rc]{52}}
\put(19.00,77.42){\makebox(0,0)[cb]{$wZ$}}
\put(21.00,62.37){\makebox(0,0)[lb]{\large ${\cal A}_2$}}
\put(39.00,62.37){\makebox(0,0)[cb]{26}}
\put(64.00,68.39){\makebox(0,0)[rb]{56}}
\put(75.00,76.13){\makebox(0,0)[cb]{$vY$}}
\put(68.00,61.08){\makebox(0,0)[lb]{46}}
\put(76.00,60.22){\makebox(0,0)[lc]{$uX$}}
\put(61.00,57.21){\makebox(0,0)[lt]{\large${\cal A}_1$}}
\put(34.00,58.07){\makebox(0,0)[ct]{$X'$}}
\put(39.00,42.16){\makebox(0,0)[rc]{36}}
\put(45.00,43.00){\makebox(0,0)[lc]{$Y'$}}
\put(24.00,19.36){\makebox(0,0)[lc]{\large ${\cal A}_3$}}
\put(21.00,11.19){\makebox(0,0)[lc]{34}}
\put(19.00,3.02){\makebox(0,0)[ct]{$Z$}}
\put(9.00,12.05){\makebox(0,0)[rb]{13}}
\put(2.00,3.02){\makebox(0,0)[rc]{$Y$}}
\put(17.00,39.14){\makebox(0,0)[rc]{$Z'$}}
\put(20.00,43.02){\makebox(0,0)[lc]{23}}
\end{picture}
\end{center}
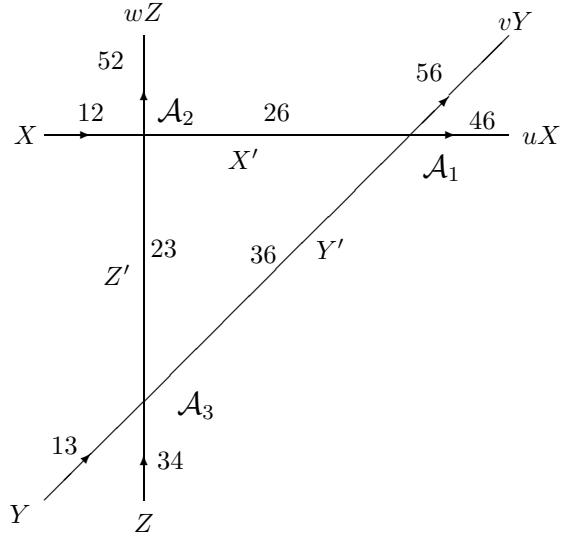
\caption{Factorization of matrix $\protect\cal A$ and the divisors}
\label{2figdecom}
\efig

The meaning of the numbers standing near the edges in this figure is
as follows: if those numbers are
$jk$, then the meromorphic vector corresponding to the edge consists of such
functions $f$ whose zero and pole divisor $(f)$ satisfies inequality
$$(f)+{\cal D} - {\cal D}_j -{\cal D}_k \geq 0.$$
Those inequalities must be in agreement with
Lemmas~\ref{2lem2} and \ref{2lem3}.
In particular, the matrix ${\cal A}_3$ will be {\em defined}\/ by equality
(notations of formulae~(\ref{203}) are used),
\be \bmat{cc}A_3&B_3\\C_3&D_3\emat \bmat{c}Y\\Z  \emat=
\bmat{c} Y'\\Z' \emat, \label{228} \ee
where the meromorphic vector  $Y$ consists of functions $f$ such that
$$(f)+{\cal D} - {\cal D}_1 - {\cal D}_3 \geq 0$$
(formulae (\ref{217}) and (\ref{214})); $Z$ of functions such that
$$(f)+{\cal D} - {\cal D}_3 - {\cal D}_4 \geq 0$$
(formulae (\ref{218}) and (\ref{214})); $Y'$ and $Z'$  consist
{\em by definition}
of such linearly independent functions that
$$(f)+{\cal D} - {\cal D}_3 - {\cal D}_6 \geq 0$$
for $Y'$, and
$$(f)+{\cal D} - {\cal D}_2 - {\cal D}_3 \geq 0$$
for $Z'$. It is easy to see that (\ref{228})  is a correct definition
for matrix ${\cal A}_3$, because the components of each of the vectors
$\bmat{c}Y\\Z \emat$ and $\bmat{c}Y'\\Z' \emat$ form a basis in the space of
meromorphic functions $f$ such that
$$(f)+{\cal D} - {\cal D}_3  \geq 0.$$

Next, let
\be  \bmat{cc} A_2&B_2\\ C_2& {\cal D}_2 \emat
\bmat{c} X\\Z' \emat = \bmat{c} X'\\wZ \emat, \label{229}  \ee
\be  \bmat{cc} A_1&B_1\\ C_1& {\cal D}_1 \emat
\bmat{c} X'\\Y' \emat = \bmat{c}uX\\vY \emat, \label{230} \ee
where $X'$ consists of functions $f$ such that
$$(f)+{\cal D} - {\cal D}_2 - {\cal D}_6 \geq 0.$$
It is shown in much the same way as above that equalities~(\ref{229}) and
(\ref{230}) do correctly define the matrices ${\cal A}_2$ and ${\cal A}_1$.
What remains is to check the validity of equality~(\ref{202}) for
${\cal A}_1$, ${\cal A}_2$ and ${\cal A}_3$ given by these definitions.
To do this, observe that (\ref{228}--\ref{230}) together yield
\be {\cal A}_1{\cal A}_2{\cal A}_3 \bmat{c} X\\Y\\Z \emat=
\bmat{c} uX\\vY\\wZ \emat . \label{231} \ee
The equality~(\ref{202}) follows from comparing~(\ref{231}) with (\ref{213}).

Note that the arbitrariness in choosing $X',Y'$ and $Z'$ corresponds,
of course, to transformations~(\ref{204}).

Now let us pass to matrix $\cal B$, a product of the same three factors
in the inverse order. The formulae (\ref{209}) and (\ref{228}--\ref{230})
together yield (if one multiplies both sides of (\ref{229}) by $u$, and
both sides of (\ref{228}) by $v$~):
\be {\cal B} \bmat{c} X'\\Y'\\uZ' \emat =
\bmat{c} uX'\\vY'\\vZ' \emat . \label{232} \ee
Compare the divisors of meromorphic vectors in L.H.S.'s of
(\ref{232}) and (\ref{213}). An easy calculation shows that
$$ {\cal D}_{X'}-{\cal D}_X = {\cal D}_{Y'}-{\cal D}_Y =
{\cal D}_{(uZ')}-{\cal D}_Z = {\cal D}_1 - {\cal D}_6.$$
One sees hence that the same curve $\Gamma$ corresponds to the operator
${\cal B}$ as to the operator ${\cal A}$, while the divisor ${\cal D}$
changes to ${\cal D}+{\cal D}_1-{\cal D}_6 $.

Thus, in this section the name ``invariant'' has been justified for the
curve $\Gamma$: it has been shown not to change under the evolution of
Section~\ref{2secdef}. At the same time, it was demonstrated how to construct
the factorization~(\ref{202}). Finally, it was shown that
the evolution is described in algebro-geometrical terms
 as a linear, with respect to discrete time,
change of (the linear equivalence class of) divisor ${\cal D}$: it changes
by ${\cal D}_1-{\cal D}_6$ per each unit of time.

\section{Algebro-geometrical objects in the orthogonal and symplectic
cases}
\label{2secsmport}

In this section, the algebro-geometrical devices developed earlier
are supplemented with necessary means for studying the evolution
of {\it orthogonal\/} and
{\it symplectic\/} matrices $\cal A$. It turns out
(Section~\ref{2secdecomp}),
that the condition of orthogonality or symplecticity
is an admissible, i.e.\ compatible with the evolution,
reduction of the dynamical system defined in Section~\ref{2secdef}.
We will start, however, with some consideration for {\it generic\/}
matrices $\cal A$, namely, looking at what happens with the invariant
curve~$\Gamma$ and the bundle of vectors~$\cal X$ (see formula (\ref{213}))
under two operations: inversion
${\cal A}\to {\cal A}^{-1}$  and transposing
${\cal A}\to {\cal A}^{\rm T}$.

As for the inversion, here virtually everything is seen from
formula~(\ref{213}), especially if one rewrites it as
\be {\cal A}^{-1} \bmat{c}uX \\ vY \\ wZ \emat=
\bmat{c} X \\ Y \\ Z \emat .  \label{239} \ee
It is clear that the curve $\Gamma$, basically, does not change;
the r\^oles of column vectors $X,Y,Z$ are played by $uX, vY$ and $wZ$,
while the r\^oles of meromorphic functions
$u, v, w$ on $\Gamma$ are played by $u^{-1}, v^{-1}$, and $ w^{-1}$.
The exact formulation is given in the following lemma.

\begin{lemma} \label{2lemAinv}
If to a matrix $\cal A$ corresponds a curve\/ $\Gamma$
given by equations $P(u,v,w)=0$ and $v=uw$ (see~(\ref{211}--\ref{212})
or (\ref{227})), and a divisor $\cal D$ in it (see Lemma~\ref{2lem3}
and the paragraph before it), then to the matrix ${\cal A}^{-1}$
corresponds the curve\/ $\Gamma^{(-1)}$ given by equations
$P(u^{-1}, v^{-1}, w^{-1})=0$ and $v=uw$,
and the divisor ${\cal D}^{(-1)}$ in it that is the image of
$\cal D$ under the natural isomorphism
\be
\Gamma \to \Gamma^{(-1)}: \quad (u, v, w)\mapsto (u^{-1},
v^{-1},w^{-1}).
\label{240}
\ee
\end{lemma}

{\it Proof}. It remains to remind that $\cal D$ is the pole divisor
common for the meromorphic column vectors in LHS and RHS of
(\ref{213})  or (\ref{239}).
Hence the validity of the statement in lemma about the divisor
${\cal D}^{(-1)}$ is clear, and thus the lemma is proved.
Let us add, however, that, for the curve $\Gamma^{(-1)}$,
the r\^oles of divisors ${\cal D}_1$, ${\cal D}_6$, ${\cal D}_2$,
${\cal D}_4$, ${\cal D}_3$ and ${\cal D}_5$  (let them be
listed in this order) are played by the images of divisors
${\cal D}_6$,
${\cal D}_1$, ${\cal D}_4$, ${\cal D}_2$, ${\cal D}_5$ and
${\cal D}_3$. This is due to the fact that the r\^ole of divisor
$(u)_{\infty}$ is played by the image of
$(u)_0$, the r\^ole of $(u)_0$ is played by $(u)_{\infty}$, and similarly
for the functions $v$  and $w$. The connection between those divisors
is given by formulae~(\ref{214}).
\medskip

Consider now the transposing ${\cal A}\to {\cal A}^{\rm T}$.
The determinant is not changed under transposing, thus it follows
from (\ref{211}) that the invariant curve of matrix
${\cal A}^{\rm T}$ is the same $\Gamma$.  Somewhat  more complicated
is to find column vectors
$\tilde X$, $\tilde Y$, $\tilde Z$,
satisfying the equation
\be {\cal A}^{\rm T} \bmat{c} \tilde X\\ \tilde Y \\ \tilde Z \emat =
\bmat{c}\tilde u \tilde X\\ \tilde v \tilde Y \\ \tilde w \tilde Z \emat.
\label{241} \ee
We will obtain them here using the direct constructive method
from the author's work~\cite{Korepanov-ortsym}. Note however that
we might act also in another way, adapting to our case the idea of
{\em divisor duality\/} and of the scalar product of meromorphic functions
on an algebraic curve associated to this duality, ascending to the
work~\cite{B-Ch} and described also in more recent works
\cite{Cherednik,Koz-Kot,Its-Kot}. We will pay tribute to this elegant
idea in Section~\ref{4seca} in the framework of the ``local'' approach
to the evolution of orthogonal matrices, while returning now
to the methods of~\cite{Korepanov-ortsym}.

Transpose (\ref{241}):
\be \left({\tilde X}^{\rm T}\;\;{\tilde  Y}^{\rm T}\;\;
{\tilde Z}^{\rm T}\right){\cal A}=
\left(\tilde u \tilde  X\vphantom X^{\rm T}\; \;
\tilde v \tilde  Y\vphantom X^{\rm T}\;\;
\tilde w \tilde  Z\vphantom X^{\rm T}\right). \label{242} \ee
Multiply both sides of (\ref{242}) from the right (scalarly) by
$\bmat{c} X\\Y\\Z \emat $, i.e.\ by  the column vector appearing in
(\ref{213}), and use the formula (\ref{213}) to exclude
$\cal A$ from the LHS. One gets, after collecting similar terms,
\be (u-\tilde u){\tilde  X}^{\rm T} X+(v-\tilde v){\tilde  Y}^{\rm T} Y+
(w-\tilde w){\tilde Z}^{\rm T}Z=0. \label{243} \ee

In formula (\ref{243}), $(u,v,w)$  and $(\tilde u, \tilde v, \tilde w)$
are so far arbitrary points in the curve~$\Gamma$ (one and the same,
as was explained above). Set now $\tilde u = u$. To a given (generic)
$u$ correspond $n+r$  different $w$'s (recall the formulae~(\ref{227})).
Choose now $w$ and $\tilde w$ different:  $w \neq \tilde w$.
Then (\ref{243}), on being multiplied by
$\displaystyle {\tilde w \over w-\tilde w}$, yields
$$ \tilde v \tilde  Y^{\rm T} Y + \tilde w \tilde  Z^{\rm T} Z=0,
\mbox{ \ \  where  \   } \tilde v=u \tilde w. $$

Let $\tilde w$ be fixed, and $w$ take other $n+r-1$ possible values
which we will denote $w_2,\ldots , w_{n+r}$. Then the components of
vector $\pmatrix{v \tilde Y \cr w \tilde Z}$ are proportional
to the cofactors of the determinant first column entries
\be
\Delta (u, \tilde w)=\left|
\matrix{Y(u,u\tilde w, \tilde w) & Y(u, uw_2, w_2) &
\ldots & Y(u, uw_{n+r}, w_{n+r} )\cr
Z(u, u\tilde w, \tilde w) & Z(u, uw_2, w_2) & \ldots &
Z(u, uw_{n+r}, w_{n+r}) } \right| , \label{244} \ee
Choose the normalization as follows:
\be \pmatrix{ \tilde v \tilde Y \cr  \tilde w \tilde Z } =
{1\over \Delta(u,\tilde w)}\cdot
\pmatrix{\hbox{vector of the}\cr
\hbox{first column entries}\cr
\hbox{cofactors of $\Delta(u,
\tilde w)$}}.
\label{245}
\ee

We have already met the determinant $\Delta(u,\tilde w)$ under the name of
$d$ (formula~(\ref{224a})).  Here, however, we consider it not as a
two-valued function of a point
$(u, u\tilde w, \tilde w)\in \Gamma$ rather than a function of $u$.
This leads to a large multiplication of this determinant zeroes,
some of them being also zeroes of all the minors---components of the
vector from (\ref{245}), while others being not. The zeroes of
$\Delta(u, \tilde w)$ are of interest for us, of course, as candidates
for being poles of the vector
$\pmatrix{ \tilde v\tilde Y\cr \tilde w \tilde Z}$.
Note that, although the enumerator and denominator in the RHS of (\ref{245})
are two-valued functions, the resulting fraction is obviously
single-valued.

Consider first the ``branch points'', i.e.\ the points where either
a)~$w_j=\tilde w$ for some $j$, or b)~$w_j=w_k$ for
some unequal $j$ and $k$. There are $b$ points of the first type
(see formula~(\ref{222})), and those points form the {\em ramification
divisor\/}
${\cal D}_{\rm ram}$, while the points of the second type are of no
interest for us, because all the minors in such points
have, too, zeroes of the same character (namely as the square root
of a local parameter in a curve), so that zeroes in the RHS of
(\ref{245}) cancel one another. Next, we know that in some points
the columns of determinant $\Delta(u,w)$ must vanish as a whole.
Such points, too, are divided into
a)~those where the {\em first\/} column vanishes and
b)~others. The first are again $m$, and they form the divisor~${\cal D}_3$,
while the second are again of no interest, because the minors also vanish
in them.

Other candidates for being the poles of the vector
$\pmatrix{\tilde v \tilde Y \cr \tilde w \tilde Z}$
might be the poles of the enumerator in the RHS of~(\ref{245}).
However, in a point where a column of minors has a pole,
the corresponding column of $\Delta (u,\tilde w)$ has a pole, too.
Recalling also the linear independence of columns of
$\Delta (u,\tilde w)$ (see the proof of Lemma~\ref{2lem4}), we get
a pole of $\Delta (u,\tilde w)$ that cancels the pole of minors.

Thus, the poles of the vector
$\pmatrix{\tilde v \tilde Y \cr \tilde w \tilde Z}$
can only be the points of the divisor
$${\cal D}_{\rm ram} + {\cal D}_3. $$
We must consider also points where  all coordinates of this vector
vanish. Of all candidates the only survivers here are
the {\em first order poles\/} of $\Delta(u, \tilde w)$.
There are $mn+mr+nr$ such poles, and they form the divisor denoted in
this paper as $\cal D$. Other candidates in zeroes of all coordinates of
the vector $\pmatrix{\tilde v \tilde Y \cr \tilde w \tilde Z}$
are sifted away by arguments like those used above when searching
for the poles.

Summarize these considerations in the following lemma.

\begin{lemma}\label{2lemdivXY}
The vector $\pmatrix{\tilde v \tilde Y \cr \tilde w \tilde Z}$
from (\ref{245}) consists of functions $f$ whose divisors obey
the relation
$$
(f)+{\cal D}_{\rm ram} +{\cal D}_3 -{\cal D} \geq 0.
$$
\end{lemma}

It remains to clarify the following. Those functions, in principle, may
obey some stronger constraints, such as
$$
(f)+{\cal D}_{\rm unknown} \geq 0, \hbox{ \ \  where \ }
{\cal D}_{\rm unknown}< {\cal D}_{\rm ram}+{\cal D}_3-{\cal D}.
$$
It turns out that they do not. This is seen from the degrees of divisors:
$$
{\rm deg} \left({\cal D}_{\rm ram} +{\cal D}_3 -{\cal D} \right)
= b+m-(mn+mr+nr)=g+n+r-1
$$
(recall (\ref{222}) and (\ref{225})), and this is exactly what we need.

Denote now as $\tilde {\cal D}$ the divisor of singularities common for
the vectors
$\pmatrix{\tilde X\cr \tilde Y \cr \tilde Z}$ and
$\pmatrix{u\tilde X\cr \tilde v \tilde Y \cr \tilde w \tilde Z}$
(we mean here that {\em poles\/} enter with the plus sign!).
As, for the divisor $\tilde {\cal D}$  and vectors $\tilde X, \tilde Y,
\tilde Z$,  the same relations as (\ref{216}--\ref{218}) hold,
one can derive that $\tilde {\cal D}$ is obtained from the divisor
of singularities of the vector
$\pmatrix{\tilde v \tilde Y \cr \tilde w \tilde Z}$,
by adding ${\cal D}_5$ (for illustration,
Fig.~\ref{2figdecom} again can be used).
Thus, we got the following: {\em under the transposing of a matrix
${\cal A}$, the divisor $\cal D$  changes to}
\be
\tilde {\cal D} = {\cal D}_{\rm ram}+{\cal D}_3+{\cal D}_5 -{\cal D}.
\label{246}
\ee
Here ${\cal D}_{\rm ram}$ is the ramification divisor with regard to
the variable $u$.
In the following theorem, this result is formulated in more symmetric
and elegant form.

\begin{theorem}\label{2thAtr}
Let to a matrix $\cal A$ correspond an algebraic curve\/ $\Gamma$
and a divisor $\cal D$  in it (as was described in Sections \ref{2seccrv}
and \ref{2secevd}), while
to the transposed matrix ${\cal A}^{\rm T}$ ---(the same curve and)
a divisor $\tilde{\cal D}$. Then
\be
\tilde{\cal D}+{\cal D}\sim {\cal D}_{\rm can} +{\cal D}_1+{\cal D}_2
+{\cal D}_3+{\cal D}_4+{\cal D}_5+{\cal D}_6, \label{247}
\ee
where ${\cal D}_{\rm can}$ is a {\em canonical\/} divisor in\/ $\Gamma$,
and divisors ${\cal D}_1,\ldots, {\cal D}_6$ are defined in
Lemma~\ref{2lem2}.
\end{theorem}

{\it Proof}. As is known \cite{Lang}, canonical divisor is the
divisor of any differential in $\Gamma$. Take the differential  $du$ and
find its divisor $(du)=(du)_0-(du)_{\infty}$.
It is not difficult to understand that the zeroes of the form $du$
are situated exactly in the branch points of
$\Gamma$ considered as a covering over the Riemann sphere of
variable $u$:
\be (du)_0={\cal D}_{\rm ram}, \label{248} \ee
while the poles of $du$ coincide with the poles of function
$u$, but are of order~2:
\be (du)_{\infty}=2(u)_{\infty}=2({\cal D}_1+{\cal D}_2), \label{249} \ee
according to Lemma \ref{2lem2}.
It follows from (\ref{248})  and (\ref{249}) that, for any canonical divisor
${\cal D}_{\rm can}$,
$$
{\cal D}_{\rm ram}\sim {\cal D}_{\rm can}+2({\cal D}_1+{\cal D}_2).
$$
For more symmetry, recall that
${\cal D}_1+{\cal D}_2\sim (u)_{\infty} \sim (u)_0 \sim
{\cal D}_4+{\cal D}_6$, after which (\ref{246})  turnes into
(\ref{247}). The theorem is proved.
\medskip

\begin{theorem}\label{2thAM}
If a block matrix $\cal A$ of the form (\ref{201}) is such that
${\cal A}^{\rm T}={\cal M}{\cal A}^{-1}{\cal M}^{-1}$,
with $\cal M$ having the form
\be
{\cal M}=\pmatrix{M_1 &\bf 0&\bf 0 \cr
\bf 0 &M_2&\bf 0 \cr
\bf 0 & \bf 0 &M_3} \label{249a}
\ee
(in other words, ${\cal A}^{\rm T}$ and ${\cal A}^{-1}$ are
{\it gauge equivalent}, cf.~the formula~(\ref{210})),
then the invariant curve\/ $\Gamma$ of matrix $\cal A$ possesses
the involution
\be
I: \quad (u,v,w)\longleftrightarrow (u^{-1}, v^{-1},w^{-1}),
\label{250}
\ee
while the divisor $\cal D$ and its image ${\cal D}^I$ under involution
$I$ obey the equivalence
\be
{\cal D}+{\cal D}^I \sim {\cal D}_{\rm can}+{\cal D}_1+
\ldots +{\cal D}_6 . \label{251}
\ee

Conversely, if a matrix $\cal A$ is such that its curve\/ $\Gamma$
possesses involution (\ref{250}), and the divisor equivalence~(\ref{251})
holds, then ${\cal A}^{\rm T}$  and ${\cal A}^{-1}$
are gauge equivalent.
\end{theorem}

{\it Proof\/} follows at once from Lemma~\ref{2lemAinv}
and Theorems \ref{2thAtr} and \ref{2thvstav}.
\medskip

Further properties of the block-diagonal matrix
$\cal M$ (\ref{249a}) providing the gauge equivalence between
${\cal A}^{\rm T}$ and
${\cal A}^{-1}$ come out during some algebro-geometrical examination,
which we will now perform.
These properties determine, roughly speaking, whether matrix
$\cal A$ is ``in essence'' symplectic or orthogonal.
Calculate $\cal M$, using meromorphic vectors
$$
\tilde {\cal X}=\pmatrix{\tilde X \cr \tilde Y \cr \tilde Z} \qquad
\hbox{  and  } \qquad \bar{\cal X}=
\pmatrix{\bar X\cr \bar Y \cr \bar Z}
$$
of matrices ${\cal A}^{\rm T}$ and ${\cal A}^{-1}$ respectively, connected,
according to Theorem~\ref{2thvstav}, by the relation
\be
\tilde{\cal X}(u,v,w)={\cal M} \bar{\cal X} (u,v,w) h(u,v,w)
\label{252}
\ee
(cf.\ formulae (\ref{227a}, \ref{227b})).
To be exact, we will now deal with blocks $M_2$ and $M_3$, so that the
already obtain formulae (\ref{244}) and (\ref{245}) will be of use.

For a given generic complex number $u$, there exist
$n+r$ points in the curve $\Gamma$ with coordinates
$(u, uw_1, w_1), \ldots,\allowbreak (u, w_{n+r}, w_{n+r})$.
To avoid bulky formulas, we will omit the middle coordinate,
and write those points as $(u,w_1), \ldots ,(u,w_{n+r})$.
One sees from (\ref{244}, \ref{245}) that $\tilde Y$ and $\tilde Z$
are given in all those points at once by the formula
(where, of course, $v_1=u w_1$ and so~on)
$$
\pmatrix{v_1 \tilde Y(u,w_1)& \ldots  &
v_{n+r} \tilde Y(u,w_{n+r}) \cr
w_1 \tilde Z(u,w_1)& \ldots & w_{n+r}\tilde Z(u,w_{n+r}) }^{\rm T}=
$$
\nopagebreak
\be
= \pmatrix{Y(u,w_1) &\ldots & Y(u,w_{n+r}) \cr
Z(u,w_1)& \ldots & Z(u, w_{n+r}) }^{-1}. \label{253}
\ee

For the vectors $\bar Y$ and $\bar Z$ corresponding to the inverse matrix,
the relations
\be
v\bar Y(u,w)= Y(u^{-1}, w^{-1}), \quad w\bar Z(u,w)=Z(u^{-1}, w^{-1}).
\label{254}
\ee
hold. This follows from (\ref{239}), if we replace there
$(u,v,w)$ by  $(u^{-1},v^{-1},w^{-1})$  (the replacement applies, of course,
to the arguments of vector functions $X,Y,Z$ implied in (\ref{239}) as
well).

 From the formula (\ref{252}) follows (as before, we omit $v$ in the
triples $(u,v,w)$)
\be
\pmatrix{v\tilde Y(u,w)\cr w\tilde Z(u,w) }=
\pmatrix{M_2 &\bf 0\cr \bf 0 & M_3 }
\pmatrix{v\bar Y(u,w) \cr w \bar Z (u,w) }
\cdot h(u,w). \label{255}
\ee
Giving $w$ all $n+r$ possible values, we obtain
$n+r$ linearly independent columns in the LHS and RHS,
which allows us to express $M_2$ and $M_3$ from (\ref{255}). It is
convenient to write the result,
with equalities (\ref{253}) and (\ref{254}) taken into account,
in the following form:
$$
\pmatrix{ M_2^{-1} &\bf 0\cr \bf 0& M_3^{-1} }=
\pmatrix{ Y\left(u^{-1}, w_1^{-1}\right) &\ldots&
Y\left(u^{-1}, w_{n+r}^{-1}\right) \cr
Z(u^{-1}, w_1^{-1}) &\ldots & Z\left(u^{-1}, w_{n+r}^{-1}\right)} \cdot
$$
\nopagebreak
\be
\cdot \pmatrix{h(u,w_1) &&\cr &\ddots &\cr
&&h(u,w_{n+r}) }
\pmatrix{ Y(u,w_1)&\ldots & Y(u, w_{n+r}\cr
Z(u,w_1)&\ldots &Z(u,w_{n+r}) }^{\rm T} . \label{256}
\ee

Recall that $h$ performs the divisor equivalence between $\tilde{\cal D}$,
given by the formula (\ref{246}), and ${\cal D}^I$ ---the image of
$\cal D$ under involution $I$, in the sense that
$$
(h)={\cal D}^I-\tilde{\cal D}.
$$
Substituting here (\ref{246}), we find
\be
(h)={\cal D}+{\cal D}^I - {\cal D}_{\rm ram} -{\cal D}_3 -{\cal D}_5 .
\label{257}
\ee
The divisor in the RHS of (\ref{257}) is obviously invariant
with respect to involution $I$, thus the divisor $(h)$ of function $h$
possesses the same property. If the curve $\Gamma$ is irreducible, and
this is exactly the fact in the general position, then
$h$ is determined by its divisor up to a constant factor.
This means that under the involution $I$ the function $h$ is multiplied
by a constant which, evidently, must equal $\pm 1$:
\be
h(u,w)=\pm h(u^{-1},w^{-1}). \label{258}
\ee

Return to the equality (\ref{256}). It has a constant matrix, not
depending on a point of the curve, in its LHS.
Take the matrix transpose of that equality,
and change $(u,w)\leftrightarrow (u^{-1},w^{-1})$. We get:
$$
\pmatrix{ M_2 & \bf 0 \cr \bf 0 & M_3 }=
\pm \pmatrix{ M_2 &\bf 0 \cr \bf 0 &M_3 }^{\rm T},
$$
where the sign coincides with that in (\ref{258}). It is clear that,
for the similar reasons, the equality $M_1=\pm M_1^{\rm T}$ also holds.
As a result, the following lemma is proved.

\begin{lemma} \label{2lempm}
If a block matrix $\cal A$ has the property
\be
{\cal A}^{\rm T}={\cal M}{\cal A}^{-1}{\cal M}^{-1}, \label{259}
\ee
where $\cal M$ is a block diagonal matrix, then $\cal M$ is symmetric
or antisymmetric if the function $h$ in the curve\/ $\Gamma$ with the
zero and pole divisor (\ref{257}) is even or odd with respect to the
involution $I$, correspondingly.
\end{lemma}

It follows from (\ref{259}) that to a gauge transformation
$$
{\cal A}\to{\cal N A N}^{-1},\qquad {\cal N}={\rm diag}\, (N_1, N_2, N_3),
$$
corresponds the transformation
\be
{\cal M}\to \left({\cal N}^{-1}\right)^{\rm T}{\cal MN}^{-1}
\label{260}
\ee
of matrix $\cal M$. A symmetrical $\cal M$ can be reduced by such a
transformation to an identity matrix, and for an antisymmetrical one
each diagonal block $M_i$ can be reduced to the standard form
$\pmatrix{\bf 0&\bf 1\cr \bf -1&\bf 0}$ consisting of half-size blocks.
The relation~(\ref{259}) means, after this transformation,
the orthogonality of $\cal A$ in the first case and its symplecticity---in
the second case.

Let us summarize these considerations in the following theorem.

\begin{theorem}\label{2thsmport}
In the notations and under the assumptions of Theorem~\ref{2thAM}, a
matrix $\cal A$ with an irreducible curve\/ $\Gamma$ is gauge
equivalent to an orthogonal or symplectic matrix, if the function~$h$
with the zero and pole divisor~(\ref{257}) is even or odd, correspondingly,
with respect to the involution $I$ (\ref{250}).
\end{theorem}

Thus, in this section we studied the algebro-geome\-tric\-al objects
in a specific case of orthogonal or symplectic matrices ${\cal A}$.
Now it would be not very hard work to show that those objects retain
their specific form under the evolution, thus proving
that a matrix~${\cal A}$ retains its property to be gauge equivalent
to an orthogonal or symplectic matrix. We will prefer, however,
to go another way, considering the decomposition~(\ref{202})
in orthogonal and symplectic cases.

\section{Matrix factorization in case of orthogonality or symplecticity,
and conservation of those properties under evolution}
\label{2secdecomp}

It is natural to expect that an orthogonal or symplectic matrix
$\cal A$ can be factorized in a product of matrices of the same kind
(orthogonal or symplectic) ${\cal A}_1$, ${\cal A}_2$, ${\cal A}_3$
(formulae (\ref{202}, \ref{203})).  Indeed, the following lemma holds.

\begin{lemma}\label{2lemosftion}
Under conditions of Lemma~\ref{2lempm}, for the matrix $\cal A$ such
a factorization (\ref{202}, \ref{203}) exists that for each matrix
${\cal A}_i$ separately, $i=1,2,3$, a relation like
(\ref{259}) is valid, i.e.
\be
{\cal A}_i^{\rm T}= {\cal M}{\cal A}_i^{-1}{\cal M}^{-1}.\label{261}
\ee
\end{lemma}

{\it Proof.}  Consider, to begin, an arbitrary factorization
${\cal A}={\cal A}_1{\cal A}_2 {\cal A}_3 $ of matrix
${\cal A}$, not requiring that (\ref{261}) hold.
Then there are obvious factorizations
\be
{\cal A}^{\rm T}={\cal A}^{\rm T}_3 {\cal A}^{\rm T}_2 {\cal A}^{\rm T}_1
\label{262}
\ee
and
\be
{\cal A}^{-1}={\cal A}^{-1}_3 {\cal A}^{-1}_2 {\cal A}^{-1}_1.
\label{263}
\ee
One more factorization of matrix ${\cal A}^{\rm T}$, besides (\ref{262}),
into a product of three matrices with zero and unity blocks in the
same places, can be obtained from
(\ref{259}) and (\ref{263}) in the following way:
$$
{\cal A}^{\rm T}={\cal M}{\cal A}^{-1}{\cal M}^{-1}=
{\cal M}{\cal A}^{-1}_3{\cal A}^{-1}_2{\cal A}^{-1}_1 {\cal M}^{-1}=
$$
\be
=\left({\cal M}{\cal A}^{-1}_3 {\cal M}^{-1}\right)
\left({\cal M}{\cal A}^{-1}_2{\cal M}^{-1}\right)
\left({\cal M}{\cal A}^{-1}_1 {\cal M}^{-1}\right).
\label{264}
\ee
Applying Lemma~\ref{2lem1} to the two factorizations (\ref{262}) and
(\ref{264}), we deduce that
the following relations, for some nondegenerate matrices $F_1$, $F_2$, $F_3$
of proper sizes, must hold:
\be
{\cal A}^{\rm T}_3=\left( {\cal M}{\cal A}^{-1}_3{\cal M}^{-1}\right)
\pmatrix{
{\bf 1}&{\bf 0}&{\bf 0} \cr
{\bf 0}&F_2&{\bf 0}\cr
{\bf 0}& {\bf 0}& F_3 },
\label{265a}
\ee
\be
{\cal A}^{\rm T}_2=
\pmatrix{
{\bf 1}&{\bf 0}&{\bf 0} \cr
{\bf 0}&{\bf 1}&{\bf 0}\cr
{\bf 0}& {\bf 0}& F^{-1}_3 }
\left( {\cal M}{\cal A}^{-1}_2{\cal M}^{-1}\right)
\pmatrix{
F_1&{\bf 0}&{\bf 0} \cr
{\bf 0}&{\bf 1}&{\bf 0}\cr
{\bf 0}& {\bf 0}&{\bf 1} },
\label{265b}
\ee

\be
{\cal A}^{\rm T}_1=
\pmatrix{
F^{-1}_1&{\bf 0}&{\bf 0} \cr
{\bf 0}&F_2^{-1}&{\bf 0}\cr
{\bf 0}& {\bf 0}& {\bf 1} }
\left( {\cal M}{\cal A}^{-1}_1{\cal M}^{-1}\right).
\label{265c}
\ee

Our aim now is to find such a transformation of type (\ref{204}) for matrices
${\cal A}_1$, ${\cal A}_2$, ${\cal A}_3$ that the relations
(\ref{261}) hold. Consider at first a matrix
\be
\tilde{\cal A}_3=
\pmatrix{
{\bf 1}&{\bf 0}&{\bf 0} \cr
{\bf 0}&K_2&{\bf 0}\cr
{\bf 0}& {\bf 0}& K_3 }{\cal A}_3
\label{266}
\ee
with some nondegenerate $K_2$ and $K_3$. Relations (\ref{266}) and
(\ref{265a}) together yield the following  connection between
${\tilde{\cal A}}^{\rm T}_3$ and ${\tilde{\cal A}}_3^{-1}$:
\be
{\tilde{\cal A}}^{\rm T}_3={\cal M}{\tilde{\cal A}}^{-1}_3
\pmatrix{
{\bf 1}&{\bf 0}&{\bf 0} \cr
{\bf 0}&K_2&{\bf 0}\cr
{\bf 0}& {\bf 0}& K_3 }
{\cal M}^{-1}
\pmatrix{
{\bf 1}&{\bf 0}&{\bf 0} \cr
{\bf 0}&K_2^{\rm T}&{\bf 0}\cr
{\bf 0}& {\bf 0}& K_3^{\rm T} }
\pmatrix{
{\bf 1}&{\bf 0}&{\bf 0} \cr
{\bf 0}&F_2&{\bf 0}\cr
{\bf 0}& {\bf 0}& F_3 }.
\label{267}
\ee
This will turn into the required relation
\be
{\tilde{\cal A}}^{\rm T}_3={\cal M}{\tilde{\cal A}}^{-1}_3 {\cal M}^{-1}
\label{268}
\ee
if we manage to obey, by a proper choice of $K_2$ and $K_3$, the following
equalities:
\be
K_2M^{-1}_2K^{\rm T}_2 F_2 = M^{-1}_2,
\label{269}
\ee
\be
K_3M^{-1}_3K^{\rm T}_3F_3=M^{-1}_3.
\label{270}
\ee
Rewrite (\ref{269}) in the form
\be
K_2M^{-1}_2K^{\rm T}_2=M^{-1}_2 F^{-1}_2.
\label{271}
\ee
In the following Lemma~\ref{2lemMF} we will demonstrate that the RHS of
(\ref{271}) is symmetric or antisymmetric in case $M_2$
is symmetric or antisymmetric correspondingly.
Assuming, for a while, this fact without proof, we find that
the equation (\ref{271})
is always solvable with respect to $K_2$ ---this is a simple consequence
from the properties of quadratic forms and antisymmetric bilinear forms
and their matrices, see, e.g.,
\S\S90 and 91 of the manual~\cite{Waerden}
(one must take into account here also the nondegeneracy of matrices
$M_2^{-1}$ and $F_2^{-1}$).

Similarly, the equation (\ref{270}) is solvable with respect to $K_3$.
Thus, we managed to obey the relation~(\ref{268}).
Then, solving the equation for $K_1$, similar to (\ref{269}) and (\ref{270}),
we see that the relations
$$
{\tilde{\cal A}}^{\rm T}_2={\cal M}{\tilde{\cal A}}^{\rm T}_2{\cal M}^{-1}
$$
and
$$
{\tilde{\cal A}}^{\rm T}_1={\cal M}{\tilde{\cal A}}^{\rm T}_1{\cal M}^{-1}
$$
also hold, for
$$
{\tilde{\cal A}}^{\rm T}_2=
\pmatrix{
K_1&{\bf 0}&{\bf 0} \cr
{\bf 0}&{\bf 1}&{\bf 0}\cr
{\bf 0}& {\bf 0}& {\bf 1} }
{\cal A}_2
\pmatrix{
{\bf 1}&{\bf 0}&{\bf 0} \cr
{\bf 0}&{\bf 1}&{\bf 0}\cr
{\bf 0}& {\bf 0}& K^{-1}_3 }
$$
and
$$
{\tilde{\cal A}}_1={\cal A}_1
\pmatrix{
K_1^{-1}&{\bf 0}&{\bf 0} \cr
{\bf 0}&K_2^{-1}&{\bf 0}\cr
{\bf 0}& {\bf 0}& {\bf 1} }.
$$
We don't need any more the initial matrices ${\cal A}_i$ without tildes in
the rest of this Proof. Rename the new matrices $\tilde{\cal A}_i$  into
${\cal A}_i$, and with this Lemma~\ref{2lemosftion} is proved. Recall,
however, that its proof was based on the following lemma
(and its analogs arising from changing the subscript
2 to 1 or 3).

\begin{lemma}\label{2lemMF}
If, in the notations of Lemma~\ref{2lemosftion},
\be
M_2=\pm M^{\rm T}_2,
\label{272}
\ee
then
\be
\left(M^{-1}_2 F^{-1}_2 \right)=
\pm\left(M^{-1}_2F^{-1}_2\right)^{\rm T} ,
\label{273}
\ee
with the same sign as in (\ref{272}).
\end{lemma}

{\it Proof.}  Take the equality (\ref{265a}) and apply to its both
sides, first, the transposing, and second, the matrix inversion.
We will get, in the LHS, ${\cal A}^{-1}_3$, while in the RHS---some
expression containing ${\cal A}^{\rm T}_3$, which we have no need to
write down here.
This done, express again ${\cal A}^{\rm T}_3$ through ${\cal A}^{-1}_3$
by multiplying the LHS and RHS of the obtained equality by suitable
matrices and the interchanging LHS and RHS.
The result will be the following:
\be
{\cal A}^{\rm T}_3=
{\cal M}{\cal A}^{-1}_3
\pmatrix{
{\bf 1}&{\bf 0}&{\bf 0}\cr
{\bf 0}&F^{\rm T}_2&{\bf 0}\cr
{\bf 0}&{\bf 0}&F^{\rm T}_3
}
{\cal M}^{-1}.
\label{274}
\ee
Comparing (\ref{274}) with (\ref{265a}), we find, in particular, that
\be
M^{-1}_2 F_2 =F^{\rm T}_2 M^{-1}_2.
\label{275}
\ee
It is easy to see that the equality (\ref{275}) is equivalent to (\ref{273}),
and with this the proof of Lemma~\ref{2lemMF} is over, and the proof
of Lemma~\ref{2lemosftion} is thus complete, too.
\medskip

The following theorem, which summarizes the results of this section,
immediately follows from Lemma~\ref{2lemosftion}.

\begin{theorem}\label{2thconsSO}
The property of a matrix ${\cal A}$  of the form~(\ref{201}) to be
gauge equivalent to an orthogonal or symplectic matrix
and, consequently, to be factorable in
a product (\ref{202}, \ref{203}) of matrices of the same kind, is conserved
under the evolution described in Section~\ref{2secdef}.
\end{theorem}

{\it Proof\/} follows at once from the definition of this evolution
given in the end of Section~\ref{2secdef}.
\medskip

\chapter{Inhomogeneous 6-vertex model}

\section{Dynamical systen connected with the 6-vertex model on the
kagome lattice}
\label{2seckag}

Consider now a reduction of the system defined in Section~\ref{2secdef}
which leads to a dynamical system in  $2+1$-dimensional fully discrete
space-time. Let the linear space in which matrix $\cal A$~(\ref{201})
acts have a basis enumerated by edges of a triangular lattice
on the torus (Fig.~\ref{2figtria}),
\bfig
\begin{center}
\unitlength=0.104606\normalbaselineskip
\special{em:linewidth 0.4pt}
\linethickness{0.4pt}
\begin{picture}(91.00,91.99)
\put(16.00,15.91){\circle{4.00}}
\put(46.00,15.91){\circle{4.00}}
\put(76.00,15.91){\circle{4.00}}
\put(76.00,46.02){\circle{4.00}}
\put(46.00,46.02){\circle{4.00}}
\put(16.00,46.02){\circle{4.00}}
\put(16.00,76.13){\circle{4.00}}
\put(46.00,76.13){\circle{4.00}}
\put(76.00,76.13){\circle{4.00}}
\put(91.00,76.13){\line(-1,0){13.00}}
\put(74.00,76.13){\line(-1,0){26.00}}
\put(44.00,76.13){\line(-1,0){26.00}}
\put(14.00,76.13){\line(-1,0){13.00}}
\put(1.00,46.02){\line(1,0){13.00}}
\put(18.00,46.02){\line(1,0){26.00}}
\put(48.00,46.02){\line(1,0){26.00}}
\put(78.00,46.02){\line(1,0){13.00}}
\put(91.00,15.91){\line(-1,0){13.00}}
\put(74.00,15.91){\line(-1,0){26.00}}
\put(44.00,15.91){\line(-1,0){26.00}}
\put(14.00,15.91){\line(-1,0){13.00}}
\put(16.00,1.29){\line(0,1){12.04}}
\put(16.00,13.33){\line(0,1){0.86}}
\put(16.00,18.07){\line(0,1){26.24}}
\put(16.00,48.17){\line(0,1){25.81}}
\put(16.00,78.28){\line(0,1){12.90}}
\put(46.00,91.18){\line(0,-1){12.90}}
\put(46.00,73.98){\line(0,-1){25.81}}
\put(46.00,44.30){\line(0,-1){26.24}}
\put(46.00,14.19){\line(0,-1){12.90}}
\put(76.00,1.29){\line(0,1){12.90}}
\put(76.00,18.07){\line(0,1){26.24}}
\put(76.00,48.17){\line(0,1){25.81}}
\put(76.00,78.28){\line(0,1){12.90}}
\put(18.00,47.31){\line(1,1){27.00}}
\put(48.00,76.99){\line(1,1){15.00}}
\put(48.00,47.31){\line(1,1){27.00}}
\put(78.00,76.99){\line(1,1){13.00}}
\put(78.00,47.31){\line(1,1){13.00}}
\put(78.00,17.21){\line(1,1){13.00}}
\put(1.00,1.29){\line(1,1){13.00}}
\put(31.00,1.29){\line(1,1){13.00}}
\put(61.00,1.29){\line(1,1){13.00}}
\put(1.00,30.97){\line(1,1){13.00}}
\put(13.00,43.01){\line(4,5){1.00}}
\put(1.00,61.08){\line(1,1){13.00}}
\put(31.00,91.18){\line(-1,-1){13.00}}
\put(18.00,18.06){\line(1,1){26.00}}
\put(48.00,18.06){\line(1,1){26.00}}
\end{picture}
\end{center}
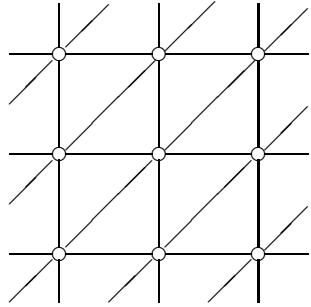
\caption{The triangular lattice}
\label{2figtria}
\bigskip
\begin{center}
\unitlength=0.104606\normalbaselineskip
\special{em:linewidth 0.4pt}
\linethickness{0.4pt}
\begin{picture}(67.00,65.61)
\put(34.00,32.70){\circle{4.00}}
\put(34.00,12.48){\vector(0,1){13.33}}
\put(34.00,25.82){\line(0,1){4.73}}
\put(34.00,34.85){\vector(0,1){9.03}}
\put(34.00,43.88){\line(0,1){10.75}}
\put(51.00,40.87){\makebox(0,0)[lc]{``outgoing'' edges}}
\put(22.00,26.68){\makebox(0,0)[rc]{``incoming'' edges}}
\put(24.00,22.81){\line(1,1){8.00}}
\put(13.00,11.62){\vector(1,1){11.00}}
\put(0.00,32.69){\vector(1,0){18.00}}
\put(18.00,32.69){\line(1,0){14.00}}
\put(36.00,32.69){\vector(1,0){14.00}}
\put(50.00,32.69){\line(1,0){17.00}}
\put(34.00,0.86){\line(0,1){12.04}}
\put(34.00,54.63){\line(0,1){10.32}}
\put(36.00,34.84){\vector(1,1){14.00}}
\put(50.00,48.61){\line(1,1){17.00}}
\put(13.00,11.61){\line(-1,-1){11.00}}
\end{picture}
\end{center}
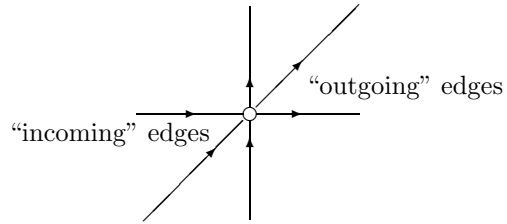
\caption{The vectors corresponding to ``incoming'' edges are transformed by
$\cal A$ into linear combinations of those corresponding to ``outgoing''
edges}
\label{2figinout}
\efig
so that those components of the vector ${\cal X}=\pmatrix{X\cr Y\cr Z}$
corresponding to horisontal edges form the vector~$X$,
while those corresponding to oblique edges form the vector~$Y$,
and the ones corresponding to vertical edges form the vector~$Z$.
As the number of each of those three types of edges is the same,
all the blocks in the matrix $\cal A $ have the same sizes:
\be
m=n=r\quad ({}={}\hbox{the number of lattice edges}).
\label{232.5}
\ee
Impose the following ``locality'' condition on matrix $\cal A $:
let the vector corresponding  to any given edge of the lattice be
transformed under the action of  $\cal A$ into a linear combination of
just three vectors, corresponding to the edges coming upwards, to the right
and  northeastwards from the vertex that is the upper, right, or
northeastern end of the considered ``incoming'' edge (Fig.~\ref{2figinout}).
Thus, only those elements of matrix $\cal A$ are not zeros that correspond to
``local'' transitions of Fig.~\ref{2figinout}.

The factorization of a ``local'' matrix  $\cal A$ into the product
(\ref{202}) corresponds to each vertex represented by a small circle in
Fig.~\ref{2figtria} being converted into a triangle of the type shown in
Fig.\ref{2figdecom}, so that the lattice transforms into a kagome lattice
(Fig.~\ref{2figkagome}). The triangles arising from the vertices-circles
are shaded in Fig.~\ref{2figkagome}.
\bfig
\begin{center}
\unitlength=0.104606\normalbaselineskip
\special{em:linewidth 0.4pt}
\linethickness{0.4pt}
\begin{picture}(80.00,80.86)
\put(10.00,0.86){\line(0,1){40.00}}
\put(10.00,40.86){\line(0,0){0.00}}
\put(10.00,40.86){\line(0,0){0.00}}
\put(10.00,40.86){\line(0,0){0.00}}
\put(10.00,40.86){\line(0,0){0.00}}
\put(10.00,40.86){\line(0,0){0.00}}
\put(10.00,40.86){\line(0,0){0.00}}
\put(10.00,40.86){\line(0,0){0.00}}
\put(10.00,40.86){\line(0,0){0.00}}
\put(10.00,40.86){\line(0,0){0.00}}
\put(10.00,40.86){\line(0,0){0.00}}
\put(10.00,40.86){\line(0,0){0.00}}
\put(10.00,40.86){\line(0,0){0.00}}
\put(10.00,40.86){\line(0,0){0.00}}
\put(10.00,40.86){\line(0,0){0.00}}
\put(10.00,40.86){\line(0,0){0.00}}
\put(10.00,40.86){\line(0,0){0.00}}
\put(10.00,40.86){\line(0,1){40.00}}
\put(40.00,80.86){\line(0,-1){80.00}}
\put(70.00,0.86){\line(0,1){80.00}}
\put(0.00,10.75){\line(1,0){80.00}}
\put(80.00,40.86){\line(-1,0){80.00}}
\put(0.00,70.97){\line(1,0){80.00}}
\put(2.00,48.60){\line(1,1){30.00}}
\put(2.00,18.93){\line(1,1){60.00}}
\put(17.00,3.44){\line(1,1){60.00}}
\put(48.00,2.58){\line(1,1){30.00}}
\put(13.00,70.97){\line(0,-1){11.18}}
\put(16.00,62.80){\line(0,1){8.17}}
\put(19.00,70.97){\line(0,-1){5.16}}
\put(13.00,40.86){\line(0,-1){11.18}}
\put(16.00,32.69){\line(0,1){8.17}}
\put(19.00,40.86){\line(0,-1){5.16}}
\put(43.00,40.86){\line(0,-1){11.18}}
\put(46.00,32.69){\line(0,1){8.17}}
\put(49.00,40.86){\line(0,-1){5.16}}
\put(43.00,70.97){\line(0,-1){11.18}}
\put(46.00,62.80){\line(0,1){8.17}}
\put(49.00,70.97){\line(0,-1){5.16}}
\end{picture}
\end{center}
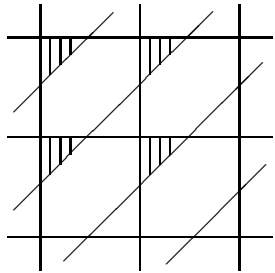
\caption{The kagome lattice}
\label{2figkagome}
\efig

One can easily see that the ``locality'' property of matrix $\cal A$
is preserved by a step of evolution, if, of course, the proper gauge is
taken. The detailed description of the
step of evolution from the ``local'' viewpoint and the description of
gauge transformations preserving the ``locality'' of $\cal A$
are given in the beginning of Section~\ref{3secqp}. Here we will
concentrated on the values conserved under the evolution.

Let us return to the triangle lattice of Figure~\ref{2figtria}.
Express the ``integral of motion''
\be
I(u,w)= \det \bigl({\bf 1}-{\cal A} \bmat{ccc} u^{-1}&0&0\\
0&u^{-1}w^{-1}&0 \\ 0&0&w^{-1} \emat \bigr) \label{233}
\ee
in terms of paths going along the edges of this lattice ($I(u,w)$
is indeed an integral of motion with any $u,w$, because the equality
$I(u,w)=0$ determines the invariant curve, and a possible
multiplicative constant is fixed by the fact that the constant term
in (\ref{233}) equals unity).

As it known, the determinant of a martix is an alternating sum of
its elements' products, each summand corresponding to some permutation
of the matrix columns, while each permutation  factorizes into a product
of cyclic ones. As applied to our matrix $\cal A$, it means that the
determinant~(\ref{233}) is a sum each term of which corresponds to a set
of closed trajectories going along the arrows according to
Fig.~\ref{2figinout} (recall that the lattice is situated on the torus!).
The trajectories of each given set can have intersections and
self-intersections, but none of the {\em edges} may be passed through twice
or more by one or several trajectories.

To be exact, to each trajectory corresponds a product of entries of
the matrix
$$
{\cal A} \bmat{ccc} u^{-1}&0&\\ 0&u^{-1}w^{-1}&0\\
0&0&w^{-1} \emat
$$
corresponding to transitions through a vertex to a neighboring edge
according to Fig.~\ref{2figinout}, multiplied (the product as a whole)
by  $(-1)$. To each set of trajectories (including, of course, the empty
set) corresponds the product of the mentioned values corresponding to
its trajectories. A direct check shows that all the minus signs,
including that in formula~(\ref{233}), have been taken into account correctly.

It is easy also to describe the determinant $I(u,w)$ in terms of the kagome
lattice obtained on factorizing the matrix  $\cal A$ into the
product~(\ref{202}).
This description almost repeats two preceding paragraphs. Let us formulate it
as the following lemma.
\begin{lemma}
\label{2lemdet}
$I(u,w)$ is a sum over sets of trajectories on the kagome lattice; the
direction of motion is upwards,  to the right, or northeastwards;
none of the edges is passed through twice by trajectories of a given set;
to the vertices of types
\unitlength=0.083685\normalbaselineskip
\special{em:linewidth 0.4pt}
\linethickness{0.4pt}
\begin{picture}(15.00,9.46)
\put(1.00,4.73){\line(1,0){14.00}}
\put(1.00,0.00){\line(3,2){14.00}}
\end{picture}
 ,
\unitlength=0.083685\normalbaselineskip
\special{em:linewidth 0.4pt}
\linethickness{0.4pt}
\begin{picture}(14.00,12.04)
\put(0.00,6.02){\line(1,0){14.00}}
\put(7.00,0.00){\line(0,1){12.04}}
\end{picture}
 ,
\unitlength=0.083685\normalbaselineskip
\special{em:linewidth 0.4pt}
\linethickness{0.4pt}
\begin{picture}(9.00,11.18)
\put(4.00,0.00){\line(0,1){11.18}}
\put(0.00,0.00){\line(4,5){9.00}}
\end{picture}
 \/,
if a trajectory passes through them, correspond the factors
equalling matrix elements of matrices  ${\cal A}_1$, ${\cal A}_2$,
${\cal A}_3$ respectively; besides, to each move to the right through
a lattice period corresponds a factor $u^{-1}$, and to each move
upwards---a factor $w^{-1}$ (and both of them to a diagonal move);
finally, to each  set corresponds one more factor,
$(-1)^{\mbox{\small\rm (number of trajectories)}}$.
\end{lemma}

Now let us link  $I(u,w)$ with the statistical sum of inhomogeneous
6-vertex model on the kagome lattice. Let each edge of the kagome lattice
be able to take one of two states, which will be depicted below
as either presence or absence of an arrow on the edge (the arrow will
always be directed upwards, to the right, or northeastwards).
A ``Boltzmann weight'' will correspond to each vertex as follows:
if there are no arrows on the edges meeting at the vertex,
the weight will be $1$; if there is exactly one arrow coming into
the vertex and exactly one going out of it, the weight will be equal to
the corresponding matrix element of  ${\cal A}_1$, ${\cal A}_2$
or ${\cal A}_3$ (e.g., to the vertex
\unitlength=0.125526\normalbaselineskip
\special{em:linewidth 0.4pt}
\linethickness{0.4pt}
\begin{picture}(20.00,12.00)
\put(0.00,2.00){\vector(1,0){5.00}}
\put(5.00,2.00){\line(1,0){15.00}}
\put(10.00,-8.00){\vector(0,1){15.00}}
\put(10.00,7.00){\line(0,1){5.00}}
\end{picture}
\vspace{3mm} \ \
corresponds the weight equal to the matrix element of ${\cal A}_2$
that is responsible for a transition between the vector
corresponding to the left edge, and the vector corresponding to the upper
edge); if there are 2 incoming and 2 outgoing arrows, the weight is
the difference between the products of weights corresponding to the
intersecting and non-intersecting paths through the vertex:
\be
\ba{c} \mbox{Weight} \left(
\unitlength=0.125526\normalbaselineskip
\special{em:linewidth 0.4pt}
\linethickness{0.4pt}
\begin{picture}(20.00,12.00)
\put(0.00,2.00){\vector(1,0){6.00}}
\put(6.00,2.00){\vector(1,0){10.00}}
\put(6.00,2.00){\line(1,0){14.00}}
\put(10.00,-8.00){\vector(0,1){6.00}}
\put(10.00,-2.00){\vector(0,1){10.00}}
\put(10.00,8.00){\line(0,1){4.00}}
\end{picture}
\right)=\mbox{Weight}
\left(
\unitlength=0.125526\normalbaselineskip
\special{em:linewidth 0.4pt}
\linethickness{0.4pt}
\begin{picture}(20.00,12.00)
\put(0.00,2.00){\vector(1,0){5.00}}
\put(5.00,2.00){\line(1,0){15.00}}
\put(10.00,-8.00){\vector(0,1){15.00}}
\put(10.00,7.00){\line(0,1){5.00}}
\end{picture}
 \right)
\cdot \mbox{Weight}\left(
\unitlength=0.125526\normalbaselineskip
\special{em:linewidth 0.4pt}
\linethickness{0.4pt}
\begin{picture}(20.00,12.00)
\put(0.00,2.00){\vector(1,0){15.00}}
\put(15.00,2.00){\line(1,0){5.00}}
\put(10.00,-8.00){\vector(0,1){5.00}}
\put(10.00,-3.00){\line(0,1){15.00}}
\end{picture}
 \right)- \\\noalign{\medskip}
-\mbox{Weight} \left(
\unitlength=0.125526\normalbaselineskip
\special{em:linewidth 0.4pt}
\linethickness{0.4pt}
\begin{picture}(20.00,12.00)
\put(0.00,2.00){\vector(1,0){6.00}}
\put(6.00,2.00){\vector(1,0){10.00}}
\put(16.00,2.00){\line(1,0){4.00}}
\put(10.00,-8.00){\line(0,1){20.00}}
\end{picture}
 \right) \cdot
\mbox{Weight}\left(
\unitlength=0.125526\normalbaselineskip
\special{em:linewidth 0.4pt}
\linethickness{0.4pt}
\begin{picture}(20.00,12.00)
\put(0.00,2.00){\line(1,0){20.00}}
\put(10.00,-2.00){\vector(0,1){10.00}}
\put(10.00,8.00){\line(0,1){4.00}}
\put(10.00,-8.00){\vector(0,1){6.00}}
\end{picture}
 \right); \ea
\label{234}
\ee
in the rest of cases the weight is zero.

A weight will also correspond to each edge of the kagome lattice:
weight $1$ to an edge without an arrow, and weights  $u^{-1/2}$,
$w^{-1/2}$ or  $u^{-1/2}w^{-1/2}$ to a horizontal, vertical or oblique
edge having an arrow. If needed, the edge weights can be included in the
vertex weights, but we will not do that here.

The statistical sum $S(u,w)$ of our 6-vertex model is, of course, a sum
of products of vertex and edge weights over all arrow configurations.
The next lemma is the key statement.

\begin{lemma} \label{2lemsta}
The statistical sum $S(u,w)$ is a sum over the same sets of trajectories as
the determinant $I(u,w)$, and to each set corresponds the same summand
up to, maybe, a minus sign. To be exact, the {\rm number of trajectories}
 in the
exponent of $(-1)$ in Lemma~\ref{2lemdet}
changes to the {\rm number of intersections} (self-intersections included)
of a given set of trajectories.
\end{lemma}

{\it Proof\/} is evident from the statistical sum definition.
\medskip

Each closed path on the torus is homologically equivalent to a linear
combination of two basis cycles ${\bf a}$ and ${\bf b}$. The same is true
for a set of paths (trajectories), regarded as a formal sum of them.
Different sets may be homologically equivalent to a given cycle
$l{\bf a}+m{\bf b}$ but, as the following lemma shows, they have something
in common.
\begin{lemma} \label{2lemtor}
For any set of trajectories on the torus homologically equivalent to a cycle
$l{\bf a}+m{\bf b}$ (${\bf a},{\bf b}$ being basis cycles, $l,m$---integers),
\be
(\mbox{\rm number of intersections})-(\mbox{\rm number of trajectories})
\equiv lm-l-m(\mbox{\rm mod}\, 2).
\label{235}
\ee
\end{lemma}

{\it Proof\/} may consist in the following simple consideration:
1)~if the set consists of $l$ trajectories going along ${\bf a}$, and $m$
ones going along ${\bf b}$, (\ref{235}) is obviously true,
2)~under deformations of trajectories, the number
of intersections changes only by even numbers,
3)~with elimination of an intersection $\left(
\unitlength=0.104606\normalbaselineskip
\special{em:linewidth 0.4pt}
\linethickness{0.4pt}
\begin{picture}(59.00,12.00)
\put(10.00,-8.00){\line(0,1){20.00}}
\put(0.00,2.00){\line(1,0){20.00}}
\put(24.00,2.00){\vector(1,0){10.00}}
\put(39.00,12.00){\oval(20.00,20.00)[rb]}
\put(59.00,-8.00){\oval(20.00,20.00)[lt]}
\end{picture}
\right)$
or inverse operation, the LHS of (\ref{235}) may change also only by an
even number. Starting from an arbitrary set and applying the
transformations 2) and 3), one can arrive at a set of type 1),
so the lemma is proved.
\medskip

Let the following products of edge weights correspond to the basis cycles:
$x=u^{\alpha_1}w^{\beta_1}$  for ${\bf a}$ and
$y=u^{\alpha_2}w^{\beta_2}$ for ${\bf b}$. Denote
\be
s(x,y)=S(u,w), \quad f(x,y)=I(u,w). \label{236}
\ee

\begin{theorem}
\label{2thstadet}
The statistical sum of the inhomogeneous 6-vertex model on the kagome lattice
defined in this section is invariant with respect to the evolution of the
reduced
$2+1$-dimensional model (for all $u,w$) and is connected with the
determinant $I(u,w)$~(\ref{233}), whose vanishing defines the invariant curve
of the model, by relations (in the notations of (\ref{236}))
\be
s(x,y)=1/2 \left( -f(x,y)+f(-x,y)+f(x,-y)+f(-x,-y) \right), \label{237}
\ee
\be
f(x,y)=1/2 \left( -s(x,y)+s(-x,y)+s(x,-y)+s(-x,-y) \right). \label{238}
\ee
\end{theorem}

{\it Proof.} It follows from Lemmas~\ref{2lemdet}, \ref{2lemsta},
\ref{2lemtor} that in the expansions of $s(x,y)$ and  $f(x,y)$ in powers of
$x$ and $y$ the coefficients  near $x^l y^m$ coincide if $lm-l-m$ is even,
and differ in their signs in the opposite case. This is exactly what the
formulae~(\ref{237},\ref{238}) are about.

\begin{remark}\label{2remdetphys}
 From the physical viewpoint, the determinant $I(u,w)$ can be regarded as
a statistical sum of the same model, but with other boundary conditions
than for $S(u,w)$. This can be explained as follows.
Let us regard our torus as obtained by indentification of ``opposite
sides'' of a plane domain that can be obtained by cutting the torus
along the basis cycles
$\bf a$ and $\bf b$. We will consider those cycles as components
of the plane domain boundary. Of course, we can determine
the numbers $l$ and $m$, for each set of trajectories from Lemmas
\ref{2lemdet}, \ref{2lemsta}, \ref{2lemtor},
if we just know how the trajectories intersect with the boundary
(and we may know nothing about the behavior of trajectories
inside the domain). And it is exactly the numbers
$l$ and $m$ that determine, according to the proof of
Theorem~\ref{2thstadet}, whether a sign must be changed of
a given summand in a ``usual'' statistical sum $S(u,w)$ in order
to get~$I(u,w)$.
\end{remark}

\section{A finite-gap theory for the model on infinite lattice}
\label{3secqp}

We considered in Section~\ref{2seckag} an inhomogeneous 6-vertex model
satisfying the ``free fermion'' condition, on the kagome lattice.
The lattice itself was situated on a torus, thus, in particular,
it was finite.
The evolution was in fact ``local'',
i.e.\  a given weight in a moment $\tau$ was influenced only
by weights in a few neighboring points in the moment $\tau -1$. However,
the evolution was described ``globally'', just as a particular case
of the evolution from Section~\ref{2secdef}.

In this section, we will describe the evolution in local terms,
the description being valid, because of its localness, for an infinite
in both spatial directions lattice as well.
The evident solitonic character of the model stimulates one to study,
in particular, the ``finite-gap''  quasiperiodic solutions on this
infinite lattice.
Recall however that the algebraic curve appearing in
Section~\ref{2seckag} was associated with a ``global''
object---the statistical sum of the model regarded as a function
of two parameters, $u$ and $w$. In this section we develop another
approach, namely, in the spirit of the usual theory of
finite-gap solutions, we start from a given algebraic curve
with some marked points in it and construct a solution out of those
objects.

It is natural to begin with a definition of evolution of matrices
corresponding to vertices of a finite or infinite kagome lattice
independent of any global objects. More precisely, we will assume,
to be in accord with the evolution definition given in
Section~\ref{2secdef}, that before a step of evolution there are
given some products of such matrices (of size $2\times 2$),
while the matrices themselves appear and disappear within the step.
Hence, at the beginning of a step,
to each triangle of the form 
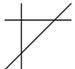
\unitlength=0.104606\normalbaselineskip
\special{em:linewidth 0.4pt}
\linethickness{0.4pt}
\begin{picture}(20.00,15.00)
\emline{0.00}{-5.00}{1}{20.00}{15.00}{2}
\emline{5.00}{15.00}{3}{5.00}{-5.00}{4}
\emline{0.00}{10.00}{5}{20.00}{10.00}{6}
\end{picture}
\vs\ contained in the kagome
lattice
(and shaded in Fig.~\ref{2figkagome} situated on
p.~\pageref{2figkagome}) a $3\times 3$-matrix of complex numbers
must correspond. The step begins with this matrix being factorized
into a product of three matrices of the following form:
\be
\pmatrix{
a_1&b_1&0\cr
c_1&d_1&0\cr
0&0&1}
\pmatrix{
a_2&0&b_2\cr
0&1&0\cr
c_2&0&d_2}
\pmatrix{
1&0&0\cr
0&a_3&b_3\cr
0&c_3&d_3}. \label{301}
\ee
This done, we will assume that in our triangle
the matrix $\pmatrix{a_1&b_1\cr c_1&d_1}$
corresponds to the vertex 
\unitlength=0.073223\normalbaselineskip
\special{em:linewidth 0.4pt}
\linethickness{0.4pt}
\begin{picture}(20.00,15.00)
\emline{0.00}{5.00}{1}{20.00}{5.00}{2}
\emline{0.00}{-5.00}{3}{20.00}{15.00}{4}
\end{picture}
 \vs,
the matrix $\pmatrix{a_2&b_2\cr c_2&d_2}$
corresponds to the vertex 
\unitlength=0.073223\normalbaselineskip
\special{em:linewidth 0.4pt}
\linethickness{0.4pt}
\begin{picture}(20.00,15.00)
\emline{0.00}{5.00}{1}{20.00}{5.00}{2}
\emline{10.00}{-5.00}{3}{10.00}{15.00}{4}
\end{picture}
 \vs,
and the matrix $\pmatrix{a_3&b_3\cr c_3&d_3}$
corresponds to the vertex 
\unitlength=0.073223\normalbaselineskip
\special{em:linewidth 0.4pt}
\linethickness{0.4pt}
\begin{picture}(20.00,15.00)
\emline{0.00}{-5.00}{1}{20.00}{15.00}{2}
\emline{10.00}{15.00}{3}{10.00}{-5.00}{4}
\end{picture}
 \vs.
The kagome lattice can also be viewed as made up of triangles
of the form
\unitlength=0.104606\normalbaselineskip
\special{em:linewidth 0.4pt}
\linethickness{0.4pt}
\begin{picture}(20.00,15.00)
\emline{0.00}{0.00}{1}{20.00}{0.00}{2}
\emline{0.00}{-5.00}{3}{20.00}{15.00}{4}
\emline{15.00}{15.00}{5}{15.00}{-5.00}{6}
\end{picture}
 \vs\quad (non-shaded in Fig.~\ref{2figkagome}).
Suppose that in such a triangle to the vertex 
\unitlength=0.073223\normalbaselineskip
\special{em:linewidth 0.4pt}
\linethickness{0.4pt}
\begin{picture}(20.00,15.00)
\emline{0.00}{5.00}{1}{20.00}{5.00}{2}
\emline{0.00}{-5.00}{3}{20.00}{15.00}{4}
\end{picture}
 \vs\quad
corresponds a matrix
$\pmatrix{
\tilde a_1&\tilde b_1\cr
\tilde c_1&\tilde d_1}$,
to the vertex 
\unitlength=0.073223\normalbaselineskip
\special{em:linewidth 0.4pt}
\linethickness{0.4pt}
\begin{picture}(20.00,15.00)
\emline{0.00}{5.00}{1}{20.00}{5.00}{2}
\emline{10.00}{-5.00}{3}{10.00}{15.00}{4}
\end{picture}
 \vs\ ---a matrix
$\pmatrix{
\tilde a_2&\tilde b_2\cr
\tilde c_2&\tilde d_2}$,
and to the vertex 
\unitlength=0.073223\normalbaselineskip
\special{em:linewidth 0.4pt}
\linethickness{0.4pt}
\begin{picture}(20.00,15.00)
\emline{0.00}{-5.00}{1}{20.00}{15.00}{2}
\emline{10.00}{15.00}{3}{10.00}{-5.00}{4}
\end{picture}
 \vs\ ---a matrix
$\pmatrix{
\tilde a_3&\tilde b_3\cr
\tilde c_3&\tilde d_3}$.
Then let the matrix
\be
\pmatrix{
1&0&0\cr
0&\tilde a_3&\tilde b_3\cr
0&\tilde c_3&\tilde d_3}
\pmatrix{
\tilde a_2&0&\tilde b_2\cr
0&1&0\cr
\tilde c_2&0&\tilde d_2}
\pmatrix{
\tilde a_1&\tilde b_1&0\cr
\tilde c_1&\tilde d_1&0\cr
0&0&1} \label{302}
\ee
correspond to the triangle as a whole.
Finally, turn inside out all the triangles of the form
\unitlength=0.104606\normalbaselineskip
\special{em:linewidth 0.4pt}
\linethickness{0.4pt}
\begin{picture}(20.00,15.00)
\emline{0.00}{0.00}{1}{20.00}{0.00}{2}
\emline{0.00}{-5.00}{3}{20.00}{15.00}{4}
\emline{15.00}{15.00}{5}{15.00}{-5.00}{6}
\end{picture}
 \vs\vs\quad in our lattice,
converting them into triangles of the form
\unitlength=0.104606\normalbaselineskip
\special{em:linewidth 0.4pt}
\linethickness{0.4pt}
\begin{picture}(20.00,15.00)
\emline{0.00}{-5.00}{1}{20.00}{15.00}{2}
\emline{5.00}{15.00}{3}{5.00}{-5.00}{4}
\emline{0.00}{10.00}{5}{20.00}{10.00}{6}
\end{picture}
 \vs\ (e.g. moving all {\em oblique\/}
lines one lattice period right), putting the same matrix (\ref{302})
in correspondence to each ``converted'' triangle as to the initial one.
The description of a step of evolution is over.

The same can be presented in a somewhat other way, starting from
the triangular lattice of Fig.~\ref{2figtria} (p.~\pageref{2figtria}).
By the beginning of an evolution step, to each {\it circle\/}---a vertex
of that lattice---a $3\times 3$-matrix must correspond that is to be
factorized in a product~(\ref{301}), which corresponds to a
``decomposition'' of the circle in a triangle 
\unitlength=0.104606\normalbaselineskip
\special{em:linewidth 0.4pt}
\linethickness{0.4pt}
\begin{picture}(20.00,15.00)
\emline{0.00}{-5.00}{1}{20.00}{15.00}{2}
\emline{5.00}{15.00}{3}{5.00}{-5.00}{4}
\emline{0.00}{10.00}{5}{20.00}{10.00}{6}
\end{picture}
\vs.
In these terms, the step of evolution ends with
the triangles 
\unitlength=0.104606\normalbaselineskip
\special{em:linewidth 0.4pt}
\linethickness{0.4pt}
\begin{picture}(20.00,15.00)
\emline{0.00}{0.00}{1}{20.00}{0.00}{2}
\emline{0.00}{-5.00}{3}{20.00}{15.00}{4}
\emline{15.00}{15.00}{5}{15.00}{-5.00}{6}
\end{picture}
 \vs\quad being ``packed'' in new circles.

Just as in Section~\ref{2secdef}, the evolution is defined up to
``gauge transformations''.  To explain what the gauge transformations
look like now,
note that each entry of the abovementioned
$2\times 2$-matrices naturally corresponds to a certain pair
``incoming edge, outgoing edge'' for the given vertex
(Figure~\ref{2figinout}  on p.~\pageref{2figinout}
will remind which edges are called incoming and outgoing).
If we mark those pairs of edges with arrows, we will obtain the following
pictures for entries of the matrix
$\pmatrix{
a_1&b_1\cr
c_1&d_1}$:
$$
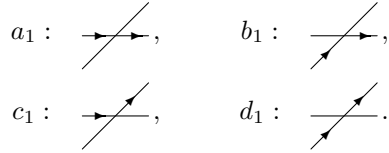
\matrix{\noalign{\vskip 0.5\baselineskip}
 a_1:& \matrix{ 
\unitlength=0.104606\normalbaselineskip
\special{em:linewidth 0.4pt}
\linethickness{0.4pt}
\begin{picture}(20.00,15.00)
\emline{0.00}{5.00}{1}{20.00}{5.00}{2}
\emline{0.00}{-5.00}{3}{20.00}{15.00}{4}
\put(5.00,5.00){\vector(1,0){2.00}}
\put(16.00,5.00){\vector(1,0){2.00}}
\end{picture}
 },\qquad&b_1:&
\matrix{
\unitlength=0.104606\normalbaselineskip
\special{em:linewidth 0.4pt}
\linethickness{0.4pt}
\begin{picture}(20.00,15.00)
\emline{0.00}{5.00}{1}{20.00}{5.00}{2}
\emline{0.00}{-5.00}{3}{20.00}{15.00}{4}
\put(4.00,-1.00){\vector(1,1){2.00}}
\put(16.00,5.00){\vector(1,0){2.00}}
\end{picture}
 },\cr
\noalign{\vskip 0.75\baselineskip}
          c_1:&  \matrix{
\unitlength=0.104606\normalbaselineskip
\special{em:linewidth 0.4pt}
\linethickness{0.4pt}
\begin{picture}(20.00,15.00)
\emline{0.00}{5.00}{1}{20.00}{5.00}{2}
\emline{0.00}{-5.00}{3}{20.00}{15.00}{4}
\put(5.00,5.00){\vector(1,0){2.00}}
\put(14.00,9.00){\vector(1,1){2.00}}
\end{picture}
 },\qquad&d_1:&
          \matrix{
\unitlength=0.104606\normalbaselineskip
\special{em:linewidth 0.4pt}
\linethickness{0.4pt}
\begin{picture}(20.00,15.00)
\emline{0.00}{5.00}{1}{20.00}{5.00}{2}
\emline{0.00}{-5.00}{3}{20.00}{15.00}{4}
\put(4.00,-1.00){\vector(1,1){2.00}}
\put(14.00,9.00){\vector(1,1){2.00}}
\end{picture}
 }.\cr
\noalign{\vskip 0.5\baselineskip}
}
$$
Now it is seen that to each edge a set of
{\em elementary gauge transformations\/} naturally corresponds
parameterized by the set of nonzero complex numbers.
Namely, we can multiply by a nonzero constant the matrix entries in
the vertex corresponding to the given edge as an outgoing one,
and divide by the same constant the entries in the neighboring
vertex corresponding to the given edge as an incoming one.
Nonzero numbers can be put in correspondence to
{\em all\/} edges, and the elementary gauge transformations
can be performed for all simultaneously.
The resulting transformation will be called simply
a {\em gauge transformation\/} on the whole lattice.

\begin{lemma}\label{3lemgct}
1) The result of the step of evolution described earlier in this section
is determined to within a gauge transformation.

2) The result of the evolution step is not changed
if a gauge transformation is applied to the initial state.

3) For a finite-dimensional system, the evolution coincides with that
defined in Section~\ref{2secdef}, and the gauge transformations considered
here are such in the sense of Section~\ref{2secdef} as well
(but there are more gauge transformations in Section~\ref{2secdef}
because they include also ``non-local'', from the standpoint of
model on the kagome lattice, transformations).
\end{lemma}

{\it Proof\/} follows immediately from the definitions given
here and in Section~\ref{2secdef}.
\medskip

Passing on to  the ``local'' algebro-geometrical description
of evolution, let us consider first the following {\em abstract divisor
evolution on an infinite in both spatial directions lattice}.
At the moment, we don't need
to know the exact structure of those divisors or an algebraic manifold
where they belong.
Thus, let us temporarily understand by ``divisors'' just elements of
some abelian group~$\cal G$. Let six elements
${\cal D}_1, \ldots ,{\cal D}_6$ be fixed in that group.

Let a divisor correspond to each {\it edge\/} of the triangular
lattice in Fig.~\ref{2figtria} by the beginning of a step of evolution,
with the following condition fulfilled:
for each lattice {\it vertex\/}  (a circle in
Fig.~\ref{2figtria}) an element
${\cal D}\in \cal G$ can be indicated such that
the divisors corresponding to edges abutting on that vertex are
as shown in Fig.~\ref{3fig03}.
\bfig
$$
\unitlength=0.104606\normalbaselineskip
\special{em:linewidth 0.4pt}
\linethickness{0.4pt}
\begin{picture}(100.00,100.00)
\put(50.00,50.00){\circle{4.00}}
\emline{0.00}{50.00}{1}{48.00}{50.00}{2}
\emline{100.00}{50.00}{3}{52.00}{50.00}{4}
\emline{50.00}{100.00}{5}{50.00}{52.00}{6}
\emline{50.00}{0.00}{7}{50.00}{48.00}{8}
\emline{0.00}{0.00}{9}{48.59}{48.59}{10}
\emline{100.00}{100.00}{11}{51.41}{51.41}{12}
\put(52.00,2.00){\makebox(0,0)[lb]{${\cal D}-{\cal D}_3-{\cal D}_4$}}
\put(100.00,54.00){\makebox(0,0)[cb]{${\cal D}-{\cal D}_4-{\cal D}_6$}}
\put(0.00,54.00){\makebox(0,0)[cb]{${\cal D}-{\cal D}_1-{\cal D}_2$}}
\put(48.00,98.00){\makebox(0,0)[rt]{${\cal D}-{\cal D}_5-{\cal D}_2$}}
\put(0.00,4.00){\makebox(0,0)[rb]{${\cal D}-{\cal D}_1-{\cal D}_3$}}
\put(100.00,96.00){\makebox(0,0)[lt]{${\cal D}-{\cal D}_5-{\cal D}_6$}}
\end{picture}
 $$
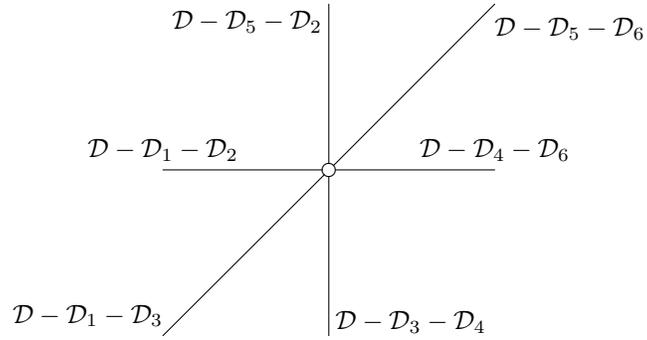
\caption{Divisors on edges abutting on a triangular lattice vertex}
\label{3fig03}
\vskip2\bigskipamount
$$
\unitlength=0.17433\normalbaselineskip
\special{em:linewidth 0.4pt}
\linethickness{0.4pt}
\begin{picture}(60.00,60.00)
\emline{0.00}{40.00}{1}{60.00}{40.00}{2}
\emline{20.00}{60.00}{3}{20.00}{0.00}{4}
\emline{0.00}{0.00}{5}{60.00}{60.00}{6}
\put(1.00,42.00){\makebox(0,0)[cb]{\small 12}}
\put(22.00,58.00){\makebox(0,0)[lt]{\small 52}}
\put(31.00,42.00){\makebox(0,0)[cb]{\small 26}}
\put(53.00,55.00){\makebox(0,0)[rb]{\small 56}}
\put(58.00,42.00){\makebox(0,0)[rb]{\small 46}}
\put(32.00,28.00){\makebox(0,0)[lt]{\small 36}}
\put(22.00,2.00){\makebox(0,0)[lb]{\small 34}}
\put(2.00,5.00){\makebox(0,0)[rb]{\small 13}}
\put(18.00,30.00){\makebox(0,0)[rc]{\small 23}}
\end{picture}
 $$
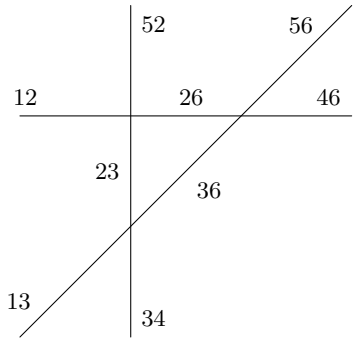
\caption{Divisors on edges of the kagome lattice}
\label{3fig04}
\efig

Thus, ${\cal D}$ depends linearly on the coordinates of a vertex,
increasing by ${\cal D}_1+{\cal D}_2-{\cal D}_4-{\cal D}_6$ when
moving one lattice period to the right, and by
${\cal D}_3+{\cal D}_4-{\cal D}_5-{\cal D}_2$ when
moving one lattice period up.

A step of evolution begins with every circle being decomposed into
a triangle such as depicted in Fig.~\ref{3fig04},
where the numbers near an edge serve as a brief notation for a divisor
corresponding to it, e.g.\
``12'' denotes
${\cal D}-{\cal D}_1-{\cal D}_2$, and so on, compare Fig.~\ref{2figdecom}.
Thus, the kagome lattice arises from the triangular one,
and then we take the triangles of the form 
\unitlength=0.104606\normalbaselineskip
\special{em:linewidth 0.4pt}
\linethickness{0.4pt}
\begin{picture}(20.00,15.00)
\emline{0.00}{0.00}{1}{20.00}{0.00}{2}
\emline{0.00}{-5.00}{3}{20.00}{15.00}{4}
\emline{15.00}{15.00}{5}{15.00}{-5.00}{6}
\end{picture}
 \vs\quad in it
and ``pack'' them in circles---the vertices of the new triangular lattice.
With this, the divisor evolution step is over. As the following lemma states,
everything is ready for a next step.

\begin{lemma} \label{3lemtrifle}
The divisors corresponding to edges abutting on each vertex of
the triangular lattice obtained in the end of the previous paragraph
are again of the form of Fig.~\ref{3fig03}, with $\cal D$ depending
on the vertex.
\end{lemma}

{\it Proof\/} follows from an easy direct calculation.
\medskip

Now let an algebraic curve ${\Gamma}_0$ of genus $g_0$ be given, and
the abelian group $\cal G$  let be the group of all divisors
on $\Gamma_0$.
Let ${\cal D}_1, \ldots, {\cal D}_6$ be divisors on $\Gamma_0$,
consisting each of one point (note in parentheses that, of course, one can
consider a ``vector'' model as well,
${\cal D}_1, \ldots, {\cal D}_6$ consisting in that case each of
several points).
Let divisors $\cal D$ corresponding to vertices of triangular lattice
(see Fig.~\ref{3fig03}) be of the degree $g_0+2$. If all those
algebro-geome\-trical objects are generic,
the Riemann--Roch
theorem shows that to each edge of triangular (Fig.~\ref{3fig03}) or
kagome (Fig.~\ref{3fig04}) lattice
corresponds a
{\it one-dimen\-sional\/} space of meromorphic functions
$f$ such that
\be
(f)+{\cal D}-{\cal D}_j-{\cal D}_k \leq 0, \label{305}
\ee
if a divisor ${\cal D}-{\cal D}_j-{\cal D}_k$ corresponds to the
given edge.

It remains to explain how the matrices are constructed corresponding
to vertices of our two lattices.
Fix a nonzero function $f$ satisfying~(\ref{305}) for every
$j$ and $k$, and denote it as $f_{jk}$.
Then, say, the matrix ${\cal A}_{\rm local}$ corresponding
to a vertex in Fig.~\ref{3fig03} is found from the relation
\be
{\cal A}_{\rm local}
\pmatrix{f_{12}\cr f_{13} \cr f_{34}}=
\pmatrix{f_{46}\cr f_{56}\cr f_{52}} \label{306}
\ee
that must hold in each point of the curve $\Gamma_0$.
Similarly, the matrix
$\pmatrix{a_1& b_1\cr c_1& d_1}$
(see~(\ref{301})  and the graphical analog of that relation,
Figure~\ref{3fig04}) is found from the relation
\be
\pmatrix{a_1 &b_1\cr c_1& d_1}
\pmatrix{f_{26}\cr f_{36}}=
\pmatrix{f_{46}\cr f_{56}} \label{307}
\ee
etc. Relations of type (\ref{306}--\ref{307}) determine the matrices
correctly due to the fact that
all meromorphic functions in every such relation lie
in a linear space of needed dimension.
For example, in relation~(\ref{307}) divisors of all functions
satisfy  the condition
$$
(f_{jk})+{\cal D}-{\cal D}_6\geq 0,
$$
hence the Riemann--Roch
theorem, together with equalities
$$
\deg{\cal D}=g_0+2,\quad \deg{\cal D}_6=1,
$$
shows that the space of such functions is two-dimen\-sional.

Note that other choice of any function $f_{jk}$, i.e.\
its multiplication by a nonzero constant, corresponds to an elementary
gauge transformation of matrices.

\begin{lemma} \label{3div-mat}
The described above divisor evolution generates, by means of formulae
of type (\ref{306}) and (\ref{307}), the matrix evolution
described in the beginning of this section.
\end{lemma}

{\it Proof}. We must show that the operation of decomposition
of a triangular lattice vertex in a triangle of the kagome lattice,
and the inverse operation of
``packing'', when applied to divisors, generate the similar
operations on matrices. E.g., if a matrix
${\cal A}_{\rm local}$ is determined by the relation (\ref{306}),
and we have to factorize that matrix in a product
(\ref{301}), then the values
$a_1,\ldots,d_3$ obtained from (\ref{307}) and two similar relations
must give the solution to the problem. It is worth while
to write out explicitely those two relations similar
to~(\ref{307}):
\be
\pmatrix{a_2 &b_2\cr c_2&d_2}
\pmatrix{f_{12}\cr f_{23}}=
\pmatrix{f_{26}\cr f_{52}},
\qquad
\pmatrix{a_3& b_3\cr c_3&d_3}
\pmatrix{f_{13}\cr f_{34}}=
\pmatrix{f_{36}\cr f_{23}}.
\label{308}
\ee

If we now let the product (\ref{301}) act on the column
from the LHS of (\ref{306}), and use (\ref{307}) and (\ref{308}),
we will get exactly the column in the RHS of (\ref{306}), which means
that the factorization of a matrix is done
with the help of divisors in our local case
with equal success as in the global case of
Section~\ref{2secevd}, cf.~formulae~(\ref{228}--\ref{230}).
Certainly, the similar considerations are valid also
for the ``packing'' operation.
The lemma is proved.
\medskip

Thus, we have shown how to construct solutions to the problem
of matrix evolution on the infinite lattice
in the spirit of standard finite-gap integration theory.
In Section~\ref{3secps} we will connect this approach
with the ``global'' approach of Sections \ref{2seccrv} and \ref{2secevd}.
Explicit expression for the solution in multidimensional
theta functions is presented in
Section~\ref{3sechgr}.

\section{Connection between ``local'' and ``global'' curves
in the periodic case}
\label{3secps}

We continue to consider divisors from Section~\ref{3secqp} corresponding
to edges of the infinite triangular lattice.
Let there exist integers ${\xi}_1$ and ${\eta}_1$ such that the divisor
equivalence holds in the curve ${\Gamma}_0$
\be
{\xi}_1({\cal D}_1+{\cal D}_2-{\cal D}_4-{\cal D}_6)
+{\eta}_1({\cal D}_3+{\cal D}_4-{\cal D}_5-{\cal D}_2)\sim 0.
\label{309}
\ee
(compare with the text between lemmas \ref{3lemgct} and \ref{3lemtrifle}).
As each of the divisors ${\cal D}_1$, \dots, ${\cal D}_6$ consists now of
just one point, it is convenient to introduce for those points
notations $P_1$, \dots, $P_6$.
Note also that ${\cal D}_j$'s in (\ref{309}) are now other divisors
than the ``global'' ones denoted by the same letters in
Sections~\ref{2seccrv}--\ref{2secsmport}.
Equivalence~(\ref{309}) means that there exists a function $g_1$ on
${\Gamma}_0$ that has a zero of multiplicity $\xi_1$ in the point $P_1$,
a zero of multiplicity ${\xi}_1-{\eta}_1$ in the point
$P_2$, \dots, a pole of multiplicity ${\xi}_1$ in the point
$P_6$. Thus, meromorphic functions $f$, out of which we will construct
matrices according to relations of type~(\ref{306}),
can be put in correspondence
to lattice edges in such a way that the multiplying of a function
$f$ by $g_1$
will correspond to a lattice translation by the vector
$({\xi}_1, {\eta}_1)$.
As for the matrices like ${\cal A}_{\rm local}$ from (\ref{306}),
they will obviously be {\it periodic\/} with period
$({\xi}_1, {\eta}_1)$.

\begin{lemma}\label{3lemtrif2}
Let there exist two linearly independent vectors with integer entries
$({\xi}_1, {\eta}_1)$ and $({\xi}_2, {\eta}_2)$
such that two following divisor equivalences in the curve\/ ${\Gamma}_0$
hold: (\ref{309}) and a similar one
\be
{\xi}_2\,({\cal D}_1+{\cal D}_2-{\cal D}_4-{\cal D}_6)+
{\eta}_2\,({\cal D}_3+{\cal D}_4-{\cal D}_5-{\cal D}_2)\sim 0.
\label{310}
\ee
Then matrices corresponding to lattice edges
are, in a proper gauge, doubly periodic in coordinates,
with periods
$({\xi}_1, {\eta}_1)$ and $({\xi}_2, {\eta}_2)$.
\end{lemma}

{\it Proof\/} is obtained by adding to the above consideration
for $({\xi}_1, {\eta}_1)$ a similar consideration for
$({\xi}_2, {\eta}_2)$.
\medskip

Thus, conditions (\ref{309}) and (\ref{310}) are sufficient
for the model on infinite lattice to become, in essence,
a model on a torus introduced in Section~\ref{2seckag}. Our
next task is to learn how to pass on from the curve
${\Gamma}_0$ to the ``global'' curve
$\Gamma$, and by that calculate the statistical sum of the  model
on the torus starting from a ``small'' curve ${\Gamma}_0$.
To be concrete, consider the problem of constructing vectors
$\pmatrix{X\cr Y\cr Z}$ such that
\be
{\cal A}\pmatrix{X\cr Y\cr Z}=
\pmatrix{uX\cr uwY\cr wZ},\label{311}
\ee
the components of vectors $X$, $Y$, and $Z$ corresponding to
horizontal, oblique, and vertical edges respectively;
$\cal A$ being a global operator, as in Section~\ref{2seckag};
and $u$ and $w$ being some so far unknown values.

It turns out that the solution of (\ref{311}) is obtained
if $u$ and $w$ are introduced as
multivalued functions on ${\Gamma}_0$ by the formulae
\be
\matrix{ u^{{\xi}_1}w^{{\eta}_1}=g_1, \cr
u^{{\xi}_2}w^{{\eta}_2}=g_2,} \label{312}
\ee
where the function $g_2$ is constructed out of (\ref{310}) in the same way
as $g_1$ out of~(\ref{309}).
The vectors $X$, $Y$ and $Z$ are built now as follows.
Choose some vertex to be the origin of integer-valued
coordinates $(\xi,\eta)$.
For each of the vectors $X$, $Y$ and $Z$, one component---the one
corresponding to the {\it incoming\/} edge abutting on the origin
of coordinates
(recall Fig.~\ref{2figinout} on p.~\pageref{2figinout})---is defined
as the value of meromorphic function~$f$ corresponding to this edge
as in Section~\ref{3secqp}, in some point of ${\Gamma}_0$. Next,
we take some values
$u$ and $w$ for this point according to~(\ref{312}), and for an
incoming edge at a point
with coordinates $(\xi,\eta)$ we take as a component of the
vector $X$, $Y$ or $Z$
the value of the corresponding function $f$
{\it multiplied by\/} $u^{-\xi}w^{-\eta}$.
As a result, as was required, the vector components are not changed
under a translation by the periods
$({\xi}_1, {\eta}_1)$ and $({\xi}_2, {\eta}_2)$
(we assume of course that a ``quasiperiodic'' gauge was chosen
for functions $f$, as in Lemma~\ref{3lemtrif2}).

\begin{lemma} \label{3lemtrif3}
The vectors $X$, $Y$ and $Z$ constructed in the previous paragraph
satisfy indeed the equation (\ref{311}) for any point of the
curve\/ ${\Gamma}_0$
and any pair  $(u,w)$ lying above it (i.e.\ satisfying (\ref{312})).
\end{lemma}

{\it Proof}. Indeed, it is easy to see that (\ref{311})
is equivalent to the totality
of relations of type (\ref{306}) for all lattice vertices,
because the factors $u$, $uw$ and $w$ in (\ref{311})
are canceled by factors arising because of multiplying the functions~$f$
by $u^{-\xi}w^{-\eta}$.
\medskip

Having constructed the solutions of equation (\ref{311}), we arrive at some
algebraic dependence between $u$ and $w$ arising from (\ref{312}) and
an algebraic dependence between the meromorphic functions $g_1$ and $g_2$
on the curve ${\Gamma}_0$. We expect that it will be, maybe under some
additional conditions, the same dependence as given by the
``global'' equations (\ref{211}, \ref{212}),
because solutions of (\ref{312}) do satisfy, of course, the system
(\ref{211}, \ref{212}). However, we don't know at the moment
whether our local construction does not lead to some
singular operators~$\cal A$, e.g.~those having an invariant curve
that contains several components.
In such case, one can conceive a situation where (\ref{312}) together with
the dependence between
$g_1$ and $g_2$ yields only a part of the curve~$\Gamma$,
while all $\Gamma$ must be known, of course, for calculating the
statistical sum.
Besides, we must attach the exact meaning to the words
``algebraic dependence between $g_1$ and $g_2$''.

It turns out that there exist simple conditions that guarantee
the {\em irreducibility\/} of $\Gamma$. They are as follows:

\smallskip\indent\llap{(i)}\label{3cond1}~irreducibility of $\Gamma_0$,

\smallskip\indent\llap{(ii)}\label{3cond2}~the parallelogram
built on vectors $(\xi_1,\eta_1)$ and $(\xi_2,\eta_2)$ must be
the {\em minimal parallelogram of periods\/} (i.e.\
the parallelogram of minimal area among the parallelograms
built on two linearly independent vectors of periods),
and

\smallskip\indent\llap{(iii)}\label{3cond3}~at least one of the pairs
$(\xi_1,\xi_2)$, $(\eta_1,\eta_2)$, $(\xi_1-\eta_1, \xi_2-\eta_2)$
consists of relatively prime numbers---the condition without which
the proof of the following lemma fails.

\begin{lemma}\label{3lem-p1}
Consider the mapping $\phi:~\Gamma_0\to{\bf C}P_1\times{\bf C}P_1$
given by the formula
$$z\mapsto\bigl(g_1(z),g_2(z)\bigr).$$
Under the above conditions (i) and (iii), $\phi$ is a birational
isomorphism of the curve\/ $\Gamma_0$ on its image~$\phi(\Gamma_0)$.
\end{lemma}

{\it Proof}.
Compose the following table of zero and pole multiplicities
for functions
$g_1$ and $g_2$ in points $P_1$, \dots, $P_6$.
$$\begin{array}{c|c|c|c|c|c|c|}
\vphantom{\bigg|}&P_1&P_2&P_3&P_4&P_5&P_6\\\hline
\vphantom{\bigg|}g_1&\xi_1&\xi_1-\eta_1&\eta_1&-\xi_1+\eta_1&-\eta_1&-\xi_1\\\hline
\vphantom{\bigg|}g_2&\xi_2&\xi_2-\eta_2&\eta_2&-\xi_2+\eta_2&-\eta_2&-\xi_2\\\hline
\end{array}\label{3tab}$$
This table is made up according to formulae (\ref{309}, \ref{310}).
An integer in a table cell is the zero multiplicity of a function
in a point if it is positive, or minus pole multiplicity if it is negative.
Functions $g_1$ and $g_2$ have no other
zeros or poles.

Suppose that the mapping $\phi$ is a $q$-sheeted covering.
Let us replace $\phi(\Gamma_0)$ by its nonspecial
model $\Gamma_{ns}$ which exists
according to \S7A of the manual~\cite{Mumford}. Somewhat
freely, we will denote the covering $\Gamma_0\to\Gamma_{ns}$
by the same letter~$\phi$.

Let us prove that a point $P_j$ in which zero or pole multiplicities
of functions $g_1$ and $g_2$ are relatively prime (such a point
exists according to the condition~(iii)) cannot be a branch point of
covering $\phi$. Indeed, due to the indicated relative primality,
a local parameter in the point $P_j$ can be chosen as a product
of some degrees of $g_1$ and $g_2$, i.e.\ is uniquely determined
by an underlying point from a neighborhood of $\phi(P_j)$, which
cannot happen in a branch point.

The point $P_j$ from the previous paragraph also cannot have the
same image as another point $P\in\Gamma_0$, i.e.\ $\phi(P_j)\not=\phi(P)$
when $P\not=P_j$. To show this, we note that $P$ can only be one of the
points $P_1,\ldots,P_6$, because only there zeros and poles of
$g_1$ and $g_2$ are situated. However, by virtue of the linear independence
of periods $(\xi_1,\eta_1)$ and $(\xi_2,\eta_2)$, and the above table,
the multiplicities of zeros or poles of $g_1$ and $g_2$ in points
$\phi(P_j)$ and $\phi(P)$ cannot coincide (even if $P$ is a branch
point). Hence $\phi(P_j)\not=\phi(P)$.

Combining the results of two preceding paragraphs,
we see that $P_j$ is the only point lying above
$\phi(P_j)$, and not a branch point. Hence, the number of sheets
$q=1$, and the lemma is proved.
\medskip

We continue to consider the algebraic dependence between functions
$g_1$ and $g_2$. In view of the irreducibility of $\Gamma_0$, this
dependence can be expressed as
\be
{\cal P}_0(g_1,g_2)=0,
\label{315p}
\ee
${\cal P}_0$ being an irreducible polynomial in two
variables over the field of complex numbers.
By Lemma~\ref{3lem-p1}, the curve given by equation~(\ref{315p})
is birationally isomorphic to~$\Gamma_0$.

\begin{theorem}\label{3th-pc}
Assume conditions (i), (iii) on p.~\cpages{3cond1}{3cond3}. Then,
equation~(\ref{315p}), which describes the algebraic dependence
between functions $g_1$ and $g_2$ realizing the equivalences
(\ref{309}, \ref{310}) of divisors on $\Gamma_0$,
after substituting in it the expressions~(\ref{312}) of $g_1$ and
$g_2$ through $u$ and $w$ gives exactly the invariant curve
of global operator $\cal A$ on a torus with periods
$(\xi_1,\eta_1)$ and $(\xi_2,\eta_2)$.
\end{theorem}

{\it Proof}. Note that the equation of curve $\Gamma$ has as well
the form
\be
{\cal P}(g_1,g_2)={\cal P}(u^{\xi_1}w^{\eta_1},u^{\xi_2}w^{\eta_2})=0,
\label{317p}
\ee
where $g_1$ and  $g_2$ correspond to going around two basis cycles
on the torus, see Section~\ref{2seckag}. We will prove that equations
(\ref{315p}) and (\ref{317p}) define the same curve.

According to the table on p.~\pageref{3tab}, functions $g_1$ and $g_2$ have
$l_1$ and $l_2$ poles respectively in $\Gamma_0$, where
\be
l_j=2\cdot\max\{|\xi_j|,|\eta_j|,|\xi_j-\eta_j|\},\quad j=1,2.
\label{316r}
\ee
Hence any generic value is taken by a variable
$g_j$ in
$l_j$ points of the curve, so that $l_j$ values of the other variable
correspond to it.
This situation can be described by equation~(\ref{315p}) only if
${\cal P}_0$ {\em has degree $l_2$ in $g_1$ and degree
$l_1$ in $g_2$}.

Passing on to equation~(\ref{317p}), note that it is natural to suppose
that the degrees of $g_1$ and $g_2$ (but not $u$ and $w$) in it
can be negative, if
the expansion of the cycle associated with a given
trajectory (see Section~\ref{2seckag}, three paragraphs starting
from the one containing formula~(\ref{233})),
in basis cycles has one or two negative coefficients. It is not hard
to check that the difference between the maximal and minimal degrees
of, say, $g_2$ does not exceed the number of lines in the
triangular lattice of Fig.~\ref{2figtria} that intersect with
a vector with coordinates $(\xi_1,\eta_1)$
(it is convenient that we imagine it as situated ``generically'', e.g.\
having a point with irrational coordinates as its origin). Indeed,
the difference between the greatest and least intersection numbers
of any trajectory going along the edges on the torus always to the
right, upwards, or to the north-east, and the basis cycle corresponding
to the vector $(\xi_1,\eta_1)$, cannot exceed, of course, the mentioned
number of lines. This number of lines is, however,
nothing else but $l_1$ from formula~(\ref{316r}), because the vector
$(\xi_1,\eta_1)$, evidently, intersects with $|\xi_1|$
vertical, $|\eta_1|$ horizontal, and $|\xi_1-\eta_1|$ oblique
lines.

Thus, we see that, having multiplied ${\cal P}(g_1,g_2)$ (if needed) by
its common denominator, we get a polynomial of the same degrees
$l_{2,1}$ in variables $g_{1,2}$ as ${\cal P}_0(g_1,g_2)$. Hence
it is clear that the equation~(\ref{315p}) together with~(\ref{312})
yields indeed the whole
``global'' curve $\Gamma$. The theorem is proved.
\medskip

The following lemma
describing more precisely
the structure of $\Gamma$
will be the last in this section. It is here that we will use the
condition~(ii), not needed for us before.

\begin{lemma}\label{3lem-n2}
Consider a covering\/ $\Gamma'$ of the curve\/ $\Gamma_0$ defined by
multivalued functions $u$ and $w$ according to~(\ref{312}). Let the
conditions (i) and (ii) on p.~\cpages{3cond1}{3cond2} hold (while
we will not use condition~(iii) that guarantees that\/ $\Gamma'$ and\/
$\Gamma$ are isomorphic).
Then the curve\/ $\Gamma'$ is irreducible.
\end{lemma}

{\it Proof}. The system~(\ref{312}) defines an extension of the field
of meromorphic functions on $\Gamma_0$. This extension is {\em normal},
because there exist automorphisms of $\Gamma'$ mapping its any sheet
into any other, defined by an evident formula
\be
(u,w)\to(\omega_1u,\omega_2w),
\label{314m}
\ee
$\omega_1$ and $\omega_2$ being some root of unity. There are
$|\xi_1\eta_2-\xi_2\eta_1|$ such automorphisms, and if they all enter
in the Galois group, then this group acts transitively on solutions
of the system~(\ref{312}), and both this system and, hence,
the curve $\Gamma'$ are irreducible (see a chapter on Galois theory
in the book \cite{Waerden} or \cite{Lang-Algebra}).
In this case, besides, the extension defined by system~(\ref{312}) is
$|\xi_1\eta_2-\xi_2\eta_1|$-dimensional over the function field
on~$\Gamma_0$.

If, otherwise, the curve $\Gamma'$ is reducible, then this extension
is less then
$|\xi_1\eta_2-\xi_2\eta_1|$-dimensional over the function field
on~$\Gamma_0$.
Consider all possible products $u^\xi w^\eta$ of integer degrees
of $u$ and $w$
for $\xi$ and $\eta$ lying within {\em one parallelogram of periods\/}
(it is implied that every such parallelogram contains
$|\xi_1\eta_2-\xi_2\eta_1|$ integer points, and that one can
cover all the plane with such parallelograms, with periods
$(\xi_1,\eta_1)$ and
$(\xi_2,\eta_2)$, while other details of their disposition
are not important). The products
$u^\xi w^\eta$ are multiplied under the action of Galois group elements
by some {\em characters\/} of that group, and as there are
$|\xi_1\eta_2-\xi_2\eta_1|$ products, and fewer characters, then
within one parallelogram of periods there are products to which
correspond the same character.
This makes clear that {\em there exists a product
$g_0=u^{\xi_0}w^{\eta_0}$ invariant under the Galois group
for which $(\xi_0,\eta_0)$ is not an integer linear combination
of periods $(\xi_1,\eta_1)$ and $(\xi_2,\eta_2)$}.
The existence of such function
$g_0$ on $\Gamma_0$ means, however, that $(\xi_0,\eta_0)$ {\em is\/}
a period, which contradicts
to the minimality of parallelogram built on
$(\xi_1,\eta_1)$ and $(\xi_2,\eta_2)$. Thus, an assumption
of reducibility of $\Gamma'$ has led to a contradiction.
The lemma is proved.
\medskip

Let us sum up the results. In this section, a situation was described when
matrices in vertices of the triangular lattice constructed
by method of Section~\ref{3secqp} depend doubly periodically on the
coordinates, so that a model on an infinite lattice becomes a model
on a torus. For such a model, a statistical sum exists
that is a polynomial in $u$ and $w$ closely connected with
the equation of invariant curve of a
``global'' matrix $\cal A$ composed of matrices situated in
lattice vertices. To deal with this situation, we presented, under some
restrictions, a principle
of constructing a ``global'' invariant curve (usually,
of a very high genus) out of a curve $\Gamma_0$ (that may be,
e.g., elliptic). The mentioned restrictions guarantee
the irreducibility of algebraic curves whose components otherwise
might be lost or confused.
A constructive realization of the mentioned principle will be
presented in Section~\ref{6sec}.

Note that the statement about obtaining the whole $\Gamma$ was deduced
{\em without\/} using condition~(ii) and Lemma~\ref{3lem-n2}.
We used only the irreducibility of $\cal P$ as a polynomial in $g_1$ and
$g_2$. Condition~(ii), guaranteeing the irreducibility of $\cal P$ as
a polynomial in $u$ and $w$, will not be needed for the explicit
calculation of statistical sum in Section~\ref{6sec}. It appears however
that a not complicated Lemma~\ref{3lem-n2}
gives a useful information on the curve~$\Gamma$. Later we will use
this lemma to calculate the number of components of some
{\em reducible\/} curves $\Gamma$ (Remark~\ref{6remab}).

\section{Expression of a finite-gap solution in theta functions
and a multilinear form of the equation}
\label{3sechgr}

In this section, we will express the meromorphic functions $f_{jk}$
entering in formulae (\ref{306}--\ref{308}) and defined on the
curve~$\Gamma_0$, through a theta function $\theta({\bf z})$ defined
on the Jacobian ${\rm Jac}\,(\Gamma_0)$ of that curve. Recall
(see~\S1 in Chapter~II of the manual~\cite{Mumford-theta}) that
\be
\theta({\bf z})=\theta({\bf z},\Omega)=\sum_{\hbox{\footnotesize$
{\bf n}\in{\bf Z}^{g_0}$}}
\exp(\pi i{\bf n}^{\rm T}\Omega{\bf n}+2\pi i{\bf n}^{\rm T}{\bf z}),
\label{340}
\ee
where $\bf z$ is a complex column vector of height~$g_0$ equal to the
genus of curve~$\Gamma_0$; the components of vector~$\bf n$ of height~$g_0$
run through all {\em integer\/} values; and $\Omega$ is the {\em matrix
of periods\/} of~$\Gamma_0$. We will obtain, from the expressions for
$f_{jk}$, very simple expressions for matrix elements of matrices
entering in LHS's of (\ref{307},
\ref{308}). This, in its turn, will lead us to constructing
of some analogs of Hirota's $\tau$-function satisfying a
{\em six-linear\/} homogeneous equation.
As is known, recently there appear more and more
integrable equations for which one cannot construct a {\em bilinear\/}
Hirota representation. It is replaced, according to papers~\cite{HGR1,HGR2},
by its {\em multilinear\/} analog.

Pass on to a detailed account.
As was explained in Section~\ref{3secqp},
meromorphic functions correspond to edges of the kagome lattice,
and out of those functions matrices can be found
corresponding to vertices of type
\unitlength=0.073223\normalbaselineskip
\special{em:linewidth 0.4pt}
\linethickness{0.4pt}
\begin{picture}(20.00,15.00)
\emline{0.00}{5.00}{1}{20.00}{5.00}{2}
\emline{0.00}{-5.00}{3}{20.00}{15.00}{4}
\end{picture}
 \hvs, 
\unitlength=0.073223\normalbaselineskip
\special{em:linewidth 0.4pt}
\linethickness{0.4pt}
\begin{picture}(20.00,15.00)
\emline{0.00}{5.00}{1}{20.00}{5.00}{2}
\emline{10.00}{-5.00}{3}{10.00}{15.00}{4}
\end{picture}
 \hvs\ \ or
\unitlength=0.073223\normalbaselineskip
\special{em:linewidth 0.4pt}
\linethickness{0.4pt}
\begin{picture}(20.00,15.00)
\emline{0.00}{-5.00}{1}{20.00}{15.00}{2}
\emline{10.00}{15.00}{3}{10.00}{-5.00}{4}
\end{picture}
 \hvs, according to the formulae (\ref{307}, \ref{308}).
We will write out a general expression for such functions through
the theta function~(\ref{340}).
Choose one of the lattice edges as ``initial'' and divide
the meromorphic functions on all edges by the function corresponding to
the initial edge. Hence, we change all divisors satisfying relations of type
\be
(f_{\rm on\;\; a\;\; given\;\; edge})+
{\cal D}_{\rm on\;\; a\;\; given\;\; edge}\geq0
\label{341}
\ee
by a constant divisor.

Now to the initial edge a function satisfies identically equal to
unity, and also an {\em effective\/} divisor ${\cal D}_{\rm ini}$,
i.e.\ a formal sum of $g_0$ points in $\Gamma_0$. According to
\S 3 in Chapter~II of the book~\cite{Mumford-theta}, we can
choose a vector
${\bf z}_0\in {\bf C}^{g_0}$ and a point $P_0\in \Gamma_0$ in such a way that
the multivalued function
\be
\varphi(P)=\theta\bigl({\bf z}_0+\intop^P_{P_0} \ombf\bigr)
\ee
of a point $P\in \Gamma_0$ will have exactly
${\cal D}_{\rm ini}$ as its zero divisor. Recall that $\ombf$ is
a vector composed of holomorphic differentials on~$\Gamma_0$.

As for the rest of the edges, to each of them a divisor of type
\be
{\cal D}_{\rm on\;\; a\;\; given\;\; edge}=
{\cal D}_{\rm ini}+\sum^k_{j=1} R_j-\sum^k_{j=1}Q_j
\label{343}
\ee
corresponds, where $k$ is some number (increasing as we are moving off
the initial edge), and each of the points $R_j$ and $Q_j$ coincides with
one of the points $P_1,\ldots, P_6$.
Correctness of~(\ref{343}) follows immediately from comparing
divisors on {\em neighboring\/} edges, see e.g.\ Fig.~\ref{3fig04}
(from where it is seen also what the points $R_j$ and $Q_j$ exactly are;
we will be concerned with this somewhat later).

The explicit expression for a function satisfying~(\ref{341}) with
a divisor~(\ref{343}) is as follows (necessary explanations
are given just below the formula):
\be
f_{\rm on\;\; a\;\; given\;\; edge}(P)=
{{\displaystyle \theta\bigl({\bf z}_0+\intop^P_{P_0}\ombf-
\sum^k_{j=1} \intop^{R_j}_{Q_j}\ombf\bigr)} \over
{\displaystyle \theta\bigl({\bf z}_0+\intop^P_{P_0}\ombf\bigr)}}
\cdot
{{\displaystyle \prod^k_{j=1} \theta\bigl({\bf e}+\intop^P_{Q_j}\ombf\bigr)}
\over
{\displaystyle \prod^k_{j=1} \theta\bigl({\bf e}+\intop^P_{R_j}\ombf\bigr)}}.
\label{344}
\ee
Here ${\bf e}\in{\bf C}^{g_0}$ is any such vector that
\be
\theta({\bf e})=0,
\label{345}
\ee
while the function
$$
E_{{\bf e}}(x,y)=\theta\bigl({\bf e}+\intop^y_x \ombf\bigr)
$$
of two points $x,y\in\Gamma$, called
{\em principal form\/},
does not identically vanish. It remains to indicate
the integration paths in (\ref{344}),
more precisely, their homological classes. Connect all points $R_j$ and
$Q_j$, $1\leq j \leq k$, and also $P$, with $P_0$ by arbitrary paths.
This done, we will assume that each path $Q_jR_j$ consists of two
parts $Q_jP_0$ and $P_0R_j$, and, similarly, that paths $R_jP$ and $Q_jP$
also ``pass through the point~$P_0$''.

One can verify by standard means that~(\ref{344}) defines a
single-valued function on the curve~$\Gamma_0$, so that the choice
of path~$P_0P$ in fact plays no r\^ole. As for the equality~(\ref{341}),
it follows from the fact that all ``superfluous'' zeros of
principal
forms in the enumerator and denominator of the right-hand fraction
in~(\ref{344}) cancel each other~\cite{Mumford-theta}.

Now pass on to calculating matrix elements corresponding to
transitions between neighboring edges of kagome lattice,
by formulas like (\ref{307}, \ref{308}),
indicating {\it en passant\/} concretely points $R_j$ and $Q_j$ in
formulas (\ref{343}, \ref{344}). To begin, consider a situation
depicted in Fig.~\ref{3pichgr1}.
\bfig
\begin{center}
\unitlength=0.209206\normalbaselineskip
\special{em:linewidth 0.4pt}
\linethickness{0.4pt}
\begin{picture}(60.00,60.00)
\emline{0.00}{40.00}{1}{60.00}{40.00}{2}
\emline{20.00}{60.00}{3}{20.00}{0.00}{4}
\emline{0.00}{0.00}{5}{60.00}{60.00}{6}
\put(1.00,42.00){\makebox(0,0)[cb]{$P_6-P_1$}}
\put(22.00,58.00){\makebox(0,0)[lt]{$P_6-P_5$}}
\put(53.00,55.00){\makebox(0,0)[rb]{$P_2-P_5$}}
\put(58.00,42.00){\makebox(0,0)[rb]{$P_2-P_4$}}
\put(32.00,28.00){\makebox(0,0)[lt]{$P_2-P_3$}}
\put(22.00,2.00){\makebox(0,0)[lb]{$P_2+P_6-P_3-P_4$}}
\put(2.00,5.00){\makebox(0,0)[rb]{$P_2+P_6-P_1-P_3$}}
\put(18.00,30.00){\makebox(0,0)[rc]{$P_6-P_3$}}
\put(20.00,39.50){\rule{20.00\unitlength}{1.00\unitlength}}
\end{picture}
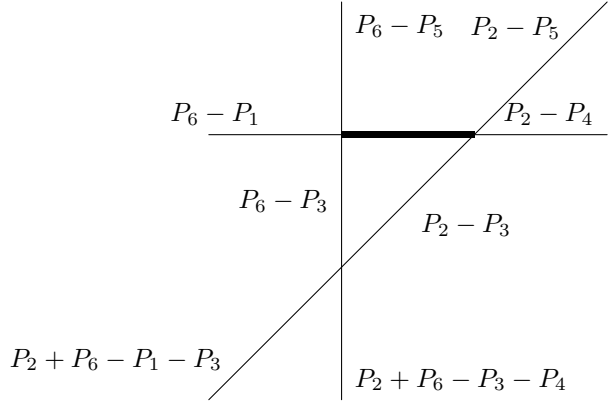
\caption{The ``initial'' edge and divisors
$\displaystyle\sum^k_{j=1}R_j-\sum^k_{j=1}Q_j$
on the neighboring edges}
\label{3pichgr1}
\end{center}
\efig
This figure is, in essence, a modification of Fig.~\ref{3fig04}.
The thick
edge is the initial one. Near other edges, the divisors
$\displaystyle\sum^k_{j=1}R_j-\sum^k_{j=1}Q_j$ corresponding to
them are written out
(recall that the poles of functions are situated in points~$R_j$,
while the zeros---in points~$Q_j$).
The system~(\ref{307}), after substituting the functions according
to~(\ref{344}) and Fig.~\ref{3pichgr1} and multiplying by
the common denominator, acquires the following form:
$$
\pmatrix{a_1&b_1\cr
c_1&d_1}
\pmatrix{\displaystyle\theta\bigl({\bf z}_0+\intop^P_{P_0}\ombf\bigr)
\,\theta\bigl({\bf e}+\intop^P_{P_2}\ombf\bigr) \cr
\noalign{\smallskip}
\displaystyle
\theta\bigl({\bf z}_0+\intop^P_{P_0}\ombf-\intop^{P_2}_{P_3}\ombf\bigr)
\,\theta\bigl({\bf e}+\intop^P_{P_3}\ombf\bigr)
} =
$$
\par\nobreak
\be
=\pmatrix{\displaystyle
\theta\bigl({\bf z}_0+\intop^P_{P_0}\ombf-\intop^{P_2}_{P_4}\ombf\bigr)
\,\theta\bigl({\bf e}+\intop^P_{P_4}\ombf\bigr)\cr
\noalign{\smallskip}
\displaystyle
\theta\bigl({\bf z}_0+\intop^P_{P_0}\ombf-\intop^{P_2}_{P_5}\ombf\bigr)
\,\theta\bigl({\bf e}+\intop^P_{P_5}\ombf\bigr)
}
\label{347}
\ee

The equality~(\ref{347}) must hold for any point $P\in \Gamma_0$.
Recalling~(\ref{345}), we see that $a_1$  and $c_1$ are easily found
if we set
$P=P_3$, while $b_1$ and $d_1$ ---if we set $P=P_2$.
It is convenient to introduce notations
$$
{\bf z}(P)=\intop^P_{P_0} \ombf,\quad
{\bf z}_j=\intop^{P_j}_{P_0} \ombf, \quad
1\leq j\leq 6.
$$

We get
\be
a_1={\theta({\bf z}_0+{\bf z}_3-{\bf z}_2+{\bf z}_4)
\,\theta({\bf e}+{\bf z}_3-{\bf z}_4) \over
\theta({\bf z}_0+{\bf z}_3)\,\theta({\bf e}+{\bf z}_3-{\bf z}_2)},
\label{348}
\ee
\be
b_1={\theta({\bf z}_0+{\bf z}_4)\,\theta({\bf e}+{\bf z}_2-{\bf z}_4) \over
\theta({\bf z}_0+{\bf z}_3)\,\theta({\bf e}+{\bf z}_2-{\bf z}_3)},
\label{349}
\ee
\be
c_1={
\theta({\bf z}_0+{\bf z}_3-{\bf z}_2+{\bf z}_5)\,\theta({\bf e}+{\bf z}_3-{\bf
z}_5) \over
\theta({\bf z}_0+{\bf z}_3)\,\theta({\bf e}+{\bf z}_3-{\bf z}_2)},
\label{350}
\ee
\be
d_1={
\theta({\bf z}_0+{\bf z}_5)\,\theta({\bf e}+{\bf z}_2-{\bf z}_5)\over
\theta({\bf z}_0+{\bf z}_3)\,\theta({\bf e}+{\bf z}_2-{\bf z}_3)}.
\label{351}
\ee

Matrix elements $a_2,\ldots,d_3$ are found similarly. We will write out for
them only systems of type~(\ref{347}), because explicit expressions
like (\ref{348}--\ref{351}) are obtained from them
in an obvious way by using~(\ref{345}). So,
$$
\pmatrix{a_2&b_2\cr
c_2&d_2}
\pmatrix{\theta({\bf z}_0+{\bf z}(P)-{\bf z}_6+{\bf z}_1)
\,\theta({\bf e}+{\bf z}(P)-{\bf z}_1)\cr
\theta({\bf z}_0+{\bf z}(P)-{\bf z}_6+{\bf z}_3)
\,\theta({\bf e}+{\bf z}(P)-{\bf z}_3)        } =
$$
\par\nobreak
\be
=\pmatrix{\theta({\bf z}_0+{\bf z}(P))
\,\theta({\bf e}+{\bf z}(P)-{\bf z}_6)\cr
\theta({\bf z}_0+{\bf z}(P)-{\bf z}_6+{\bf z}_5)
\,\theta({\bf e}+{\bf z}(P)-{\bf z}_5)},
\label{352}
\ee
$$
\pmatrix{a_3&b_3\cr
c_3&d_3}
\pmatrix{
\theta({\bf z}_0+{\bf z}(P)-{\bf z}_2-{\bf z}_6+{\bf z}_3+{\bf z}_1)
\,\theta({\bf e}+{\bf z}(P)-{\bf z}_1)\cr
\theta({\bf z}_0+{\bf z}(P)-{\bf z}_2-{\bf z}_6+{\bf z}_3+{\bf z}_4)
\,\theta({\bf e}+{\bf z}(P)-{\bf z}_4) } =
$$
\par\nobreak
\be
=\pmatrix{
\theta({\bf z}_0+{\bf z}(P)-{\bf z}_2+{\bf z}_3)
\,\theta({\bf e}+{\bf z}(P)-{\bf z}_6)\cr
\theta({\bf z}_0+{\bf z}(P)-{\bf z}_6+{\bf z}_3)
\,\theta({\bf e}+{\bf z}(P)-{\bf z}_2)}.
\label{353}
\ee

No we can extend Figure~\ref{3pichgr1} in any direction and calculate
matrix elements in some more vertices.
Fortunately, not complicated calculations like those done above
show that
{\em the formulae (\ref{347}--\ref{353}) remain valid under translations
by lattice periods, if a change
\be
{\bf z}_0\to{\bf z}_0+\xi(P_4+P_6-P_1-P_2)+
\eta(P_2+P_5-P_3-P_4),
\label{354}
\ee
is done in them, where the integers $\xi$ and $\eta$ show
show how many lattice periods horizontally and vertically
we have moved off the ``initial'' triangle in Fig.~\ref{3pichgr1}}.

Perfectly similar considerations, with using the divisor evolution
from Section~\ref{3secqp}, are valid for the dependence of matrix
elements on the time. The result consists, of course, in adding
in (\ref{354}) a linear dependence of ${\bf z}_0$ on the time~$\tau$.
We will formulate this as follows:
{\em if divisors on edges around two vertices of the same type
(for example, of type $\matrix{
\unitlength=0.073223\normalbaselineskip
\special{em:linewidth 0.4pt}
\linethickness{0.4pt}
\begin{picture}(20.00,15.00)
\emline{0.00}{5.00}{1}{20.00}{5.00}{2}
\emline{0.00}{-5.00}{3}{20.00}{15.00}{4}
\end{picture}
 }$) differ by
$\displaystyle \sum^l_{j=1}\tilde R_j-
\sum^l_{j=1}\tilde Q_j$, then the corresponding values
${\bf z}$ differ by
$\displaystyle \sum^l_{j=1}\intop^{\tilde Q_j}_{\tilde R_j}\ombf$
\/}.

Let us end this section by some simple remarks about an analog
of Hirota's bilinear representation for our dynamical system.
As was indicated in the beginning of this section,
only a {\em six-linear\/} representation could be found,
which is obtained if we write the entries of matrices like
$\pmatrix{a_1&b_1\cr c_1& d_1}$ in the form
\be
a_1={\alpha_1\over t_1},\quad b_1={\beta_1\over t_1},\quad
c_1={\gamma_1\over t_1},\quad d_1={\delta_1\over t_1}
\label{355}
\ee
etc.\ and then multiply every equality of the form
$$
\pmatrix{1&0&0\cr 0&a'_3&b'_3\cr 0&c'_3&d'_3}
\pmatrix{a'_2&0&b'_2\cr 0&1&0\cr c'_2&0&d'_2}
\pmatrix{a'_1&b'_1&0\cr c'_1&d'_1&0\cr 0&0&1} =
$$
\par\nobreak
\be
=\pmatrix{a_1&b_1&0\cr c_1&d_1&0 \cr 0&0&1}
\pmatrix{a_2&0&b_2\cr 0&1&0 \cr c_2&0&d_2}
\pmatrix{1&0&0\cr 0&a_3&b_3 \cr 0&c_3&d_3}
\label{356}
\ee
(where the ``shaded'' matrices belong to the time less by a
time unit than the non-shaded ones) by the common
denominator~$t'_1t'_2t'_3t_1t_2t_3$.
The reasonableness of the obtained equation is substantiated by the
following. Recall the formulae (\ref{348}--\ref{351}), where only
${\bf z}_0$ depends on coordinates and time, and set
$t_1=\theta({\bf z}_0+{\bf z}_3)$.
Similarly, choose the rest of $t_j$ equal to theta functions of
${\bf z}_0+{\rm const}$ situated in the denominators of formulae of type
(\ref{348}--\ref{351}). We get for each of the values $\alpha_j,
\beta_j, \gamma_j, \delta_j, t_j$ that it is in the ``finite-gap'' case
a theta function multiplied by a harmless
constant, which is a distinguishing feature of a Hirota
$\tau$-function.

\section{Invariant curve equation in the thermodynamical limit}
\label{6sec}

As soon as we began in Section~\ref{3secqp} to consider an inhomogeneous
6-vertex model on the infinite lattice whose Boltzmann weights
are constructed out of an algebraic curve~$\Gamma_0$ (of not very high genus)
and some divisors in it and depend on coordinates quasiperiodically,
we naturally come to the problem of existence and calculation of
the thermodynamic limit of specific free energy, i.e.\ the limit of
the logarithm of statistical sum divided by lattice area.
In this work, some steps in this direction are proposed.
Just now we will continue to consider the
{\em doubly periodic\/} case of Section~\ref{3secps}, assuming conditions
(\ref{309}, \ref{310}) of linear dependence with integer coefficients
for divisors
$P_1-P_6$, $P_2-P_4$ and $P_3-P_5$. The vectors of periods $(\xi_1,\eta_1)$
and $(\xi_2,\eta_2)$ can be, however, as big as desired.

We will see that there exists an elegant integral representation for the
real part of the limit of the ratio of the logarithm of
determinant~(\ref{233}) to
the area of the lattice ``covering'' the lattice on the torus
with periods
$(\xi_1,\eta_1)$ and $(\xi_2,\eta_2)$, with both dimensions of the
covering lattice tending to infinity.  It follows from
Theorem~\ref{2thstadet} on the ``physical level of rigor'' that this limit
coincides with the specific free energy (see also
Remark~\ref{2remdetphys}).
We will present some confirmation for this in a simple particular case
in the concluding part of this section, while now returning to
the mathematical level of rigor.

Let meromorphic functions $g(z)$ and $h(z)$ be given
on a smooth irreducible algebraic curve $\Gamma_0$. Then those functions
obey an algebraic dependence
\be
{\cal P}(g,h)=0,
\label{619}
\ee
${\cal P}$ being an irreducible polynomial. Standard methods of the theory
of functions of a complex variable
provide in a not complicated way an integral representation
for the {\em absolute value\/} of ${\cal P}(g,h)$.

Introduce the following notations:
$k$ and $l$ will be the numbers of poles, with regard to their
multiplicities, of functions $g(z)$ and $h(z)$ respectively. By $g_j(h_0)$,
where $1\leq j\leq l$, we will denote the values of function~$g$ in those
points where
$h(z)=h_0$ (we take those points in any order and also with regard to
their multiplicities).
Similarly, $h_j(g_0)$, with $1\leq j\leq k$, will be the values of $h$
in points where $g(z)=g_0$. The following integral representation
(a version of the Cauchy integral formula) takes place:
$$
2\sum^l_{j=1} \ln \left| {g_0-g_j(h_0) \over g_0-g_j(\infty)}\right|=
2\sum^k_{j=1}\ln\left|{h_0-h_j(g_0)\over h_0-h_j(\infty)}\right|=
$$
\par\nobreak
\be
={1\over 2\pi i}\iint_{\Gamma_0}
{{\rm d} g(z)\wedge {\rm d} \overline{h(z)} \over
\bigl(g(z)-g_0\bigr)\bigl(\overline{h(z)}-\overline{h\vphantom(}_0\bigr)
\vphantom{\big|}
}.
\label{620}
\ee

Let us prove, for the sake of completeness,
the formula~(\ref{620}). Consider a real-valued function
$$
\varphi(z)=2\ln\left|g(z)-g_0\right|
$$
defined on $\Gamma_0$, except, of course, the points where $g(z)=g_0$ or
$\infty$.
As the {\em holomorphic part\/}
of the function $\varphi$ differential coincides with ${\rm d} g(z)/
\bigl(g(z)-g_0\bigr)$, the integral in
(\ref{620}) can be rewritten as
\be
{1\over 2\pi i} \iint_{\Gamma_0}
{{\rm d} \varphi(z)\wedge {\rm d}\overline{h(z)} \over
\overline{h(z)}-\overline{h\vphantom(}_0
\vphantom{\big|}
}.
\label{622}
\ee
By Stokes theorem, the integral~(\ref{622}) equals
$$
-{1\over 2\pi i}\sum_{\scriptstyle\rm over\;\; singular
\atop\scriptstyle\rm points}
\oint\varphi(z)
{{\rm d}\overline{h(z)}\over \overline{h(z)}-\overline{h\vphantom(}_0
\vphantom{\big|}
},
$$
where the sum is taken over all singular points of the
integrand in (\ref{622}),
while the integrals under summation sign---along
infinitesimal contours around those points in the anticlockwise
direction.
As the singularities of function $\varphi$ have the character of
modulus of local parameter logarithm, the integrals along
infinitesimal contours around them
vanish unless the singularities of $\varphi$ coincide with
the points where $h(z)=h_0$ or $\infty$. The integrals around those
points, divided by $-2\pi i$, equal $2\ln\left|g_j(h_0)-g_0\right|$
and $2\ln\left|g_j(\infty)-g_0\right|$ respectively. Hence
$$
2\sum^l_{j=1} \ln \left| {g_0-g_j(h_0) \over g_0-g_j(\infty)}\right|
={1\over 2\pi i}\iint_{\Gamma_0}
{{\rm d} g(z)\wedge {\rm d} \overline{h(z)} \over
\bigl(g(z)-g_0\bigr)\bigl(\overline{h(z)}-\overline{h\vphantom(}_0\bigr)
\vphantom{\big|}
}.
$$
The remaining equality in (\ref{620}) follows from the obvious equalness
in status of functions $g$ and $h$.
\medskip

To link the equalities~(\ref{620}) to the polynomial ${\cal P}(g,h)$ from
formula~(\ref{619}), note that
$$
\lim_{\scriptstyle g'\to\infty\atop\scriptstyle h'\to\infty}
{{\cal P}(g,h){\cal P}(g',h')\over {\cal P}(g',h){\cal P}(g,h')}=
\prod^l_{j=1}{g-g_j(h)\over g-g_j(\infty)}=
$$
\par\nobreak
\be
=\prod^k_{j=1}{h-h_j(g)\over h-h_j(\infty)}.
\label{625}
\ee
Equalities~(\ref{625}) are proved by directly comparing the zeros and
poles of LHS and RHS's in variable $g$ or~$h$.
Let ${\cal P}(g,h)$
be normalized so that the coefficient at the term
of highest degrees in both variables,
i.e.\ $g^lh^k$, equals unity. We have
\be
{{\cal P}(g',h)\over (g')^l}
\mathrel{\displaystyle\mathop{\longrightarrow}_{g'\to\infty}}
\prod^k_{j=1}\bigl(h-h_j(\infty)\bigr),
\label{626}
\ee
\be
{{\cal P}(g,h')\over (h')^k}
\mathrel{\displaystyle\mathop{\longrightarrow}_{h'\to\infty}}
\prod^l_{j=1}\bigl(g-g_j(\infty)\bigr).
\label{627}
\ee

Formulae (\ref{620}) and (\ref{625}--\ref{627}) together yield
$$
\left|{\cal P}(g_0,h_0)\right|^2=
\exp \left(
{1\over 2\pi i}\iint_{\Gamma_0}
{
{\rm d}g(z)\wedge {\rm d}\overline{h(z)} \over
\bigl(g(z)-g_0\bigr)\bigl(\overline{h(z)}-\overline{h\vphantom(}_0\bigr)
\vphantom{\big|}
}
\right)
\times
$$
\par\nobreak
\be
\times\left|
\prod^k_{j=1}\bigl(h_0-h_j(\infty)\bigr)\cdot
\prod^l_{p=1}\bigl(g_0-g_p(\infty)\bigr)
\right|^2.
\label{628}
\ee

We will see later that, in order to pass to a ``thermodynamical
limit'' in formula (\ref{628}), we must consider the following situation.
Let functions $g(z)$ and $h(z)$ on $\Gamma_0$ be represented in
the form
\be
g(z)=u^{\xi},\quad h(z)=w^{\eta},
\label{629}
\ee
where $\xi$ and $\eta$ are integers, while $u$ and $w$ are
functions on some {\em covering\/} of $\Gamma_0$ (this can always
be done, of course).
Let $a$ and $b$ be positive integers divisible by $\xi$ and $\eta$
respectively. Introduce functions
\be
G(z)=u^a=g(z)^{a/\xi}, \quad H(z)=w^b=h(z)^{b/\eta}.
\label{630}
\ee
Consider a limit
\be
\lim_{\scriptstyle a\to\infty \atop\scriptstyle b\to\infty}
{1\over ab}
\iint_{\Gamma_0} {{\rm d}G(z)\wedge{\rm d}\overline{H(z)}\over
\bigl(G(z)-G_0\bigr)\bigl(\overline{H(z)}-\overline{H\vphantom(}_0\bigr)
\vphantom{\big|}
},
\label{631}
\ee
where
\be
G_0=u^a_0, \quad H_0=w^b_0
\label{631a}
\ee
with some $u_0$ and $w_0$ not depending on $a$ and $b$. Changing the
order of passage to limit and integration
(the legitimacy of which operation is verified by standard means), we find
that the limit~(\ref{631}) equals
\be
\iint_{\Delta}{\rm d}\ln u(z)\wedge{\rm d}\,\overline{\ln w(z)},
\label{632}
\ee
$\Delta$ being that part of $\Gamma_0$ where
\be
|u|\geq |u_0|, \quad |w|\geq|w_0|.
\label{633}
\ee

Let us now link the above considerations
to Section~\ref{3secps} in the following way.
Let us take functions $g_1(z)$ and $g_2(z)$ from Section~\ref{3secps}
as $g(z)$ and $h(z)$ respectively. Assume for simplicity that
in formulae~(\ref{309},~\ref{310})
\be
\eta_1=\xi_2=0,
\label{635}
\ee
and denote
$$
\xi=\xi_1, \quad \eta=\eta_2.
$$
Thus, the torus becomes ``rectangular'', with periods $(\xi,0)$ and
$(0,\eta)$, while formulae (\ref{312}) turn into~(\ref{629}). Assume also
conditions (i--iii) on p.~\cpages{3cond1}{3cond3}, observing that the last
of them will mean just that $\xi$ and $\eta$ are relatively prime.

It follows from the table on p.~\pageref{3tab} and formula (\ref{635})
that $g_1(z)$ has zeros of multiplicity
$\xi$ in points  $P_1$ and $P_2$, and poles of multiplicity $\xi$
in $P_4$ and $P_6$, while $g_2(z)$ has zeros of multiplicity $\eta$
in points $P_3$ and $P_4$,
and poles of multiplicity $\eta$ in $P_2$ and $P_5$. Hence, it is clear
that functions $u$ and $w$
on a $\xi\eta$-sheeted  {\em nonramified covering\/} of~$\Gamma_0$
possess the following zeros and poles {\em of multiplicity one\/}:
\begin{center}
\begin{tabular}{c|c|c|}
& \vphantom{$\big|$}zeros & poles\\
\hline
$u\vphantom{\big|}$ & $P_1$, $P_2$ & $P_4$, $P_6$ \\
\hline
$w\vphantom{\big|}$ & $P_3$, $P_4$ & $P_2$, $P_5$\\
\hline
\end{tabular}
\label{6tab}
\end{center}
\bigskip
\penalty0

\begin{remark}\label{6rem}
Functions $u$ and $w$, thus, can be expressed in a simple way through\/
{\em principal forms\/} on\/ $\Gamma_0$ \cite{Mumford-theta}, but we
will not need those expressions here.
\end{remark}

As will be clear soon, the limit
$$
\lim_{\scriptstyle a\to\infty \atop \scriptstyle b\to\infty}
{1\over ab} \ln\left|\tilde{\cal P}(G_0, H_0)\right|,
$$
is important for us. Here the polynomial $\tilde{\cal P}(G,H)$
defines the algebraic dependence between functions~(\ref{630})
which can be written as
\be
\tilde{\cal P}(G,H)=\prod^{a/\xi}_{p=1}\prod^{b/\eta}_{q=1}{\cal P}
\left(g\exp{2\pi ip\xi\over a},\,h\exp{2\pi iq\eta\over b}\right)=0.
\label{638}
\ee
Substituting $G$ and $H$ in place of $g$ and $h$ in~(\ref{628}), taking
into account the equality of expressions (\ref{631}) and (\ref{632}), and
knowing the poles of functions $G(z)$ and $H(z)$
(see~(\ref{630}) and the table before Remark~\ref{6rem}), we find:
$$
\lim_{\scriptstyle a\to\infty \atop \scriptstyle b\to\infty}
{1\over ab} \ln\left|\tilde{\cal P}(G_0, H_0)\right|=
$$
\par\nobreak
$$
={1\over 4\pi i}\iint_{\Delta}{\rm d}\,\ln u(z)\wedge {\rm d}\,
\overline{\ln w(z)} +\ln|u_0|+\ln|w_0|+
$$
\par\nobreak
\be
+\ln\left(\max\{|u_0|,|u(P_5)|\}\right)+
\ln\left(\max\{|w_0|,|w(P_4)|\}\right),
\label{639}
\ee
where domain $\Delta\subset\Gamma_0$ is, as before, defined by
inequalities~(\ref{633}),
while $u_0$ and $w_0$, of course, obey~(\ref{631a}).
\medskip

Pass now on to explaining why the polynomial~(\ref{638}) arises
when we take a lattice on the torus $a\times b$ that covers torus
$\xi\times \eta$.

\begin{theorem}\label{6thab}
Let the invariant curve corresponding to the model on ``rectangular''
torus of sizes $\xi\times\eta$ be given by equality
\be
{\cal P}(g,h)={\cal P}\left(u^{\xi},w^{\eta}\right)=0.
\label{640}
\ee
Let numbers $a$ and $b$ be divisible by $\xi$  and $\eta$ respectively.
Then the invariant curve corresponding to model on the lattice
on torus $a\times b$,
with matrix elements in the vertices
``raised''
from the lattice $\xi\times\eta$ according to the natural covering,
is given by equation~(\ref{638}), where (\ref{629}) must be substituted.
\end{theorem}

\begin{remark}\label{6remab}
Thus, the invariant curve in this case consists of
$ab\over\xi\eta$ copies of the covering of\/~$\Gamma_0$ defined by
multifunctions $u$ and $w$~(\ref{629}). Recalling conditions (i--ii)
on p.~\cpages{3cond1}{3cond2} and Lemma~\ref{3lem-n2}, we see that under
these conditions the invariant curve contains exactly $ab\over \xi\eta$
components.
\end{remark}

{\it Proof of Theorem \ref{6thab}}. Recall that an invariant curve
is birationally isomorphic to a set of such pairs
$(u,w)$ for which there exists a vector
${\cal X}=\pmatrix{X\cr Y\cr Z}$ such that
\be
{\cal A}\pmatrix{X\cr Y\cr Z}=
\pmatrix{uX\cr uwY\cr wZ}
\label{641}
\ee
(see Definition~\ref{2dfn1} and formula (\ref{213}) after it). If we deal
with the torus $a\times b$, then each of column vectors $X$, $Y$, and $Z$
has $ab$ components. Recall that when considering equalities of
type~(\ref{641}), we pass from the kagome lattice to the triangular one,
``packing'' triangles in circles, as was explained in Section~\ref{2seckag}.
Let $(x,y)$  be integer-valued coordinates of a given vertex of
triangular lattice.
The equality~(\ref{641}) remains valid under the transformation
\be
{\cal X}=\pmatrix{X\cr Y\cr Z}\longrightarrow
\pmatrix{\exp (2\pi ic_1x/a)X\cr
\exp(2\pi ic_1x/a+2\pi c_2y/b)Y\cr
\exp(2\pi ic_2y/b)Z},
\label{642}
\ee
\par\nobreak
\be
\matrix{u\to \exp(-2\pi ic_1/a)u, \cr
w\to \exp(-2\pi ic_2/b)w,}
\label{643}
\ee
where $c_1$ and $c_2$ are any integers. If matrix elements of
$\cal A$ were raised
from the lattice $\xi\times\eta$, then among the vectors
$\cal X$ there are periodic ones with periods $(\xi,0)$ and $(0,\eta)$,
and pairs $(u,w)$ corresponding to such vectors form
the curve~(\ref{640}).
Clearly, {\em all\/} vectors~$\cal X$ must be periodic with periods
$(a,0)$ and $(0,b)$, and they all can be obtained from vectors with
periods $(\xi,0)$ and $(0,\eta)$ by transformation~(\ref{642}) with
proper $c_1$ and~$c_2$. It is not difficult to see that
the corresponding transformations~(\ref{643})
of pairs $(u,w)$ provide exactly $ab\over \xi\eta$ isomorphic components
of the invariant curve according to~(\ref{638}). The theorem is proved.
\medskip

Note that ``integral of motion'' $I(u,w)$
(\ref{233}) from Section~\ref{2seckag} differs from polynomial
$\tilde{\cal P}(G,H)$ by a factor whose absolute value is $|uw|^{-2ab}$
(see (\ref{638}) and a remark about nomalization of ${\cal P}(g,h)$ before
formula~(\ref{626})). Hence we will get
$$
\lim_{\scriptstyle a\to\infty \atop\scriptstyle b\to\infty}
{1\over ab}\,\ln \left| I(u_0,w_0)\right|
$$
if we subtract $\left(2\ln |u_0|+2\ln |w_0|\right)$ from the RHS
of~(\ref{639}).
Thus, the desired thermodynamic limit has been calculated.

\begin{remark}\label{6remstrings}
Logarithmic differentials of multifunctions $u(z)$ and $w(z)$
on the curve $\Gamma_0$, that occupy a key position in formula~(\ref{639}),
astonishingly belong to the same type as differentials playing
a key r\^ole in the algebro-geometrical string
theory~\cite{Kr-Nov-1,Kr-Nov-2,Kr-Nov-3}. Namely, they are differentials
of the third kind (i.e.\ possessing only simple poles) on $\Gamma_0$,
with all their periods pure imaginary (because absolute values
of  $u(z)$ and $w(z)$ are single-valued). The algebraic techniques
used in this work forced us, however,
to impose the integer linear dependence conditions
(\ref{309}, \ref{310}) on the divisors.

Besides, formula~(\ref{639}) reminds of string theory because
the double integral in it can be interpreted as
the area of domain~$\Delta$ in K\"ahler metric corresponding to
imaginary part of the 2-form
${\rm d} \ln u(z) \wedge {\rm d}\,\overline{\ln w(z)}$
(changing the form to its imaginary part
does not change the integral in (\ref{639}), because both sides
in that formula are real). As is known, the
``Nambu--Hara--Goto'' 
action in string theory is proportional to the
``world sheet'' area~\cite{Brink-H}.
\end{remark}

\medskip

In the remaining part of this section we will show how
our method works in case of a {\em homogeneous\/} model,
choosing a curve of genus~0 as~$\Gamma_0$:
\be
\Gamma_0={\bf C}P^1.
\label{650}
\ee
Clearly, conditions (\ref{309}, \ref{310}) of divisor equivalence
in this case always satisfied, and any integer vector is a period
(in the sense of Lemma~\ref{3lemtrif2}).

To simplify the task still more, we will consider a model on
{\em square\/} lattice. We will obtain it by ``removing
oblique lines'' from the triangular lattice. For this purpose, we will set
\be
P_2=P_4
\label{651}
\ee
and see how this will affect the form of matrix ${\cal A}_{\rm local}$
corresponding to a vertex of Fig.~\ref{3fig03} type and determined from
the relation~(\ref{306}). The latter is worth being written out here
once more:
\be
{\cal A}_{\rm local}
\pmatrix{f_{12}\cr f_{13} \cr f_{34}}=
\pmatrix{f_{46}\cr f_{56}\cr f_{52}} \label{652}
\ee
(recall that divisors ${\cal D}_1,\ldots,{\cal D}_6$ on Fig.~\ref{3fig03}
consist each of one point $P_1,\ldots,P_6$, and each meromorphic
function~$f_{jk}$ equals zero in points $P_j$ and $P_k$).

\begin{lemma}\label{6lemx}
Assume condition (\ref{651}) and normalize functions $f_{jk}$ on\/
{\em all\/} oblique edges by a condition
$$
f_{jk}(P_2)=1.
$$
Then, matrices ${\cal A}_{\rm local}$ in all vertices of triangular lattice
have the form
\be
{\cal A}_{\rm local}=\pmatrix{a&0&c\cr d&1&g\cr h&0&k\cr}.
\label{654}
\ee
\end{lemma}

{\it Proof\/} follows immediately from considering
relation~(\ref{652}) in the point $P_2$, taking into account
the fact that function $f_{jk}$ equals zero in that point if
2 or 4 is contained among the numbers $j$ and $k$, while otherwise equals
unity. Note that we do not use conditions~(\ref{650}) here.
\medskip

It follows from the form (\ref{654}) of matrix ${\cal A}_{\rm local}$
that every closed path on triangular lattice
to which a zero weight (~=~product of all matrix elements, associated with
that path, of matrices  ${\cal A}_{\rm local}$) corresponds, goes
either entirely through vertical and horizontal edges or
is just an oblique line (wound around the torus, of course).
The determinant~$I(u,w)$ (\ref{233}) factorizes in a product of
a factor corresponding to paths through the square lattice
and depending on $a,c,h,k$ and, certainly, on $u$ and $w$, and a factor
corresponding to oblique paths and depending only on
$u$~and~$w$.

It follows from the table on p.~\pageref{6tab} together with (\ref{651})
that functions $u$ and $w$ on $\Gamma_0={\bf C}P_1$ have each
{\em one\/}  zero and one pole.
The matrix $\pmatrix{a&c\cr h&k}$ is obtained from a relation
like the first of relations~(\ref{308}), which in our case can be written
as
$$
\pmatrix{a&c\cr h&k}
\pmatrix{f_{12}\cr f_{23}}=
\pmatrix{uf_{12}\cr wf_{23}},
$$
whence
\be
\left|\matrix{a-u & c \cr h & k-w}\right|=0.
\label{656}
\ee

We will simplify our task still more by considering the ``usual''
6-vertex model, assuming conditions
$$
a=k,\quad c=h,\quad c^2-a^2=1
$$
(see Section \ref{2seckag}, the text between Lemmas~\ref{2lemdet}
and~\ref{2lemsta},
on the connection between Boltzmann weights and matrix elements).
Then~(\ref{656}) turns into
\be
u={aw+1\over w-a}.
\label{657}
\ee

Let us assume that $u_0$ and $w_0$ in~(\ref{639}) are real and positive.
For concreteness, let $u_0$ and $w_0$ be not very different from unity,
and
$$
-1< a< 1.
$$
The integration domain $\Delta$ in (\ref{639}) for this case
(a lune between two circumferences)
is depicted, in the complex plane of variable $w$,
in Fig.~\ref{6fig}.
\typeout{Here must be a special figure, the sole one in this paper.
Please contact the author.}
\bfig
\begin{center}
\unitlength=1.00mm
\special{em:linewidth 0.4pt}
\linethickness{0.4pt}
\begin{picture}(67.00,80.00)

\end{picture}
\end{center}
\caption{Integration domain $\Delta$ in the double integral in
formula~(\protect\ref{639}) in case of a homogeneous model
on square lattice, depicted in the complex plane of variable $w$}
\label{6fig}
\efig

Transform the double integral in (\ref{639}) into
a single integral. We have
$$
\iint_{\Delta} {\rm d} \ln u \wedge {\rm d}\, \overline{\ln w}=
\iint_{\Delta} {\rm d} \ln\left|{u\over u_0}\right|^2 \wedge {\rm d}\,
\overline{\ln w}=
$$
\par\nobreak
\be
\oint\limits_{\partial\Delta}\ln\left|{u\over u_0}\right|^2 {\rm d}\,
\overline{\ln w}=
\intop_{\smile ABC}\ln\left|{u\over u_0}\right|^2 {\rm d}\,\overline{\ln w}.
\label{660}
\ee
Here we used the fact that
$\displaystyle \ln\left|{u\over u_0}\right|^2$ ---
is a single-valued function, moreover equalling zero in the arc $CDA$;
$\partial\Delta$
is the boundary of $\Delta$, i.e.\  $DCABC$. Divide the integral~(\ref{660})
by $4\pi i$, as required in formula~(\ref{639}), and transform it further,
using, in particular,~(\ref{657}):
\be
{1\over4\pi i}
\intop_{\smile ABC}\ln\left|{u\over u_0}\right|^2 {\rm d}\,\overline{\ln w}=
\intop_{-\phi_0}^{\phi_0} \ln \left|{aw_0{\rm e}^{i\phi}+1 \over
u_0(w_0{\rm e}^{i\phi}-a)}\right| {\rm d}\phi,
\label{661}
\ee
where $\phi_0$ found from conditions
$$
0<\phi_0<\pi, \quad \left|{aw_0{\rm e}^{i\phi}+1 \over
u_0(w_0{\rm e}^{i\phi}-a)}\right| =1.
$$

The integral~(\ref{661}) is the essential, i.e.\ depending on
$a$, part of the limit in the LHS of~(\ref{639}).
Of course, it coincides, to within an additive constant
(to be exact, a function depending only on
$u_0$ and $w_0$), with the specific free energy of the homogeneous
``free fermionic'' 6-vertex model on the square lattice,
see e.g.\ references \cite{Baxterbook,FT}.

\section{A ``local'' approach to orthogonal matrices}
\label{4seca}

In this section we will show what symmetry conditions
must be imposed on the
``local'' algebro-geometrical objects from Section~\ref{3secqp}
in order to obtain the description of evolution of {\em orthogonal\/}
matrices.
In contrast to direct calculations of Section~\ref{2secsmport},
here we simply present constructions made in the spirit of
works~\cite{B-Ch,Cherednik,Koz-Kot,Its-Kot}, and prove their
necessary properties.

Let, as in Section~\ref{3secqp}, there be given an algebraic curve
$\Gamma_0$ and six divisors in it consisting each of one point,
namely $P_1, \ldots ,P_6$. Let, besides, the curve~$\Gamma_0$, like
the ``global''  curve~$\Gamma$ from Section~\ref{2secsmport}, possess
an involution under which $P_1, \ldots ,P_6$ are transformed
as follows: $P_1 \leftrightarrow P_6$, $P_2 \leftrightarrow P_4$,
$P_3 \leftrightarrow P_5$. Consider {\em meromorphic differentials\/}
$\psi$ on $\Gamma_0$ possessing the following properties:
a)~$\psi$ has poles (of not higher order than 1) only in points
$P_1, \ldots, P_6$ and b)~$\psi$ transforms into $-\psi$ under the action
of involution~$I$. It follows from those properties and from the fact that
the dimension of the linear space of {\em holomorphic\/} differentials
on $\Gamma_0$ equals $g_0$ (the genus of the curve) that
the dimension of linear space of differentials~$\psi$ equals $g_0+3$.

Fix now a concrete differential $\psi$ with non-zero residues
in all points $P_1, \ldots, P_6$, and consider a divisor $\cal D$ on
$\Gamma_0$ such that
${\cal D}+{\cal D}^I$ is exactly the zero divisor of~$\psi$
(here ${\cal D}^I$ is the image of ${\cal D}$ under involution~$I$).
Thus,
\be
{\cal D}+{\cal D}^I\sim {\cal D}_{\rm can}+P_1+\ldots +P_6,
\label{401}
\ee
where ${\cal D}_{\rm can}$ is any canonical divisor on~$\Gamma_0$.
The relation~(\ref{401}) is a ``local'' analog
of relation~(\ref{251}). Divisor~${\cal D}$ consists of $g_0+2$ points,
while the dimension of linear space of meromorphic functions~$f$
satisfying condition
\be
(f)+{\cal D}\geq 0,
\label{402}
\ee
generically, equals~3.

Define the scalar product $\langle f,g \rangle$ of two functions
satisfying~(\ref{402}) as the sum of residues of differential $fg^I\psi$
in points $P_1,P_2,P_5$ ($g^I$  is, of course, the image of $g$ under
evolution~$I$):
\be
\langle f,g \rangle =\sum_{P_1,P_2,P_5} {\rm Res}\, fg^I\psi.
\label{403}
\ee
We will study the properties of this scalar product step by step.
First let us verify that it is symmetric, i.e.\ that
\be
\langle f,g \rangle = \langle g,f \rangle .
\label{404}
\ee
Apply involution $I$ to  {\em all\/} objects entering in RHS of
(\ref{403}), which will cause no change to that RHS as a whole.
We get:
\be
\langle f,g \rangle =\sum_{P_6,P_4,P_3} {\rm Res}\,(-f^Ig\psi)=
\sum_{P_1,P_2,P_5} {\rm Res}\, f^Ig\psi.
\label{405}
\ee
The right-hand equality in (\ref{405}) follows from the fact that the sum
of differential $f^Ig\psi$ residues, taken over {\em all\/} its poles
$P_1, \ldots,P_6$, equals zero. Further, the rightmost side of
(\ref{405}) obviously equals
$\langle g,f \rangle$, so (\ref{404}) is proved.

Consider now scalar products of functions $f_{jk}$ satisfying conditions
\be
f_{jk}+{\cal D} -P_j-P_k\geq 0.
\label{406}
\ee
In other words, function $f_{jk}$ has a pole divisor $\cal D$ and must
have zeros in points $P_j$ and $P_k$. We will prove the equalities
\be
\langle f_{26},f_{36} \rangle=\langle f_{56},f_{46} \rangle=0,
\label{407}
\ee
\be
\langle f_{12},f_{23} \rangle = \langle f_{26},f_{52} \rangle =0,
\label{408}
\ee
\be
\langle f_{13},f_{34} \rangle = \langle f_{36},f_{23} \rangle =0.
\label{409}
\ee

Let us prove, for example, that $\langle f_{26},f_{36} \rangle=0$. We have
$f^I_{36}=f_{51}$, so meromorphic differential $f_{26}f^I_{36}\psi $
{\em has no poles at all\/} in points $P_1,P_2,P_5$, whence the
scalar product (\ref{403}) indeed equals~0. The remaining equalities
(\ref{407}--\ref{409}) are proved similarly.

If now we manage to normalize all functions $f_{jk}$ entering in equalities
(\ref{407}--\ref{409}) so that
\be
\langle f_{jk},f_{jk} \rangle = 1,
\label{410}
\ee
and turn to
relations (\ref{307}, \ref{308}) defining ``local''
matrices $\pmatrix{a_l& b_l \cr c_l& d_l}$ for $l=1,2,3$, we will see that
those matrices are orthogonal ((\ref{407}), (\ref{408}) and (\ref{409})
are responsible, together with (\ref{410}), for orthogonality of
local matrices with
$l=1,\,2$ and $3$ respectively). At the same time, (\ref{410}) will show
the non-degeneracy of scalar product~(\ref{403}). Thus consider,
e.g., a scalar square of function~$f_{12}$. We have $f^I_{12}=f_{64}$,
so there remains only a residue in point~$P_5$ in formula~(\ref{403}).
Generically, $f_{12}(P_5)f_{64}(P_5)\neq 0$, so this residue does not
vanish, and we can divide $f_{12}$ by its square root.

Similarly, the scalar squares of other functions entering in
(\ref{407}--\ref{408}) are non-zero, hence those functions can be
normalized according to~(\ref{410}).

As a result, we have proved so far the orthogonality of three matrices
$\pmatrix{a_l & b_l\cr c_l&d_l}$ defined by relations
(\ref{307}, \ref{308}) and corresponding to vertices of some chosen
triangle in kagome lattice, depicted in Fig.~\ref{3fig04}
(p.~\pageref{3fig04}). To all (nine) edges in Fig.~\ref{3fig04}
meromorphic functions corresponded whose divisors satisfied
equality~(\ref{402}). We have introduced a scalar product
in the three-dimensional space of such functions
by formula~(\ref{403}), and normalized the functions according
to~(\ref{410}).

We can deal similarly with any other triangle of the form of
Fig.~\ref{3fig04}. If such a triangle is
$\xi$ horizontal lattice periods and $\eta$ vertical periods far
from the one that we have considered, then
divisor $\cal D$ must be replaced by
$$
{\cal D}_{(\xi,\eta)}= {\cal D} + \xi\,(P_1+P_2-P_4-P_6)+
\eta\,(P_3+P_4-P_5-P_2).
$$
Remarkably, the scalar product in the space of meromorphic functions~$f$
such that
$$
(f)+{\cal D}_{(\xi,\eta)}\geq 0
$$
may (and must) be introduced again by formula~(\ref{403}) with
the {\em same\/}
differential~$\psi$ as for functions satisfying~(\ref{402}).
The point is that ``superfluous'' poles of function~$f$ are exactly
compensated by zeros of function~$g^I$ and vice versa,
so that differential~$fg^I\psi$ can, as before, have only
first order poles in points $P_1, \ldots, P_6$.

The analogs of orthogonality relations (\ref{407}--\ref{409}) for
{\em all\/}
lattice vertices are proved by as before.
In the same way it is proved
that we can normalize functions on all edges by setting their scalar
square equal to unity. The ``outer'' edges of any diagram of type
as in Fig.~\ref{3fig04} enter also in one of neighboring diagrams.
It is important that the normalizing of a function on such edge
does not depend on a diagram to which we consider the edge belongs.

We will formulate the results of this section as a following theorem.
\begin{theorem}\label{4tha}
If a curve~$\Gamma_0$ in the situation of Section~\ref{3secqp}
possesses an involution~$I$
mapping the points $P_1,P_2,P_3$ in $P_6,P_4,P_5$ respectively, while
a divisor~$\cal D$ satisfies condition~(\ref{401}), then under a proper
gauge all square matrices in LHS's of equalities
(\ref{306}--\ref{308}) are orthogonal.
\end{theorem}

{\it Proof}. It remains to give two simple clarifications. First,
the gauge mentioned in the theorem is fixed by normalization conditions
of type~(\ref{410}). Second, we considered in this section only
a situation at one moment of discrete time. However, a simple calculation
based on divisor evolution described between Lemmas
\ref{3lemgct}  and
\ref{3lemtrifle} shows that our considerations are valid for any
time (``superfluous'' poles of each function
in~(\ref{403}) are compensated by zeros of another function as well
as before).

\section{Reduction to Ising model}
\label{4secb}

It is known that Onsager's ``star--triangle'' transformation
(\cite{Onsager},
see also manual~\cite{Baxterbook}), which can be graphically represented
as
\be
\matrix{
\unitlength=0.104606\normalbaselineskip
\special{em:linewidth 0.6pt}
\linethickness{0.2pt}
\begin{picture}(32.00,8.00)
\emline{24.00}{0.00}{1}{32.00}{0.00}{2}
\emline{32.00}{0.00}{3}{28.00}{8.00}{4}
\emline{28.00}{8.00}{5}{24.00}{0.00}{6}
\put(12.00,4.00){\vector(1,0){7.00}}
\emline{0.00}{0.00}{7}{4.00}{3.00}{8}
\emline{4.00}{3.00}{9}{8.00}{0.00}{10}
\emline{4.00}{3.00}{11}{4.00}{8.00}{12}
\end{picture}
 },
\label{420}
\ee
converts the statistical mechanical Ising model on a plane
hexagonal lattice to the Ising model on a triangular lattice.
Imagine now the obtained triangular lattice as made up of
triangles of the form $\matrix{
\unitlength=0.104606\normalbaselineskip
\special{em:linewidth 0.6pt}
\linethickness{0.2pt}
\begin{picture}(8.00,10.00)
\emline{0.00}{8.00}{1}{8.00}{8.00}{2}
\emline{8.00}{8.00}{3}{4.00}{0.00}{4}
\emline{4.00}{0.00}{5}{0.00}{8.00}{6}
\end{picture}
 }$.
If we apply the triangle--star
\be
\matrix{
\unitlength=0.104606\normalbaselineskip
\special{em:linewidth 0.6pt}
\linethickness{0.2pt}
\begin{picture}(32.00,8.00)
\emline{28.00}{0.00}{1}{28.00}{5.00}{2}
\emline{28.00}{5.00}{3}{24.00}{8.00}{4}
\emline{32.00}{8.00}{5}{28.00}{5.00}{6}
\put(12.00,4.00){\vector(1,0){7.00}}
\emline{0.00}{8.00}{7}{8.00}{8.00}{8}
\emline{8.00}{8.00}{9}{4.00}{0.00}{10}
\emline{4.00}{0.00}{11}{0.00}{8.00}{12}
\end{picture}
 },
\label{421}
\ee
transformation to those triangles, we will come to hexagonal lattice again.
If Ising model is inhomogeneous, i.e.\
a coupling along a given edge depends on the edge,
then transformations (\ref{420}) and (\ref{421}) applied alternately
lead to some evolution of those coefficients
on alternating hexagonal and triangular lattices. A hypothesis looks natural
that this evolution is completely integrable.

In this, not very large, section we will show that the mentioned evolution
of Ising model coupling coefficients is isomorphic to the evolution of
orthogonal matrices in vertices of triangular lattice. The latter evolution
was considered from a ``local'' algebro-geometrical viewpoint
in Section~\ref{4seca}.

The direct way to Ising model turns out to go via considering
the orthogonal matrices {\em with determinant~$-$1}. As we will see,
this is connected with the fact that such matrices are
{\em symmetric\/} (recall that they are of size
$2\times 2$).

So, consider a relation
$$
\pmatrix{a_1& b_1 &0 \cr b_1 &-a_1 &0 \cr 0&0&1}
\pmatrix{a_2& 0& b_2\cr 0&1&0 \cr b_2 &0& -a_2 }
\pmatrix{1&0&0 \cr 0&a_3 &b_3 \cr 0&b_3&-a_3}=
$$
\par\nobreak
\be
=\pmatrix{1&0&0\cr 0&a'_3&b'_3 \cr 0& b'_3&-a'_3}
\pmatrix{a'_2&0&b'_2\cr 0&1&0 \cr b'_2&0&-a'_2}
\pmatrix{a'_1&b'_1&0\cr b'_1&-a'_1 &0\cr 0&0&1},
\label{422}
\ee
where for all $j$
$$
a^2_j+b^2_j=1, \quad (a'_j)^2+(b'_j)^2=1.
$$
The relation (\ref{422}) corresponds to description of
matrix evolution given in Section~\ref{3secqp}: at every step
a $3\times 3$ matrix obtained as a product of the form as
in RHS of~(\ref{422}) is factorized in a product of the form as
in the LHS of that formula.

Introduce Ising model coupling coefficients $K_1$, $K_2$, $K_3$,
$L_1$, $L_2$, $L_3$ by formulae
\begin{eqnarray}
\exp(\pm 2K_1)=b_1\mp ia_1,\quad
\exp (\pm 2K_2)={ i(b_2\mp 1)\over a_2}, \nonumber \\
\exp (\pm 2K_3)=b_3\mp ia_3,
\label{424}
\end{eqnarray}
\begin{eqnarray}
\exp (\pm 2L_1)={i(b'_1 \pm 1) \over a'_1},\quad
\exp (\pm 2L_2)=b'_2\mp ia'_2, \nonumber \\
\exp(\pm 2L_3)={i(b'_3\pm 1)\over a'_3}.
\label{425}
\end{eqnarray}
The following lemma is principal in this section.

\begin{lemma} \label{4lemb}
If (\ref{422}) holds, the coefficients $K_1,
\ldots, L_3$
satisfy the following star--triangle relations (we reproduce
formulae \hbox{(6.4.8a--d)} from the book~\cite{Baxterbook}; coefficients
$K_j$ belong to a triangle, $L_j$ ---to a star, $R$ is some numeric
factor):
\begin{eqnarray}
2\cosh (L_1+L_2+L_3) &=& R\exp (K_1+K_2+K_3), \label{426} \\
2\cosh (-L_1+L_2+L_3)&=& R\exp (K_1-K_2-K_3), \label{427} \\
2\cosh (L_1-L_2+L_3)&=& R\exp (-K_1+K_2-K_3), \label{428} \\
2\cosh (L_1+L_2-L_3)&=& R\exp (-K_1-K_2+K_3). \label{429}
\end{eqnarray}
\end{lemma}

Lemma \ref{4lemb} is illustrated by Fig.~\ref{4figb}.
\bfig
\begin{center}
\unitlength=0.104606\normalbaselineskip
\special{em:linewidth 0.1pt}
\linethickness{0.1pt}
\begin{picture}(175.00,75.00)
\emline{5.00}{10.00}{1}{65.00}{70.00}{2}
\emline{65.00}{25.00}{3}{5.00}{25.00}{4}
\emline{50.00}{10.00}{5}{50.00}{70.00}{6}
\emline{115.00}{45.00}{13}{175.00}{45.00}{14}
\emline{130.00}{60.00}{15}{130.00}{0.00}{16}
\emline{175.00}{60.00}{17}{115.00}{0.00}{18}
\put(75.00,35.00){\vector(1,0){15.00}}
\special{em:linewidth 1.0pt}
\emline{0.00}{15.00}{7}{40.00}{35.00}{8}
\emline{40.00}{35.00}{9}{60.00}{15.00}{10}
\emline{40.00}{35.00}{11}{60.00}{75.00}{12}
\emline{100.00}{15.00}{19}{160.00}{15.00}{20}
\emline{160.00}{15.00}{21}{160.00}{75.00}{22}
\emline{160.00}{75.00}{23}{100.00}{15.00}{24}
\end{picture}
\end{center}
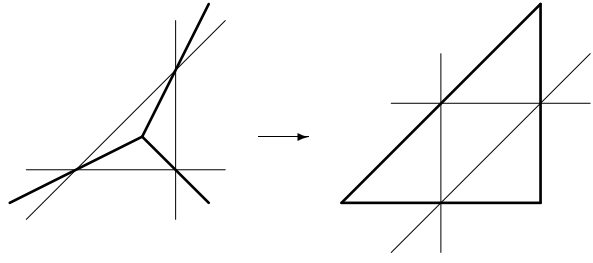
\caption{Star--triangle transformation is shown by thick lines,
while the corresponding matrix transformation---by  thin lines}
\label{4figb}
\bigskip
\begin{center}
\unitlength=0.104606\normalbaselineskip
\special{em:linewidth 0.1pt}
\linethickness{0.1pt}
\begin{picture}(175.00,75.00)
\emline{0.00}{15.00}{1}{60.00}{75.00}{2}
\emline{60.00}{30.00}{3}{0.00}{30.00}{4}
\emline{45.00}{15.00}{5}{45.00}{75.00}{6}
\emline{110.00}{50.00}{13}{170.00}{50.00}{14}
\emline{125.00}{65.00}{15}{125.00}{5.00}{16}
\emline{170.00}{65.00}{17}{110.00}{5.00}{18}
\put(85.00,40.00){\vector(1,0){15.00}}
\special{em:linewidth 1.0pt}
\emline{15.00}{60.00}{19}{75.00}{60.00}{20}
\emline{75.00}{60.00}{21}{15.00}{0.00}{22}
\emline{15.00}{0.00}{23}{15.00}{60.00}{24}
\emline{135.00}{40.00}{7}{175.00}{60.00}{8}
\emline{115.00}{60.00}{9}{135.00}{40.00}{10}
\emline{115.00}{0.00}{11}{135.00}{40.00}{12}
\end{picture}
\end{center}
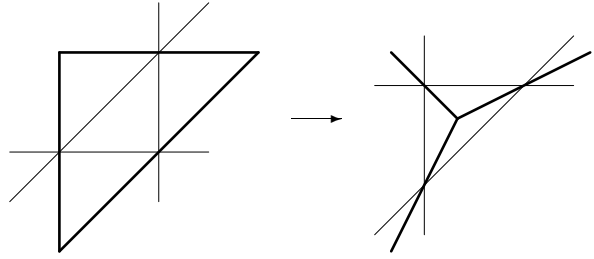
\caption{Triangle--star transformation is shown by thick lines,
while the corresponding matrix transformation---by  thin lines}
\label{4figbb}
\efig
\medskip

{\it Proof of Lemma \ref{4lemb}\/}  will be done using the following
parameterization of relations~(\ref{422}) with variables
$k,\lambda,\mu$, where $k$ is the modulus of all elliptic functions:
\be
a_1={\rm sn}\, \lambda,\quad a_2=k\,{\rm sn}\, (\lambda+\mu),\quad
a_3={\rm sn}\, \mu,
\label{430}
\ee
\be
b_1={\rm cn}\, \lambda,\quad b_2={\rm dn}\, (\lambda+\mu),\quad
b_3={\rm cn}\, \mu,
\label{431}
\ee
\be
a'_1=k\,{\rm sn}\, \lambda,\quad a'_2={\rm sn}\, (\lambda+\mu),\quad
a'_3=k\,{\rm sn}\, \mu,
\label{432}
\ee
\be
b'_1={\rm dn}\, \lambda,\quad b'_2={\rm cn}\, (\lambda+\mu),\quad
b'_3={\rm dn}\, \mu.
\label{433}
\ee
The fact that (\ref{430}--\ref{433}) is really a parameterization
of (\ref{422}) is verified directly using elementary properties
of elliptic functions. Then, from the same properties it follows that
formulae for
$K_2$ and $L_2$ from (\ref{424}) and (\ref{425}) may be rewritten as
follows:
\be
\exp(\pm2K_2)={\rm cn}\, \kappa\mp i{\rm sn}\, \kappa,\quad
\exp(\pm2L_2)={i({\rm dn}\, \kappa\pm1)\over k\,{\rm sn}\, \kappa},
\label{434}
\ee
where
\be
\lambda+\kappa+\mu=iI'.
\label{435}
\ee
Here we have denoted the half-period of elliptic functions by $I'$,
following R.~Baxter's book~\cite{Baxterbook}, although
it is usually denoted~$K'$. Our formulas
(\ref{424}) and (\ref{425}) coincide now, with regard to (\ref{434}),
with parameterization (7.8.5) from \cite{Baxterbook} for relations
(\ref{426}--\ref{429}), while (\ref{435}) coincides with formula (7.13.4)
from \cite{Baxterbook} if our $\lambda,\kappa,\mu $ equal Baxter's
$iu_1, iu_2, iu_3$. The lemma is proved.
\medskip

We have described the transformation (\ref{420}) by means of
orthogonal matrices, and illustrated this with Figure~\ref{4figb}.
Similarly, transformation~(\ref{421}) together with
the transformation of orthogonal matrices
is illustrated by Figure~\ref{4figbb}. The latter
differs from Fig.~\ref{4figb} in the fact that now to a kagome
lattice triangle
\unitlength=0.104606\normalbaselineskip
\special{em:linewidth 0.4pt}
\linethickness{0.4pt}
\begin{picture}(20.00,15.00)
\emline{0.00}{0.00}{1}{20.00}{0.00}{2}
\emline{0.00}{-5.00}{3}{20.00}{15.00}{4}
\emline{15.00}{15.00}{5}{15.00}{-5.00}{6}
\end{picture}
 \vs \quad  corresponds an Ising triangle,
while to a triangle 
\unitlength=0.104606\normalbaselineskip
\special{em:linewidth 0.4pt}
\linethickness{0.4pt}
\begin{picture}(20.00,15.00)
\emline{0.00}{-5.00}{1}{20.00}{15.00}{2}
\emline{5.00}{15.00}{3}{5.00}{-5.00}{4}
\emline{0.00}{10.00}{5}{20.00}{10.00}{6}
\end{picture}
 \vs\  corresponds a star.
One can reduce this situation to the previous one by {\em transposing\/}
the products of orthogonal matrices
(recall that matrices themselves are symmetric),
corresponding to LHS and RHS of Fig.~\ref{4figbb} and then changing
their places. Then it is clear that relations of the form
(\ref{424}, \ref{425}) between matrix elements and coefficients
$K_j$  and
$L_j$ for triangle and star respectively work in the case of
Fig.~\ref{4figbb} as well.

\end{document}